\documentclass[epj,nopacs]{svjour}
\usepackage{graphicx,color,amssymb}
\usepackage[numbers,sort&compress,merge]{natbib}
\usepackage{ulem}
\usepackage{soul}

\begin{document}

\newcommand{\lsim}{\raisebox{-0.13cm}{~\shortstack{$<$ \\[-0.07cm] $\sim$}}~} 
\newcommand{\gsim}{\raisebox{-0.13cm}{~\shortstack{$>$ \\[-0.07cm] $\sim$}}~} 
\newcommand{\beq}{\begin{eqnarray}} 
\newcommand{\eeq}{\end{eqnarray}} 
\newcommand{\tb}{\tan \beta}
\newcommand{\ee}{e^+e^-}
\newcommand{\lra}{\longrightarrow}
\newcommand{\ra}{\rightarrow}
\newcommand{\non}{\nonumber}
\newcommand{\s}{\smallskip} 

\pagestyle{plain}
\title{Higgs Properties and Supersymmetry}
\subtitle{Constraints and Sensitivity from the LHC to an $e^+e^-$ Collider}
\author{A. Arbey\inst{1,2},
        M. Battaglia\inst{2,3},
        A. Djouadi\inst{4,5},
        F. Mahmoudi\inst{1,2},
        M. M\"uhlleitner\inst{6} and
        M. Spira\inst{7}
}                     
%
%

\institute{Universit\'e de Lyon, Universit\'e Claude Bernard Lyon 1, CNRS/IN2P3,\\
Institut de Physique des 2 Infinis de Lyon, UMR 5822, F-69622, Villeurbanne, France \and
CERN, CH--1211 Geneva 23, Switzerland \and
Santa Cruz Institute of Particle Physics, University of California, Santa Cruz, CA 95064, USA \and
CAFPE and Departamento de F\'isica Te\'orica y del Cosmos, Universidad de Granada, E--18071 Granada, Spain \and
NICPB, R{\"a}vala pst. 10, 10143 Tallinn, Estonia \and
Institute for Theoretical Physics, Karlsruhe Institute of Technology,
76131 Karlsruhe, Germany \and
Paul Scherrer Institut, CH--5232 Villigen PSI, Switzerland}
\date{}

\abstract{ The study of the Higgs boson properties offers compelling perspectives for testing the effects of  physics beyond the Standard Model and has deep implications for the LHC program and future colliders. Accurate determinations of the Higgs boson properties can provide us with a distinctively precise picture of the Higgs sector, set tight bounds, and predict ranges for the values of new physics model parameters. In this paper, we discuss the constraints on supersymmetry that can be derived by a determination of the Higgs boson mass and couplings. We quantify these constraints by using scans of the 19-parameter space of the so-called phenomenological minimal supersymmetric Standard Model. The fraction of scan points that can be excluded by the Higgs measurements is studied for the coupling measurement accuracies obtained in LHC Run 2 and expected for the HL-LHC program and $e^+e^-$ colliders and contrasted with those derived from missing transverse energy searches at the LHC and from dark matter experiments.} 
\maketitle

\section{Introduction}
\label{sec:1}

The discovery \cite{discovery} of the Higgs boson \cite{hi64} at the LHC
has opened a vast program of studies of its fundamental properties
\cite{couplings}, allowing new and intensive tests of the Standard Model
(SM) of particle physics as well as indirect and tight constraints on
models of new physics beyond it. In this context, supersymmetric models
\cite{SUSY0,SUSY,HaberKane} were considered for a long time as the
most interesting benchmarks for new physics. In these models the
particle spectrum is more than doubled as every SM particle has a
partner of different spin and the Higgs sector is extended to contain
more states than the sole SM--like Higgs boson that has been observed at
the LHC.
Although present LHC studies set stringent bounds on the masses of the
new particles, to the extent where supersymmetry (SUSY) appears now to
be less ``natural" than initially thought, it is nevertheless still
worthwhile to keep using and studying it as it remains among the best
benchmarks for new physics searches and provides a rich laboratory
for testing the SM.
The determination of the Higgs boson mass and the measurement of its
couplings to SM fermions and gauge bosons with sufficient accuracy have
crucial implications for supersymmetry.
 
Indeed, while in the SM the properties of the Higgs particle are fixed
once its mass is determined, the contributions from the extended Higgs
sector and those of the additional SUSY particles may shift the
couplings of the SM--like neutral Higgs state and hence, its production
rates and decay branching fractions. A precision study of the mass and the production and decay rates is thus essential for
establishing the mechanism of electroweak symmetry breaking and of mass
generation, for exploring the contributions of new physics models to the
Higgs sector and for eventually setting constraints on their parameter
spaces. These constraints need to be compared to those obtained from
direct searches for the heavier Higgs bosons of the theory and for the
SUSY particles in channels with missing transverse energy (MET), as the lightest SUSY particle that always appears at the end of the decay chains is stable and undetectable in the model's most popular versions with R-parity conservation.

The bounds from these searches by the LHC experiments are already significant and will extend to heavier and heavier SUSY particles, if no signal is observed in the next LHC run (Run~3)  and in the high--luminosity LHC (HL-LHC) program. It is essential to assess the impact of the constraints derived from the Higgs property measurements, which will also improve in accuracy, on those scenarios that survive the tests of the SUSY direct searches. Conversely, it is important to understand how a given accuracy in the Higgs measurements may enable the reconstruction of the new physics model parameters, in the case where deviations from the SM predictions are observed.

The experience gained with the current analyses from the LHC Run~2 data provides us with firm guidelines for the evolution of the accuracy of the Higgs measurements and of the bounds from new particle searches with the larger datasets that are expected in the future. In defining the accuracy of the determination of the Higgs properties, systematic uncertainties, both theoretical and parametric, will play an important role and need to be properly accounted for. With results for most of the Higgs decay and production channels of interest now in hand and mass bounds set by a broad variety of SUSY searches, the time for a detailed assessment of the interplay between Higgs physics and SUSY at the LHC and beyond has come.   

After the LHC era, including the HL-LHC program, a new $e^+e^-$
collider promises to deliver measurements that are inherently more
precise and cover virtually all the Higgs decay channels. The importance
of these data and the requirements for their accuracy, to be considered
with the full set of LHC measurements and bounds already in hand, need
to be precisely evaluated.  Several analyses, some comprehensive and
others more focused, that address this important issue have appeared
quite recently \cite{recent-discussions}. 

In this paper, we attempt to answer these questions by considering two approaches.
First, we study the relation between the Higgs coupling modifiers, $\kappa_i$, and the fundamental SUSY parameters. Then, we explore the sensitivity of the Higgs measurements to SUSY by quantifying the fraction of the scenarios excluded by the Higgs measurements, the constraints on its parameters, but also the sensitivity to their values in case deviations are observed. The study is conducted in the framework of the so-called phenomenological Minimal Supersymmetric extension of the Standard Model (pMSSM) \cite{Djouadi:1998di}. The reduction of the viable pMSSM parameter space obtained by imposing the Higgs properties is compared to that derived from direct SUSY searches in the MET channels through the different stages of the LHC program as well as to the bounds derived from flavour physics and dark matter searches.

\section{Study of Higgs Properties in the pMSSM}
\label{sec:2}
\subsection{The pMSSM}

In the MSSM, two doublets of complex scalar fields of opposite hypercharge,  
$H_u$ and $H_d$, are required to break spontaneously the electroweak
symmetry leading to the presence of five Higgs states, two $CP$-even
Higgs bosons $h$ and $H$, where the former is considered to be the
lightest, a $CP$-odd Higgs state $A$, and two charged Higgs bosons
$H^\pm$. Because of SUSY constraints, the tree-level masses of the various Higgs bosons and their couplings depend only on two input parameters generally taken to be the pseudoscalar Higgs mass $M_A$ and the ratio of the two vacuum expectation values $\tb$. However, many other MSSM parameters will enter the radiative corrections 
to the Higgs sector, which are known to play an extremely important
role~\cite{CR-1loop,subh,CR-eff,CR-2loop,rad-cor,reviews-cor,Wagner+Lee}. In principle, all soft SUSY--breaking parameters, which in general are of ${\cal O}(100)$  in addition to those of the SM, become relevant. 

Hence, in the most general MSSM, the analysis of the Higgs sector is
tremendously complicated.  A phenomenologically more viable MSSM framework, the pMSSM, that is easier to use in practice, can be defined by adopting the following three assumptions:  first, all soft SUSY--breaking parameters are real, and there is no new source of $CP$-violation, second, the matrices for the sfermion masses and for the trilinear couplings are all diagonal implying no flavour change at tree level, and third, the soft SUSY--breaking masses and trilinear couplings of the first and second sfermion generations are the same at the electroweak symmetry breaking
scale. Making these assumptions will lead to only 22 input parameters in the pMSSM: 

\begin{itemize} 
\vspace*{-2mm}
\item[--] $\tan \beta$: the ratio of the two vacuum expectation values (vevs) of the two Higgs doublet fields, which is expected to lie in the range $1 \lsim \tb \lsim m_t/m_b$;

\item[--] $M_A$: the mass of the pseudoscalar Higgs boson that ranges from $M_Z$ to the
SUSY--breaking scale;

\item[--] $\mu$: the Higgsino (supersymmetric) mass parameter\- (which can have both
signs); 

\item[--] $M_1, M_2, M_3$: the bino, wino, and gluino mass parameters;

\item[--] $m_{\tilde{Q}}, m_{\tilde{t}_R}, m_{\tilde{b}_R},  m_{\tilde{L}},
 m_{\tilde{\tau}_R}$: the third generation sfermion mass parameters;

 \item[--] $A_t, A_b, A_\tau$: the third generation trilinear
couplings. 

\item[--] $m_{\tilde{q}}, m_{\tilde{u}_R}, m_{\tilde{d}_R},  m_{\tilde{l}}, m_{\tilde{e}_R}$: the first and second generation
  sfer\-mion mass parameters;

\item[--]  $A_u, A_d, A_e$: the first and second generation trilinear couplings;
\end{itemize} 

The first and second generation trilinear couplings $A_u$, $A_d$, and $A_e$ will only play a minor role in general and can be ignored in most
cases (and, if it is not the case, they can be equated to those of the third generation) so that at the end, one would have 19 basic parameters in practice.  This gives the model more predictability and offers an adequate framework for extensive phenomenological studies. 

\subsection{Higgs Masses and Couplings}

Let us now come back to the MSSM Higgs sector and discuss the Higgs masses and mixing angles. In the basis $(H_d,H_u)$, the $CP$-even Higgs  mass matrix  can be written  as\footnote{If the SUSY scale is very large, the evolution from this very high scale down to the electroweak scale could mix the quartic couplings of the MSSM Higgs sector in a non trivial way such  that the structure of the mass matrix at the low energy scale could be different from this expression. However, detailed studies in an effective  two Higgs doublet model, that is renormalisation group improved to resum the large logarithms involving the SUSY--breaking scale,  suggest that this assumption is justified in most cases \cite{Wagner+Lee}.} 
\beq
{\cal M}^2 &\! =\! &M_{Z}^2
\left(
\begin{array}{cc}
  c^2_\beta & -s_\beta c_\beta \\
 -s_\beta c_\beta & s^2_\beta \\
\end{array}
\right)
\!+\! M_{A}^2
\left(
\begin{array}{cc}
 s^2_\beta & -s_\beta c_\beta \\
 -s_\beta c_\beta& c^2_\beta \\
\end{array}
\right) 
\nonumber \\ &+&
\left(
\begin{array}{cc}
 \Delta {\cal M}_{11}^2~~ &  \Delta {\cal M}_{12}^2 \\
 \Delta {\cal M}_{12}^2~~ &\Delta {\cal M}_{22}^2 \\
\end{array}
\right)\;,~~~
\label{mass-matrix}
\eeq
where we use the short-hand notation $s_\beta \equiv \sin\beta$
etc.\ and introduce the radiative corrections through the general
$2 \! \times\! 2$ matrix $\Delta {\cal M}_{ij}^2$. The masses of the
neutral $CP$-even $h,H$ bosons and the mixing angle $\alpha$ that
diagonalises the two states can then be written as
\begin{eqnarray}
\hspace{-1.0cm}
M_{h/H}^2&=&\frac{1}{2} \big( M_{A}^2+M_{Z}^2+ \Delta {\cal M}_{+}^2  \mp  N \big)\;, \\
\hspace{-1.0cm}
\tan \alpha&=&\frac{2\Delta {\cal M}_{12}^2 - (M_{A}^2 + M_{Z}^2) s_{\beta}}
{ \Delta {\cal M}_{-}^2  + (M_{Z}^2-M_{A}^2)
c_{2\beta} + N
}\;,
\end{eqnarray}
with\\[-3mm] 
\begin{eqnarray}
&& \Delta {\cal M}_{\pm}^2  \! = \!  \Delta {\cal M}_{11}^2 \pm  
\Delta {\cal M}_{22}^2 \;, \nonumber  \\
&& N \!=\! \sqrt{M_{A}^4 + M_{Z}^4 - 2 M_{A}^2 M_{Z}^2 c_{4\beta} + C}\;, \nonumber \\
C &\!=\!&  4 \Delta {\cal M}_{12}^4\! + \!( \Delta {\cal M}_{-}^2)^2 \!- \! 
 2 (M_{A}^2 \! - \! M_{Z}^2)  \\ \nonumber  &\times & 
 \Delta {\cal M}_{-}^2 
 c_{2\beta} \!   - \!
 4 (M_{A}^2 \! + \!M_{Z}^2)  \Delta {\cal M}_{12}^2 s_{2\beta}\;. 
\end{eqnarray}
The leading radiative corrections to the Higgs mass matrix of
Eq.~(\ref{mass-matrix}) are controlled by the top Yukawa coupling,
$\lambda_t =  m_t/v \sin\beta$ with $v=246$ GeV, which appears with the
second power accompanied by two additional powers of the top mass. We obtain a very simple analytical expression for the correction matrix $\Delta {\cal M}_{ij}^2$ at one-loop if only this contribution 
is taken into account \cite{CR-1loop},
\beq
\label{higgscorr}
\Delta {\cal M}_{11}^2 \sim  \Delta {\cal M}_{12}^2 \sim 0 \ , \hspace*{3cm}\nonumber \\
\Delta {\cal M}_{22}^2 \sim   \frac{3 \overline{m}_t^4}{2\pi^2 v^2\sin^
2\beta} \left[ \log \frac{M_S^2}{\bar{m}_t^2} \!+ \! \frac{X_t^2}{M_S^2} \left( 1 \! -
\! \frac{X_t^2}{12M_S^2} \right) \right], 
\eeq
where $M_S$ is the geometric average of the two stop masses $M_S =\sqrt{ m_{\tilde{t}_1}
m_{\tilde{t}_2}} $ defined to be the SUSY--breaking scale and $X_{t}$ is the stop mixing 
parameter given by $X_t\!= \! A_t\! - \! \mu/\tb$ and
  $\bar{m}_t$ is the running ${\rm \overline{MS}}$ 
top quark mass at the scale $M_S$ to account for the leading two-loop QCD  corrections in a 
renormalisation-group improved approach. 

Other SUSY parameters than $X_t$ such as $\mu$
and $A_b$ and, in general, the corrections controlled by the bottom
Yukawa coupling $\lambda_b\! = \! m_b/v \cos\beta$ as well as the
gaugino mass parameters $M_{1,2,3}$, provide a small but non-negligible
correction to $\Delta {\cal M}_{ij}^2$ and can also have an impact on
the loop corrections \cite{subh,CR-eff,CR-2loop,reviews-cor}. 

At tree level, the lightest $h$ boson mass is bounded by $M_h \leq M_Z
|\cos 2 \beta| \leq M_Z \approx 91$ GeV, and is thus far from the
measured value at the LHC, $M_h=125.09 \pm 0.24$ GeV \cite{couplings}. The radiative  corrections have therefore to be rather large in order to attain this value. In the leading one-loop approximation above, 
the maximal value mass $M_h^{\rm max}$ is given  by 
\begin{eqnarray}
\label{Mh-decoup}
M_h^2 \stackrel{M_A \gg M_Z} \longrightarrow  M_Z^2 \cos^2 2 \beta +
\Delta {\cal M}_{22}^2 s_\beta^2\;,
\end{eqnarray}
and is obtained for the following  choice of SUSY parameters \cite{Mh-max}:
$i)$ a decoupling regime with heavy $A$ states, $M_A\! \sim \mathcal{O}$(TeV) in order to minimise 
Higgs mixing; 
$ii)$ large values of the parameter $\tb$, $\tb \gsim 10$, in order to maximise the tree-level contribution $M_Z |\cos 2 \beta|$; 
$iii)$ heavy stop squarks i.e. large $M_S$ values to enhance the logarithmic contributions;
$iv)$ a stop trilinear coupling of $X_t=\sqrt{6}M_S$, the so-called ma\-xi\-mal mixing scenario 
that maximises the stop loops \cite{Mh-max}.
If the parameters are optimised as above, the maximal $M_h$ value can then reach the level of the measured value $M_h\!=\!125\;$GeV for $M_S\! > \! 1\;$TeV.   

The basic feature of the $h$MSSM approach \cite{habemus,Rome,full-covering}, that we will adopt in most cases for our effective Higgs coupling study, is that we can trade the radiative correction $\Delta {\cal M}_{22}^2$ of Eq.~(\ref{higgscorr}) for the measured Higgs mass value $M_h\!=\!125$ GeV. In this case, the MSSM Higgs sector with solely the dominant
radiative corrections included, can be again described with only two
unknown parameters, such as $\tb$ and $M_A$ as it was the case at
tree-level. The dominant radiative corrections involving the SUSY
parameters are fixed by the value of $M_h$. This observation leads to a
rather simple and accurate\footnote{If the SUSY scale is not extremely high, 
the approach would need some refinements
at high values of $\tan\beta$ and $\mu$ to describe properly the Yukawa
couplings to bottom quarks and $\tau$ leptons as will be discussed
later. In addition, there are
subleading contributions to the Higgs mass matrix other than
$\Delta {\cal M}^2_{22}$, but these have been shown to be rather small as will also be discussed.}  parametrisation of the
MSSM Higgs sector and, more specifically, the heavier $CP$-even Higgs mass and the $CP$-even mixing angle can be expressed in terms of $M_A$, $M_h$, and $\tb$ as:
\begin{equation}
\begin{array}{l} 
\label{eq:MH}
M_{H}^2 = \displaystyle\frac{(M_{A}^2+M_{Z}^2-M_{h}^2)(M_{Z}^2 c^{2}_{\beta}+M_{A}^2
s^{2}_{\beta}) - M_{A}^2 M_{Z}^2 c^{2}_{2\beta} } {M_{Z}^2 c^{2}_{\beta}+M_{A}^2
s^{2}_{\beta} - M_{h}^2}\;, \\
\ \ \  \alpha = \displaystyle-\arctan\left(\frac{ (M_{Z}^2+M_{A}^2) c_{\beta} s_{\beta}} {M_{Z}^2
c^{2}_{\beta}+M_{A}^2 s^{2}_{\beta} - M_{h}^2}\right)\;.
\label{eq:alpha}
\end{array}
\label{wide} 
\end{equation}
The mass of the charged Higgs state $M_{H^\pm}$ is simply given by the tree-level relation 
\begin{eqnarray}
M_{H^\pm}^2 = 
 { M_A^2 + M_W^2}\;,
\label{eq:MH+}
\end{eqnarray}
as the SUSY radiative corrections in this particular case are known to
be small \cite{H+mass}. We can now discuss the production and decay
rates of the MSSM Higgs bosons, restricting for the moment to the
SM-like one $h$.

\subsection{$h$ Production and Decays}

In many respects, we are fortunate enough as the mass value of $M_h = 125$~GeV of the SM-like $h$ particle allows us to produce the state in several redundant channels and to detect it in a  variety of decay modes.

First, many production processes have significant rates for a light SM--like Higgs boson. 
The by far dominant gluon fusion mechanism, 
$$gg\to h$$ 
that we will denote ggh, develops a large cross section for $M_h=125$ GeV, $  \sigma^{\rm tot}_{\rm ggh} \approx\! 50$ pb at $\sqrt s\!=\! 14$ TeV. For such a Higgs mass,  the subleading channels, i.e. the vector boson fusion (VBF) process, 
$$qq \to hqq$$
and the Higgs--strahlung (hV) mechanisms,  
$$q\bar q \to hV$$ 
with $V=W,Z$, have cross sections at $\sqrt s\!=\! 14$ TeV that are of
the order of, respectively, $\sigma^{\rm tot}_{\rm VBF} \approx\! 4$ pb and $\sigma^{\rm tot }_{\rm hV} \approx\! 2.5$ pb when the two channels $hZ$ and $hW$ are combined. These rates would lead to large samples that would allow for a detailed  study of the Higgs particle with the large amount of integrated luminosity, ${\cal L} \approx 3000$ fb$^{-1}$, that is expected to be collected at the high--luminosity option of the LHC (HL--LHC). Even the associated Higgs production with top quark pairs (tth), 
$$p p\to t\bar t h$$ 
with a cross  section of $\sigma_{\rm tth}^{\rm tot} \approx\! 0.6$ pb and, to a much lesser extent, double Higgs production in the dominant gluon--fusion channel (gghh), 
$$gg \to hh$$ 
with a cross section of $\sigma^{\rm tot}_{\rm gghh} \approx\! 50$ fb
could be probed with such a high luminosity. 

Second, for $M_h = 125$ GeV, the Higgs boson mainly decays into $b \bar b$ pairs,
$$h \to b\bar b$$  with a branching ratio of $\approx 60\%$, but the decays into massive gauge boson final states, 
$$ h\to WW^* \ {\rm and}\  ZZ^*$$
before allowing the gauge bosons to decay leptonically $W \! \to \! \ell \nu$ and 
$Z\! \to \! \ell \ell$ ($\ell\! =\! e,\mu$), are also significant with branching ratios of  $\approx 20\%$ and 2.5\%, respectively. The leptonic decay channel, 
$$h\! \to \! \tau^+\tau^-$$
is also of significance with a branching fraction of $\approx 5\%$ as is the case for the $h\to gg$ ($\approx 8\%$) and $h \to c\bar c$ ($\approx 3\%$) decay modes (that are not detectable at the LHC to first approximation). The clean loop induced decay mode,
$$h\to \gamma  \gamma$$ 
can be easily detected albeit its small branching ratio of
$2 \times 10^{-3}$.  Even the rare $h\to \mu^+\mu^-$ decay with a
branching fraction of order  $2 \times 10^{-4}$ has provided first
evidence in the ATLAS~\cite{Aad:2020xfq} and CMS~\cite{CMS-PAS-HIG-19-006} latest searches, and the $h\to Z\gamma$ channel should be accessible at the HL-LHC.

\begin{figure}[!h]
\begin{center}
\begin{tabular}{cc}
\includegraphics[width=0.225\textwidth]{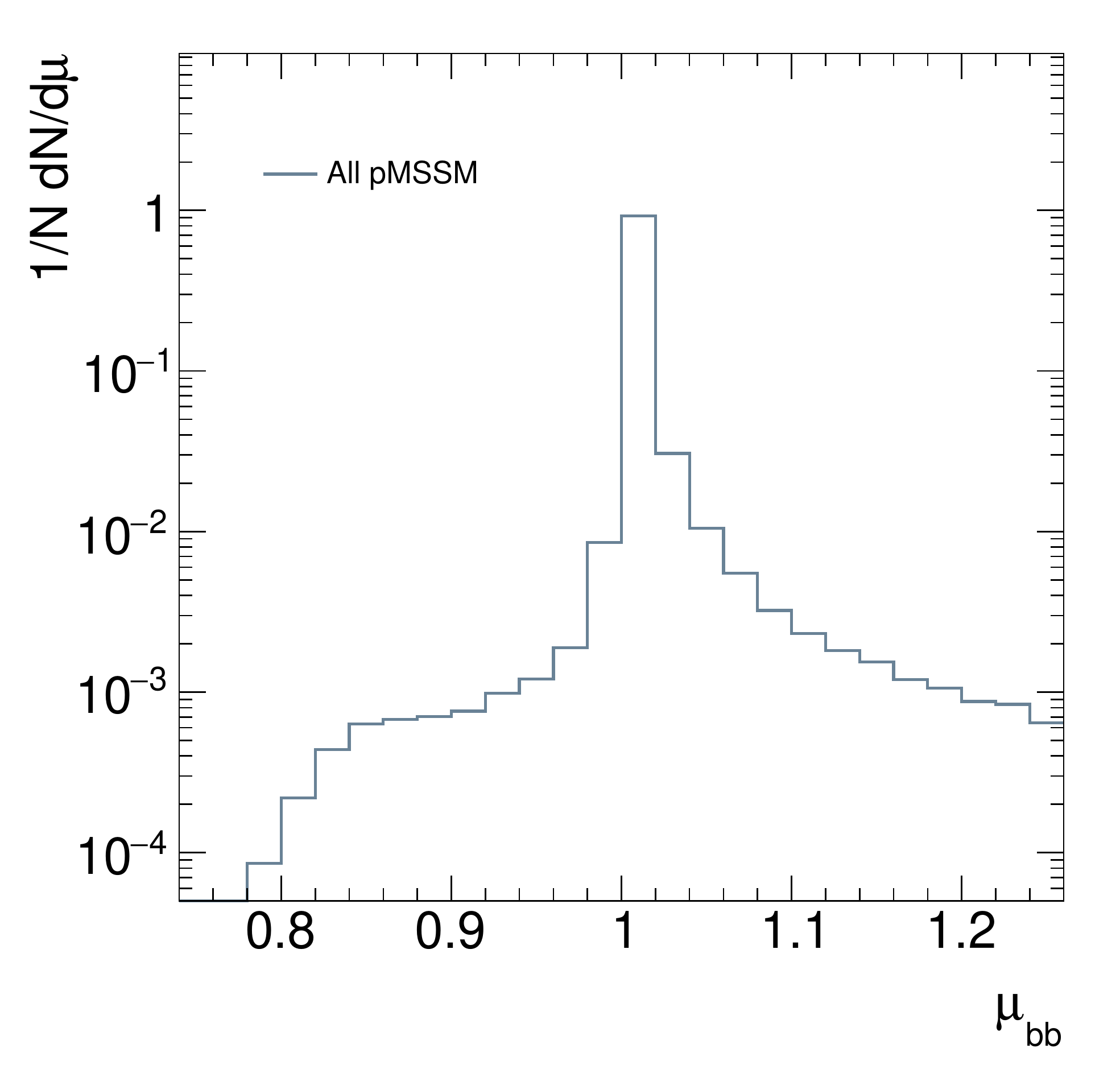} &
\includegraphics[width=0.225\textwidth]{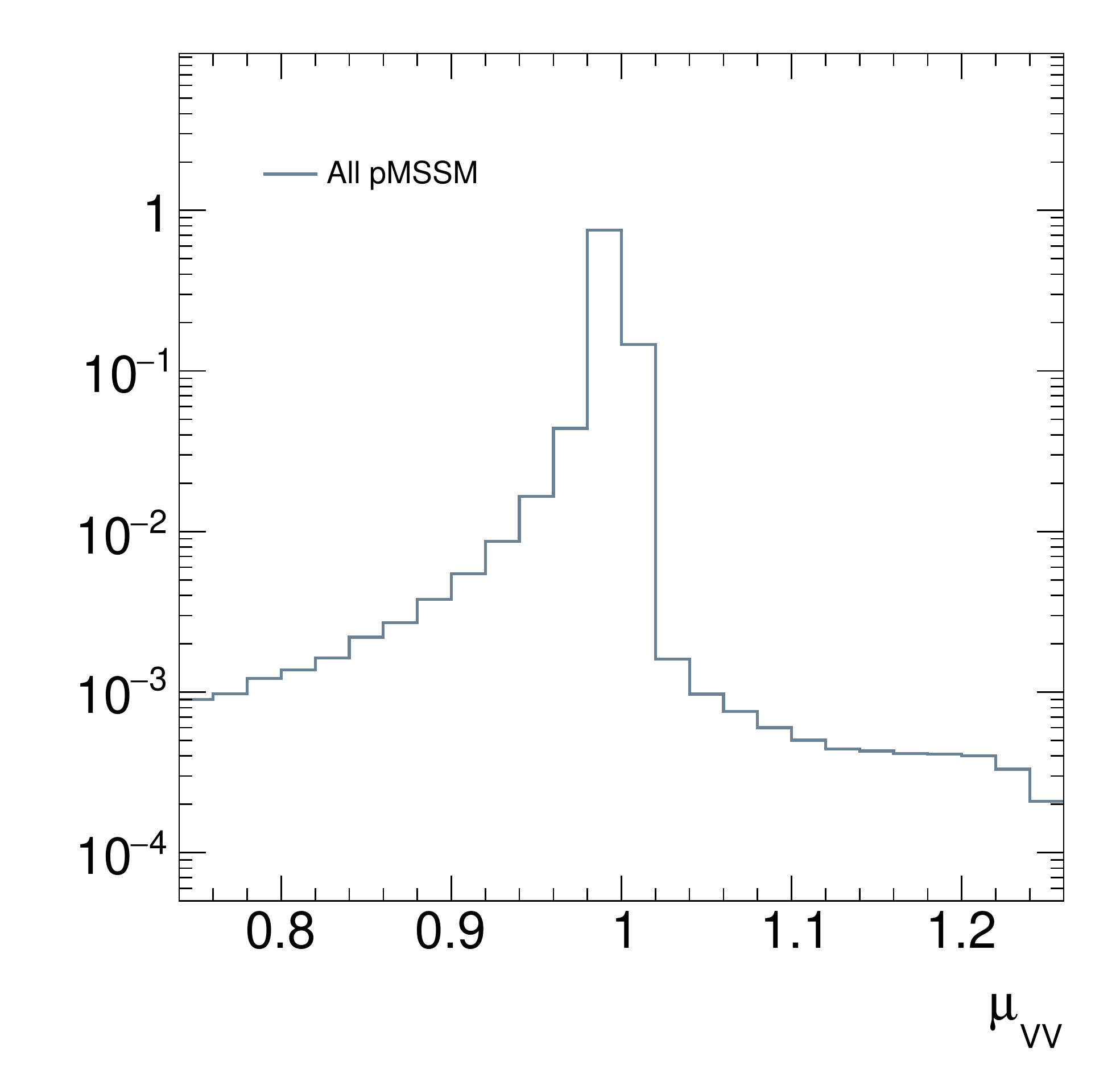} \\
\includegraphics[width=0.225\textwidth]{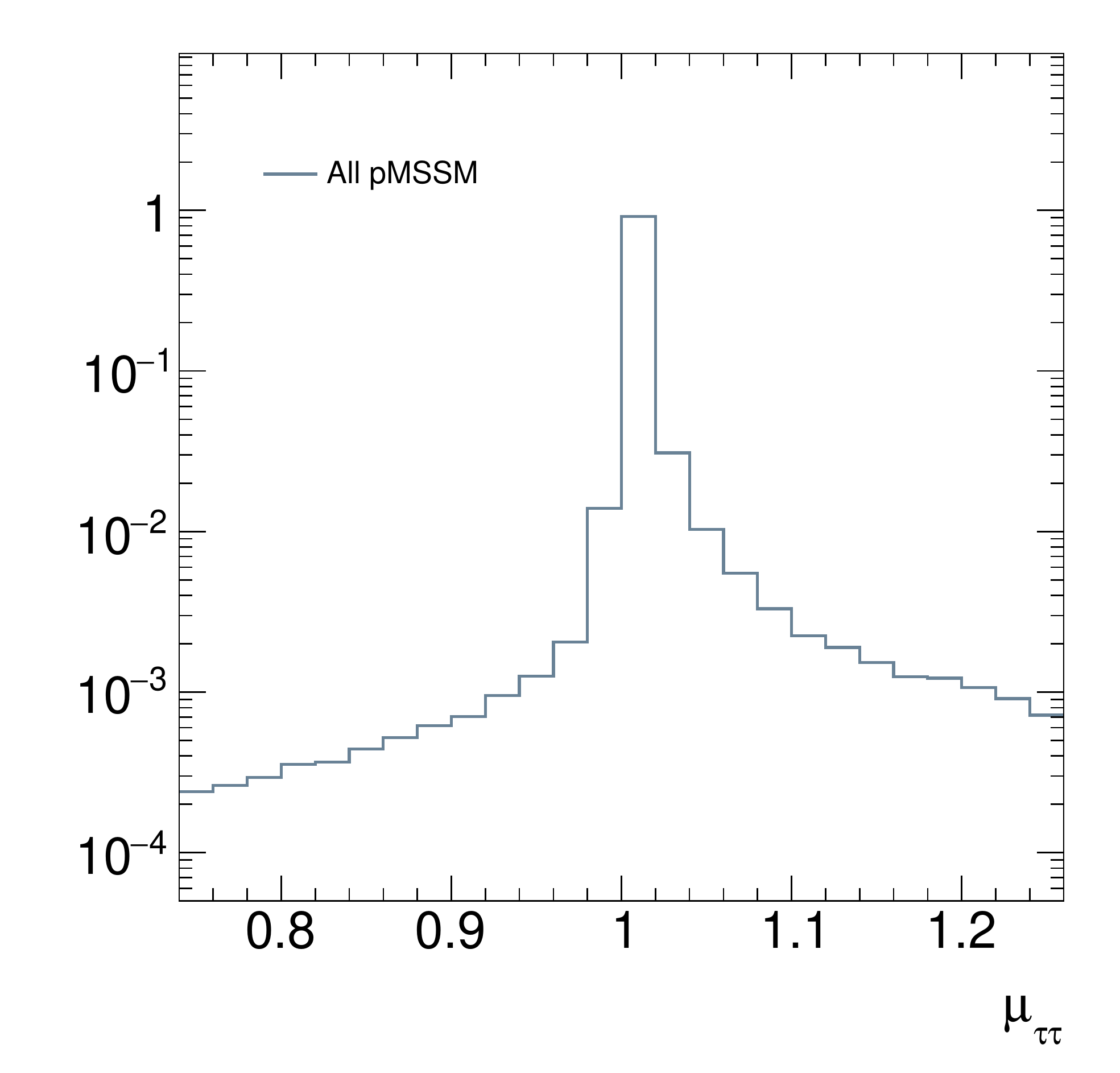} &
\includegraphics[width=0.225\textwidth]{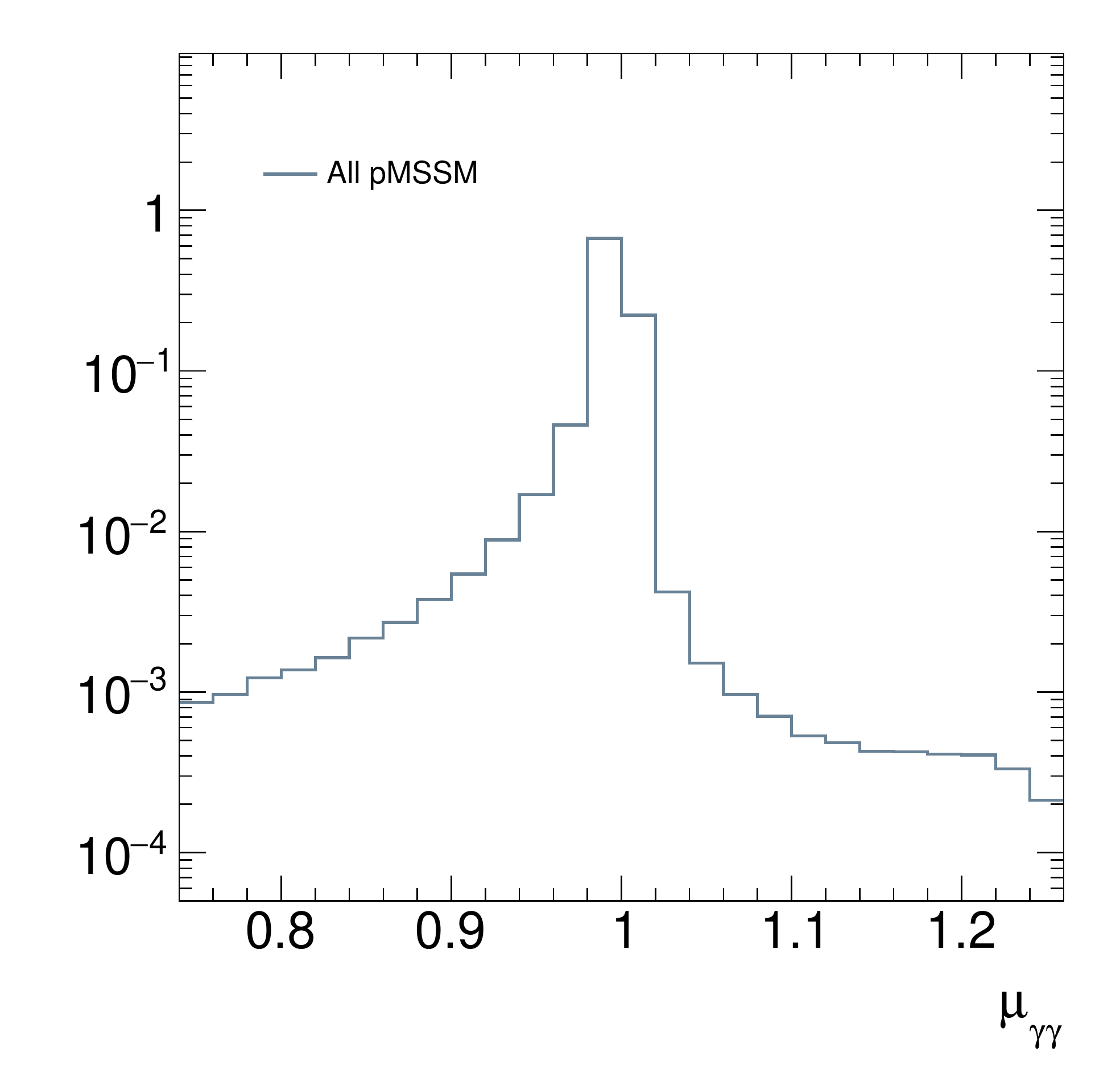} \\  
\end{tabular}
\end{center}
\begin{minipage}{8cm}
\caption{Distributions of $h$ decay branching fractions normalised to their SM prediction, $\mu$, for the $b \bar b$ (top), $W^+W^-$ and $ZZ$ (second from top), $\tau^+ \tau^-$ (bottom) and $\gamma \gamma$ (second from bottom) for pMSSM points.}
\label{fig:hbr1}
\end{minipage}
\end{figure}

\begin{figure}[!ht]
\begin{center}
\includegraphics[width=0.3\textwidth]{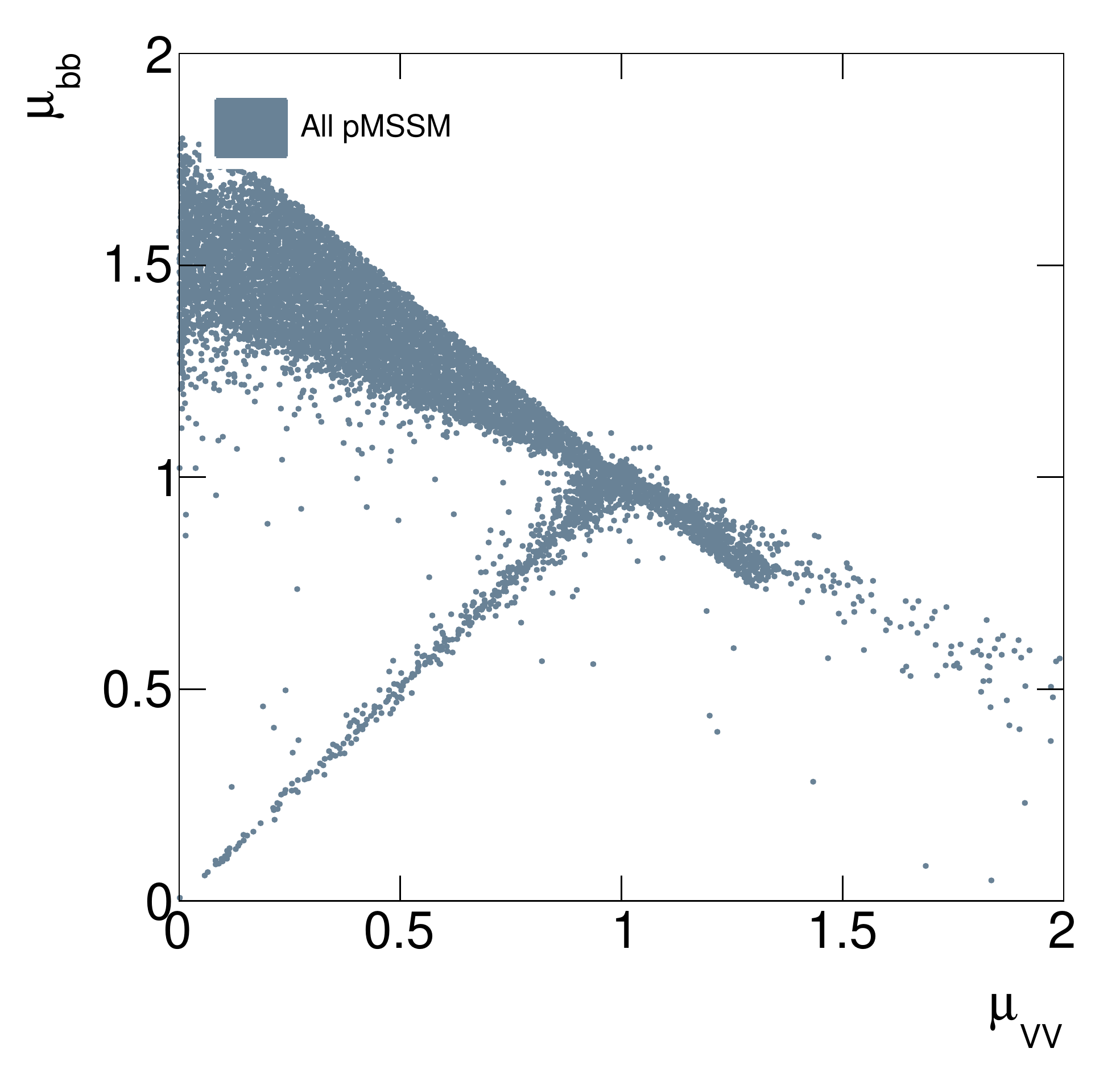} \\
\includegraphics[width=0.3\textwidth]{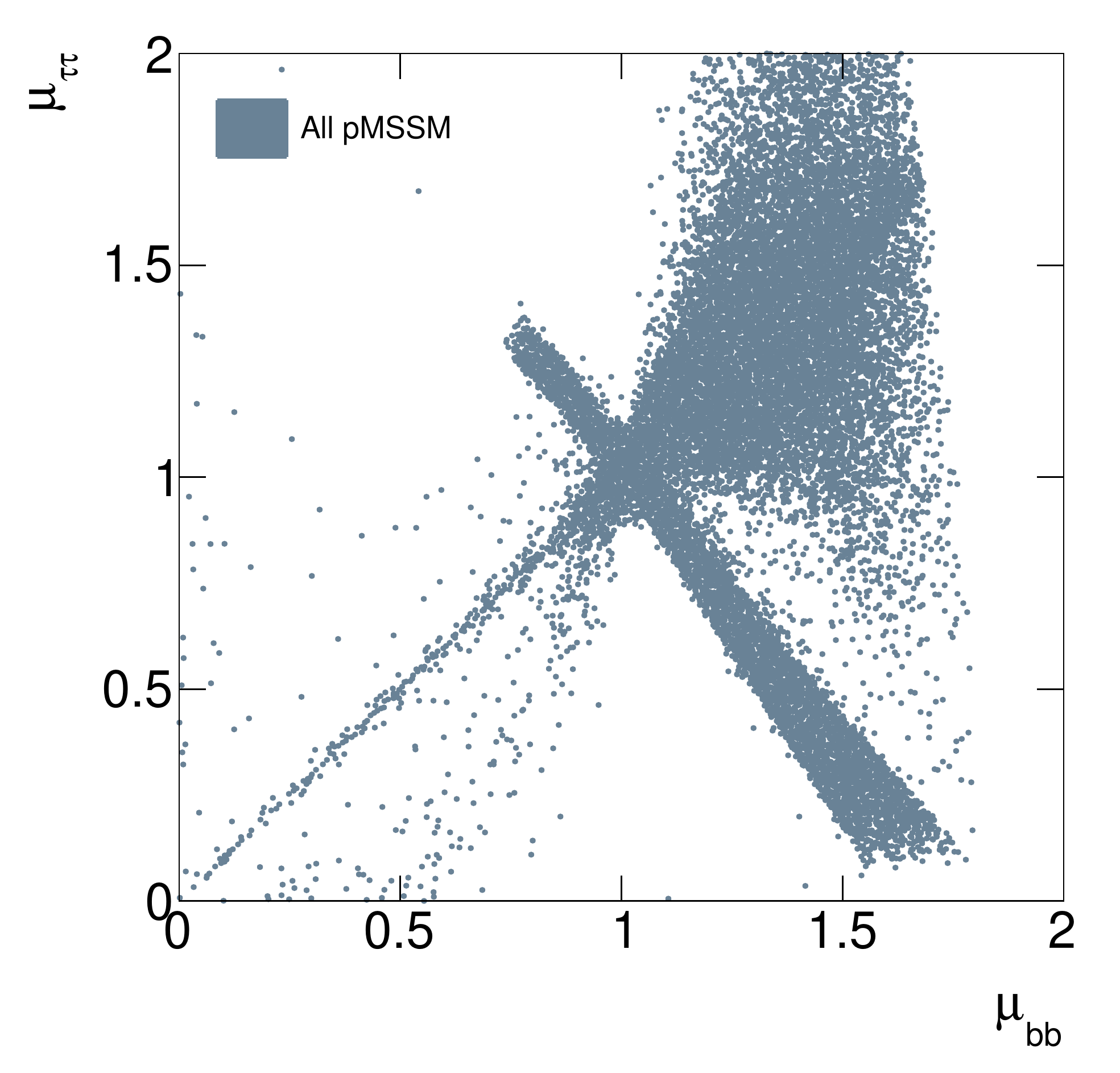} \\
\includegraphics[width=0.3\textwidth]{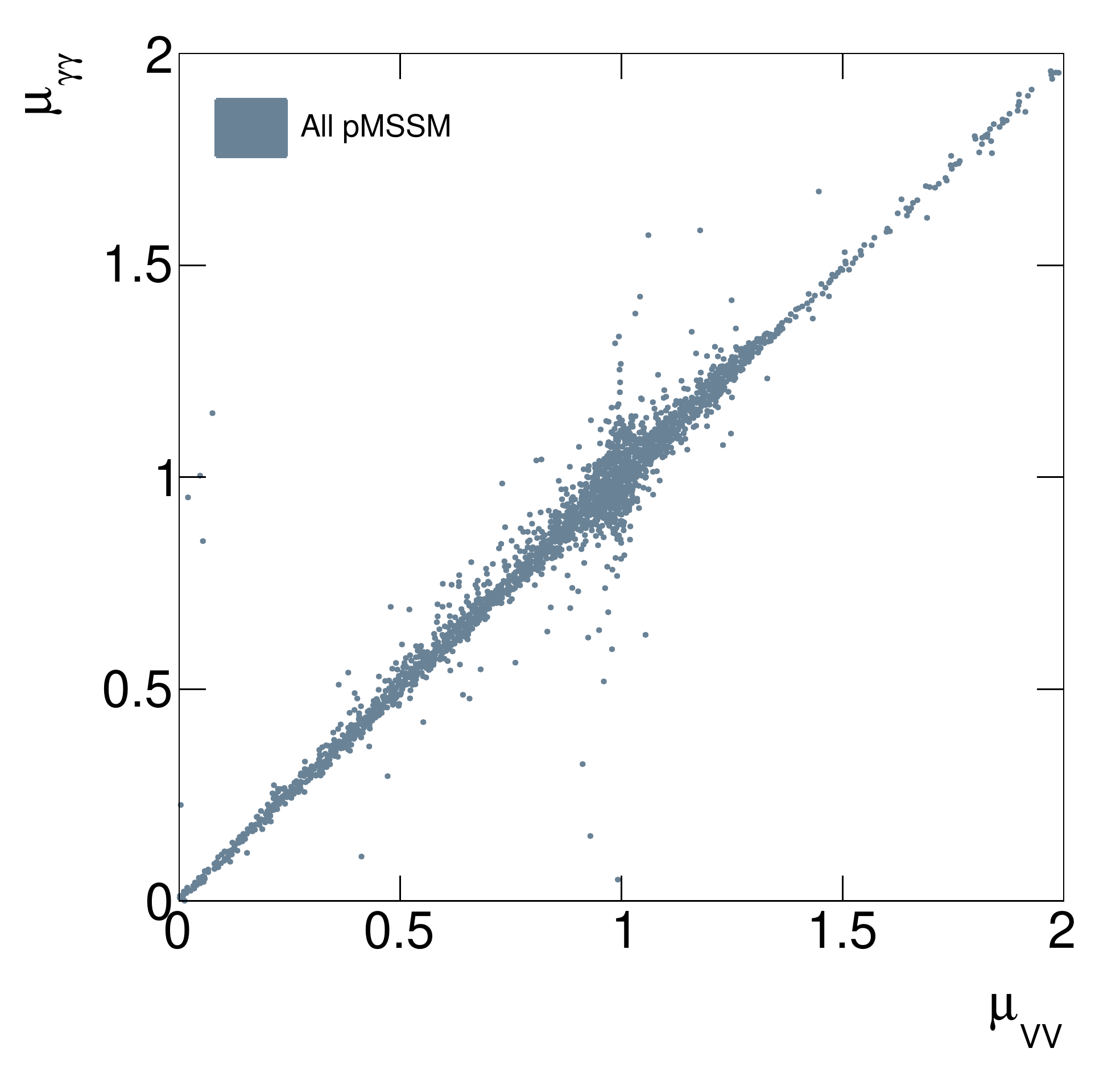} 
\end{center}
\begin{minipage}{8cm}
\caption{Distributions of the correlations between pairs of the $h$ decay branching fractions normalised to their SM predictions for pMSSM points.}
\label{fig:hbr2}
\end{minipage}
\vspace*{-7mm}
\end{figure}

The branching fractions of the $h$ bosons in our pMSSM scans in the $b
\bar b$,  $W^+W^-$ and $ZZ$ and $\tau^+ \tau^-$ decay channels
normalised to the SM predictions are shown in Figure~\ref{fig:hbr1}.
Some of their correlations are given in Figure~\ref{fig:hbr2}. They have
been obtained with the program {\tt HDECAY} \cite{hdecay}, used in this study to precisely evaluate the various Higgs partial decay widths and branching ratios and show a broad range of variation away from the SM predictions.

\subsection{The Invisible Higgs Decay Width}
\label{sec:2-5}

In many extensions of the SM, the light scalar Higgs boson can decay
into pairs of non-SM particles. In the case of the MSSM, Higgs decays
into squarks and gluinos are kinematically excluded as present
experimental bounds constrain these particles to be much heavier than
$\frac12 M_h$. These decays are also kinematically closed for the
charged sleptons, the two charginos, and the three heavier neutralinos
as searches from the LEP experiment have set limits beyond 100 GeV on
the masses of these particles. Invisible $h$ boson decays are still
possible in the context of a fully unconstrained MSSM in which the
various soft SUSY-breaking parameters are unrelated and they are of two types.

The first kinematically still possible SUSY mode for the $h$ boson is
the decay into a pair of the lightest neutralino $\chi_1^0$ which, in
most cases, is the lightest supersymmetric particle (LSP) and is stable
if R-parity is conserved. This is largely allowed by present
experimental constraints in particular in non-constrained or
non-unified models in which the soft SUSY-breaking gaugino mass
parameters $M_1, M_2, M_3$ are not related and the relatively strong
experimental bounds on the masses of the charginos, $m_{\chi_1^\pm }
\gsim 100$ GeV from LEP2, and the gluino, $m_{\tilde g} \approx M_3
\gsim 1$ TeV from present LHC searches, do not affect the invisible neutralino, and one could have $m_{\chi_1^0} < \frac12 M_h$. 

The partial width for the invisible Higgs decay into neutralinos is given by
\beq 
\Gamma (h \to \chi^0_1 \chi^0_1)= \frac{G_F M_W^2 M_h}{2 \sqrt{2} \pi}
\ g^2_{h \chi^0_1 \chi^0_1} \ \beta^3_{\chi}\;, 
\eeq 
with the neutralino velocity being $\beta_\chi=  (1 - 4m^2_{\chi_1^0}/M_h^2)^{1/2}$ and the normalised Higgs-neutralino coupling given by 
$$g_{h \chi_1^0 \chi_1^0} = (Z_{12}-\tan\theta_W Z_{11})(\sin \beta Z_{14} -\cos \beta Z_{13})\;,$$
where $Z$ is the $4\times 4$ matrix that diagonalises the neutralino
mass matrix \cite{HaberKane}. The decay is important only for
moderate and comparable values of the bino mass parameter $M_1$ and the Higgsino parameter $\mu$:~moderate $M_1 \lsim {\cal O}(60~{\rm GeV})$  to have a light enough LSP as one has $m_{\chi_1^0} \approx  M_1$ for $M_1 \lsim |\mu|, M_2$ and comparable values $|\mu| ={\cal O}( M_1)$ as the $h$ boson prefers to couple to neutralinos, which are a mixture of gauginos and Higgsinos\footnote{Note, however, that the LEP2 bound on the lightest chargino mass forces the parameters $M_2$ and $\mu$ to be somewhat large, min$(|\mu|, M_2) \approx m_{\chi_1^\pm} \gsim 100$ GeV.}, $Z_{13},Z_{14} = {\cal O}(1)$.  In this parameter range, the decay $h \to \chi^0_1 \chi^0_1$ can be substantial if $M_h$ is above the $2m_{\chi^0_1}$ threshold; close to this value, the width is strongly suppressed by the $\beta^3_\chi$ velocity factor. 

The neutralino LSP with such a mass  would have the relic density
required by PLANCK results for $\Omega_{\mathrm{CDM}}$,
since it will annihilate efficiently through the exchange of the $h$ boson \cite{DM-Hpole}. However, in this case, the invisible branching fraction should be relatively small, BR$(h\to \chi_1^0 \chi_1^0) \lsim {\cal O}(10\%)$ as discussed, for instance, in Ref.~\cite{Arbey-etal-previous}.  

Scenarios in which $\mu$ and $M_1$ are small enough to lead to a light
LSP with reasonable couplings to the $h$ boson would lead to large rates for chargino-neutralino pair production at the LHC (in particular, a large amount of trilepton events from the process $q \bar q  \to W^* \to \chi_1^\pm \chi_i^0 \to WZ \chi_1^0 \chi_1^0 \to 3\ell^\pm+ E_T^{\rm mis}$). However, because these states have rather compressed spectra, with the $\chi_1^\pm, \chi_i^0$ masses too close to that of the LSP, the missing energy is small and the process could escape observation at the LHC.

Another, albeit less likely, possibility would be that the Higgs boson decays into a pair of sneutrinos which, if they are lighter than the chargino $\chi_1^\pm$ and the second neutralino $\chi_2^0$ and the sleptons, would have as only possible decay the channel $\tilde{\nu} \to \nu \chi_1^0$ with $\chi_1^0$ the LSP and the decay is thus also invisible. As the experimental lower bound on the $\tilde{\nu}$ masses is rather low, $m_{\tilde{\nu}} \gsim 45$ GeV from the invisible $Z$ 
decay width \cite{pdg}, there is a tight room $\frac12 M_Z \lsim
m_{\tilde{\nu}} \lsim \frac12 M_h$, for the decay $h \to \tilde{\nu}
\tilde{\nu}$ to occur. However, as a result of SU(2)$_{\rm L}$
invariance, the soft SUSY-breaking sneutrino and left-handed charged
slepton mass parameters are related and to cope with the bound on
slepton masses from LEP2 searches, $m_{\tilde{l}} \gsim 90$ GeV, the
mass  $m_{\tilde{\nu}_L}$ should be high enough. Nevertheless, a small
room is still possible for relatively light sneutrino as a splitting
between $\tilde{\nu}$ and $\tilde{l}_L$ masses can be generated by the
D~terms which, for small values of the common scalar mass $\tilde{m}$
and for large $\tb$ values for which they become maximal, govern the
slepton masses (note that additional significant contributions to the D~terms could arise beyond the MSSM). As these D~terms tend to increase $m_{\tilde{\ell}_L}$ and decrease $m_{\tilde{\nu}}$, sneutrino masses $m_{\tilde{\nu}} \lsim \frac12 M_h$ can be obtained while keeping  the $m_{\tilde{\ell}} \gsim$ 90~GeV experimental constraint still valid.

The partial width for the decay mode, summing over the three possible sneutrinos, is given by
 \beq 
\Gamma (h \to \tilde{\nu} \tilde{\nu} )= \frac{3 G_F M_Z^4}{8 \sqrt{2} \pi M_h}
g^2_{h \tilde{\nu} \tilde{\nu}}  \ \beta_{\tilde{\nu}} \ , \ \ \beta_{\tilde{\nu}} = 
\left[ 1 - \frac{4 m^2_{\tilde{\nu}}}{M_h^2} \right]^{1/2} 
\label{h-sneut}
\eeq 
The $h$ boson coupling to sneutrinos, $g_{h \tilde{\nu} \tilde{\nu}} \propto \cos 2\beta$ in the decoupling limit, is also maximal at high $\tb$ and is much larger than the bottom-quark Yukawa coupling, making the partial width huge. When the decay is not kinematically suppressed, it would dominate all other decays in contrast to current LHC observation. If the possibility of light sneutrinos is to hold, one thus needs to strongly suppress this decay channel and bring it at the few 10\% level at most. Two possibilities are then at hand. A first one is that the sneutrino mass is close to the kinematical threshold $m_{\tilde{\nu}} \approx \frac12  M_h^2$ so that the partial decay width in Eq.~(\ref{h-sneut}) is strongly suppressed by the sneutrino velocity $\beta_{\tilde \nu}$. A second possibility would be to suppress the $h \tilde{\nu} \tilde{\nu}$ couplings and hence, adopt a value of $\tb$ that is very close to unity for which $\cos 2\beta \to 0$. 
Both options can coexist of course.\smallskip 

Another possibility would be Higgs decays into the very light gravitinos
and the next-to-LSP neutralino $\chi_1^0$ in gauge mediated
SUSY-breaking  models, $h \to \chi_1^0 \tilde G$, with a very long-lived next-to-LSP that decays outside the detector making the photon in
the decay $\chi_1^0 \to \tilde G \gamma$ unobservable. However, for the $h \to \chi_1^0 \tilde G$ rate to be substantial, the scale for SUSY-breaking should be rather low, and this is already excluded in most realistic scenarios \cite{hgravitino}. Nevertheless, one should keep this possibility in mind in more complex models.   

Concretely, all these cases can be parameterised by an invisible Higgs partial width: 
\beq
\Delta_{\rm inv} \! = \! \Gamma_h^{\rm inv}/\Gamma_h^{\rm tot}= \!
\Gamma (h \to {\rm invisible})/\Gamma_{\rm SM}(h\to {\rm all})\;. \quad
\eeq
For scenarios where the decay channel is open, the rate into invisible particles is expected to be mostly at the few percent to the 10\% level, and they cannot yet be excluded by the LHC data~\cite{Sirunyan:2018owy,Aaboud:2019rtt,ATLAS-CONF-2020-008, ATLAS-CONF-2020-052}.

\subsection{Direct Corrections to the Higgs Couplings}
\label{sc:hcoup}

Higgs decays into quark pairs are strongly affected by (SUSY-)QCD
corrections, while (SUSY-)electroweak corrections are in general
of moderate size \cite{hffelw,hffselw}. The dominant part of the QCD corrections to the decays into bottom and charm quarks can be absorbed in the running Yukawa coupling if it is evaluated at the scale of the Higgs mass \cite{hffqcd}. This requires the introduction of the running
quark masses that are defined in the $\overline{\rm MS}$ scheme in the
program {\tt HDECAY} \cite{hdecay}. The pure QCD corrections beyond the
running-mass effects are included up to N$^4$LO \cite{hffqcd}. The
SUSY-QCD corrections mediated by gluino-squark exchange are fully
known up to NLO \cite{hffselw,hffsqcd} and included in {\tt HDECAY}. The dominant part of the SUSY-QCD and SUSY-electroweak corrections in the Higgs decays into
bottom quarks can be approximated by the $\Delta_b$ terms that at
one-loop order are given by
\cite{deltab,deltabr,deltabr1}
\begin{eqnarray}
\Delta_b & = & \Delta_b^{QCD} + \Delta_b^{EW,t} + \Delta_b^{EW,1} +
\Delta_b^{EW,2} \;,
\end{eqnarray}
with the individual contributions,
\begin{eqnarray}
\Delta_b^{QCD} & = &
\frac{2}{3}~\frac{\alpha_s}{\pi}~m_{\tilde g}~\mu~\tan\beta~
I(m^2_{\tilde b_1},m^2_{\tilde b_2},m^2_{\tilde g})\;, \nonumber \\
\Delta_b^{EW,t} & = &
\frac{\lambda_t^2}{(4\pi)^2}~A_t~\mu~\tan\beta~
I(m_{\tilde t_1}^2,m_{\tilde t_2}^2,\mu^2)\;, \nonumber \\
\Delta_b^{EW,1} & = &
-\frac{\alpha_1}{12\pi}~M_1~\mu~\tan\beta~\bigg\{ \frac{1}{3}I(m_{\tilde b_1}^2,
m_{\tilde b_2}^2,M_1^2) \nonumber \\
& & +\bigg( \frac{c_b^2}{2}+s_b^2\bigg)I(m_{\tilde b_1}^2,M_1^2,\mu^2) \nonumber \\
&& +\bigg( \frac{s_b^2}{2}+c_b^2\bigg) I(m_{\tilde b_2}^2,M_1^2,\mu^2)
\bigg\}\;, \nonumber \\
\Delta_b^{EW,2} & = & -\frac{\alpha_2}{4\pi}~M_2~\mu~\tan\beta~\bigg\{
c_t^2 I(m_{\tilde t_1}^2,M_2^2,\mu^2) \nonumber \\
&& \hspace*{2.8cm}+ s_t^2 I(m_{\tilde t_2}^2,M_2^2,\mu^2) \;,
 \nonumber \\
& & +\frac{c_b^2}{2} I(m_{\tilde b_1}^2,M_2^2,\mu^2)
+\frac{s_b^2}{2} I(m_{\tilde b_2}^2,M_2^2,\mu^2)
\bigg\} \;, \qquad
\end{eqnarray}
where $\alpha_s$ denotes the strong coupling,
$\lambda_t=\sqrt{2}m_t/(v\sin\beta)$ the top Yukawa coupling,
$\alpha_1 = {g'}^2/4\pi$ and $\alpha_2 = {g}^2/4\pi$ the electroweak gauge couplings. The masses $m_{\tilde g}, m_{{\tilde b}_{1,2}}$, and $m_{{\tilde t}_{1,2}}$ are the gluino, sbottom, and stop masses. The terms $s/c_{t,b} = \sin/\cos \theta_{t,b}$ are related to the stop/sbottom mixing angles $\theta_{t,b}$. The generic function $I$ is
defined as
\begin{equation}
I(a,b,c) = \frac{\displaystyle ab\log\frac{a}{b} + bc\log\frac{b}{c}
+ ca\log\frac{c}{a}}{(a-b)(b-c)(a-c)} \;.
\end{equation}

The corresponding $\Delta_\tau$ term for the tau-lepton couplings acquires contributions from the EW~gauge couplings $\alpha_1$ and $\alpha_2$ only. These are given by \cite{deltab}
\begin{eqnarray}
\Delta_\tau & = & \Delta_\tau^{EW,1} + \Delta_\tau^{EW,2} \;,
\label{eq:deltatau0}
\end{eqnarray}
with the individual contributions,
\begin{eqnarray}
\label{eq:deltatau}
\Delta_\tau^{EW,1} & = &
\frac{\alpha_1}{4\pi}~M_1~\mu~\tan\beta~\bigg\{
I(m_{\tilde \tau_1}^2,m_{\tilde \tau_2}^2,M_1^2) + \bigg( \frac{c_\tau^2}{2}-s_\tau^2 \bigg) \nonumber \\ &&
\times I(m_{\tilde \tau_1}^2,M_1^2,\mu^2)
+\bigg( \frac{s_\tau^2}{2}-c_\tau^2\bigg)
I(m_{\tilde \tau_2}^2,M_1^2,\mu^2)
\bigg\}\;, \nonumber \\
\Delta_\tau^{EW,2} & = &
-\frac{\alpha_2}{4\pi}~M_2~\mu~\tan\beta~\bigg\{
I(m_{\tilde \nu_\tau}^2,M_2^2,\mu^2)  \nonumber \\
& &+ \frac{c_\tau^2}{2} I(m_{\tilde \tau_1}^2,M_2^2,\mu^2)
+ \frac{s_\tau^2}{2}
I(m_{\tilde \tau_2}^2,M_2^2,\mu^2) \bigg\} \;,
\end{eqnarray}
where $s/c_\tau = \sin/\cos \theta_\tau$ is related to the $\tilde \tau$ mixing angle $\theta_\tau$ and $m_{\tilde{\tau}_{1,2}}, m_{\tilde{\nu}_\tau}$ denote the stau and tau sneutrino masses, respectively.

These $\Delta_f~(f=b,\tau)$ terms modify the effective bottom and $\tau$
Yukawa couplings $\tilde g^\phi_f~(\phi=h,H,A)$ of the two neutral $CP$-even states $h,H$ and the $CP$-odd state $A$, as follows
\cite{deltabr,deltabr1}:\footnote{Analogous corrections emerge for the muon and strange Yukawa couplings, but they do not play a role in our analysis.}
\begin{eqnarray}
\tilde g^h_f & = & \frac{g^h_f}{1+\Delta_f}\left[ 1 -
\frac{\Delta_f}{\tan\alpha\tan\beta}  \right]\;, \nonumber \\
\tilde g^H_f & = & \frac{g^H_f}{1+\Delta_f}\left[ 1 + \Delta_f
\frac{\tan\alpha}{\tan\beta} \right]\;, \nonumber \\
\tilde g^A_f & = & \frac{g^A_f}{1+\Delta_f}\left[ 1 -
\frac{\Delta_f}{\tan^2\beta} \right] \;,
\label{eq:gqresum}
\end{eqnarray}
in terms of the original Yukawa couplings $g^\phi_f$,
\begin{eqnarray}
g^h_u & = \displaystyle \frac{\cos\alpha}{\sin\beta}\;,\quad g^h_d = &
-\frac{\sin\alpha}{\cos\beta}\;, \nonumber \\
g^H_u & = \displaystyle \frac{\sin\alpha}{\sin\beta}\;,\quad g^H_d = &
\frac{\cos\alpha}{\cos\beta}\;, \nonumber \\
g^A_u & = \cot\beta\;,\quad g^A_d = & \tan\beta\;,
\label{eq:gfhlo}
\end{eqnarray}
for up- and down-type fermions,
where $\alpha$ is the mixing angle of the neutral $CP$-even Higgs states.
It has been shown that the effective bottom Yukawa couplings absorb the
bulk of the SUSY-QCD and -EW~corrections to most of the production and
decay processes mediated by these couplings up to a remainder of a few
percent, while the full SUSY-QCD corrections can reach about 100\%. The
two-loop QCD corrections to $\Delta_b$ have been calculated and shown to
add a moderate correction of about 10\% \cite{deltab2}. They are
included in {\tt HDECAY}
\cite{hdecay} thus increasing the reliability of the predictions even in cases of large $\Delta_b$ terms.

Similarly to the isospin down-type fermions, direct corrections also
affect the $ht\bar t$ couplings. Contrary to the terms discussed so far, these are suppressed by $\tb$ and hence, can be important only at very low $\tb$, $\tb \approx 1$. The corresponding $\Delta_t$ correction is given by
\begin{eqnarray}
\Delta_t & = &  \mu \cot \beta \left[ \frac{2\alpha_s}{3\pi}m_{\tilde{g}}
 I(m_{\tilde{t}_1}^2,m_{\tilde{t}_2}^2,m_{\tilde{g}}^2) \right.
\nonumber \\
& & \hspace*{1cm} \left. + \frac{\lambda_b^2}{(4\pi)^2} A_b
I(m_{\tilde{b}_1}^2,m_{\tilde{b}_2}^2,\mu^2) \right]\;,
 \label{eq:deltat}
\end{eqnarray}
and hence, only the first term would contribute since $\lambda_b
=\sqrt{2} m_b/(v\cos\beta)$ is
small as there is no relative enhancement by $\tb$ values.
However, for the same reason, these
corrections turn out to be small in total as they are not enhanced by
$\tan\beta$ factors at all.

For the decays into photon pairs, the additional SUSY-loop contributions mediated by chargino, sfermion, and charged Higgs loops are taken into account. In this way, a significant dependence on the SUSY parameters is induced that can lead to sizeable modifications of the corresponding partial width from the SM expression, if SUSY particles are relatively light. The QCD and EW corrections are small for the light scalar Higgs decay into photons \cite{hgagaqcd,hgagaelw}. Similar features also apply to the less
relevant Higgs decays into $Z\gamma$ \cite{hzgaqcd} and, for the EW
corrections, into $gg$ \cite{hggelw}. In the latter case of gluonic
decays, the QCD corrections are large \cite{hggqcd}, however, and are
included in {\tt HDECAY} \cite{hdecay}. The QCD and EW corrections to
the $H\to Z\gamma$ decay width are not included in {\tt HDECAY}.

Within the MSSM there are novel Higgs decays as, e.g.,~the heavy scalar
Higgs decay into a pair of two light $CP$-even Higgs bosons $H\to hh$ that plays a role for small values of $\tan\beta$ below the $t\bar t$ threshold \cite{H2hh}. Higher order corrections to this decay mode are only taken into account for the effective trilinear Higgs coupling,
obtained in the framework of the RG-improved effective potential, while process-dependent corrections are not included \cite{H2hhselw}.

Finally, the MSSM Higgs bosons can also decay into SUSY particles where
the partial decay widths into final states with charginos and
neutralinos may be significant for the heavier Higgs particles
\cite{hsusy}. The possibility of the light scalar Higgs decay into a
pair of neutralinos is reduced to very small and exceptional regions of
the MSSM parameter space and is thus not of relevance for our study
as discussed in Sec.~\ref{sec:2-5}.
Higgs decays into sfermion pairs may only play a role for the heavy
Higgs bosons but not for the light scalar $h$ \cite{hsusy} in the context of the present LHC bounds on the sfermion masses (except for sneutrinos as discussed before).

The MSSM Higgs sector is implemented in {\tt HDECAY} \cite{hdecay} within the renormalisation group-improved effective potential approach, including the dominant top- and bot\-tom-Yukawa coupling induced two-loop
corrections. The residual uncertainties of this approach on the light scalar Higgs mass amount to about 3--5 GeV, depending on the MSSM
scenario. The same perturbative level is also extended to the
trilinear and quartic Higgs self-couplings by modifying the official
version of the {\tt subh} subroutine of Refs.~\cite{subh}
accordingly. The present state-of-the-art calculations for the Higgs
masses include partial three-loop corrections that reduce the
uncertainty on the light scalar Higgs mass to a level of 1--2 GeV
\cite{hmass}. These latter contributions are not included in {\tt HDECAY}.

\subsection{Coupling Modifiers and Effective Higgs Couplings}

\label{sec:3-1}
The deviations of Higgs production cross sections and decay branching fractions from the SM predictions due to new physics contributions can be analysed in terms of coupling-modifier terms, $\kappa_X$, in the context of the so-called  $\kappa$-framework
formalism~\cite{Heinemeyer:2013tqa}. The factors $\kappa_X$ for a particle $X$ are defined as
\begin{equation} 
\kappa_X = g_{hXX}^{\rm MSSM}/g_{H_{\rm SM}XX}^{\rm SM} \;,
\end{equation}
so that the $h$ cross sections and partial decay widths to the particle type $X$, normalised to the ones of the SM Higgs particle $H_{\rm SM}$, scale as $\kappa_X^2$.

It must be stressed that, under these assumptions, only the Higgs boson couplings of processes existing in the SM are modified by new physics and possible modifications to the kinematics of the production and decay processes are not considered. This is justified in the context of the pMSSM with the lightest neutralino $\chi^0_1$ being the LSP, provided that $M_{\chi^0_1} > \frac12 M_{h}$ so that decays to any SUSY particle are kinematically forbidden.

The relations between the coupling modifiers $\kappa_X$ and the SUSY
parameters can also be discussed in the context of the so-called $h$MSSM
approach. This has been shown to be a good approximation to the full
MSSM, since its basic underlying assumption,  that the radiative
corrections to the Higgs mass matrix of Eq.~(\ref{mass-matrix}) are
dominated by the $\Delta {\cal M}_{22}^2$ entry, is verified in most
cases.\footnote{In Ref.~\cite{habemus}, the impact of the subleading
corrections $\Delta {\cal M}^{2}_{11}$ and $\Delta {\cal M}^{2}_{12}$
has been proven to be small via a scan of the MSSM parameter space (in
particular the parameters $\mu, A_t, A_b, M_1, M_2, M_3$ and $M_S$) in
which the full radiative corrections to the Higgs sector up to
two loops are implemented. Several other independent analyses \cite{full-covering,Bagnaschietal} have reached the 
same conclusions.}
  
The MSSM couplings normalised to their SM-like values, $c_X^0=g_{hXX}^{\rm MSSM}/g_{hXX}^{\rm SM}$, correspond to the $\kappa_X$ modifiers when all direct radiative corrections are neglected. These values for the couplings of the light $h$ state to third generation $t,b$ fermions and $V\! = \!W/Z$ gauge bosons, including the radiative corrections entering in the MSSM Higgs masses and mixing only, are given by
\begin{eqnarray}  
c_V^0 = \sin(\beta- \alpha) \, ,\quad c_t^0 = g_u^h \, ,\quad c_b^0 = g_d^h\,,
\quad \label{Eq:MSSMlaws} 
\end{eqnarray}
with the couplings $g_u^h$ and $g_d^h$ of Eq.~(\ref{eq:gfhlo}) and
the angle $\alpha$ given by Eq.~(\ref{eq:alpha}). In the decoupling
regime with a heavy pseudoscalar Higgs boson, $M_A \! \gg \! M_Z$, the
mixing angle $\alpha$ gets close to  $\alpha \approx \beta -
\frac{\pi}{2}$ making the $h$ couplings to fermions and massive gauge bosons SM-like, namely 
$c_V^0, c_t^0 , c_b^0  \to 1$.  

In this limit, the heavier $CP$-even and the charged Higgs states
become almost degenerate in mass with the $CP$-odd $A$ boson, $M_H
\approx  M_{H^\pm} \approx M_A \gg M_h$, while the couplings of the
neutral $H$ and $A$ states become similar. In particular, there are
no more $H$ couplings to the weak bosons as is the case for the state
$A$ by virtue of $CP$ invariance:  
\beq
g_{HVV} & = & \sqrt {1- (c_V^0)^2 } = \cos(\beta- \alpha)  \nonumber \\
&& \stackrel{M_A \gg M_Z}  
\longrightarrow 0 \equiv g_{AVV} \;.
\eeq
In fact, the magnitude of the coupling $g_{HVV}$ is a very good measure
of the decoupling limit in which the $h$ couplings are SM-like.
Performing an expansion in terms of the inverse pseudoscalar Higgs mass, one obtains in the approach to this limit, 
\beq
g_{HVV}   \stackrel{\small M_A \gg M_Z} \longrightarrow \ \chi \equiv \frac12 \frac{M_Z^2}{M_A^2} \; 
\sin4 \beta - \frac12 \frac{ \Delta {\cal M}^{2}_{22}}{M_A^2} \;  \sin 2
\beta \;,\quad
\label{gHVVdecoup}
\eeq
where the first term is due to the tree-level contribution and the second one to the dominant contribution of the radiative corrections. A look at the tree-level component shows that for both large $\tb$ and $\tb$ values close to unity, the decoupling limit is reached  more quickly as the expansion parameter involves the factor $\sin 4\beta$ which, in the two limiting cases, behaves as: 
\beq
\sin 4\beta &= &\frac{4 \tb (1- \tan^2\beta)}{(1+ \tan^2\beta)^{2} } \nonumber \\
&\to&   
\bigg\{ \begin{array}{l} - 4/\tb~~      {\rm for}~\tb \gg 1 \\ 
1- \tan^2\beta~   {\rm for}~\tb \; \sim \, 1 
\end{array}
\ \to 0 \
\eeq
and the $g_{HVV}$ coupling is doubly suppressed by both $M_Z^2/$
$M_A^2$ and $\tb$ terms in these limits. In the radiatively generated
component, the one-loop correction $\Delta {\cal M}_{22}^2$ of
Eq.~(\ref{higgscorr}) involves a term that goes like $1/\sin^2 \beta$
which makes it proportional to $-\Delta {\cal M}_{22}^2
/M_A^2 \times \cot \beta$ and thus, vanishes at high $\tb$ values. This leads to the well-known fact that the decoupling limit $g_{HVV} \to 0$  is reached very quickly in this case, in fact as soon  as $M_A \gsim 150$--200 GeV.  

Instead, for $\tb \approx 1$, this radiatively generated component is maximal.  However, when both the tree-level and radiative components are included, the largest departure of the coupling $g_{HVV}$ from zero for a fixed $M_A$ value occurs when $\sin4\beta \approx -1$. This  corresponds to an angle $\beta = 3\pi/8$ and hence  to the value $ \tb \approx 2.4$.

\begin{figure}[h!]
\begin{center}
  \includegraphics[width=0.40\textwidth]{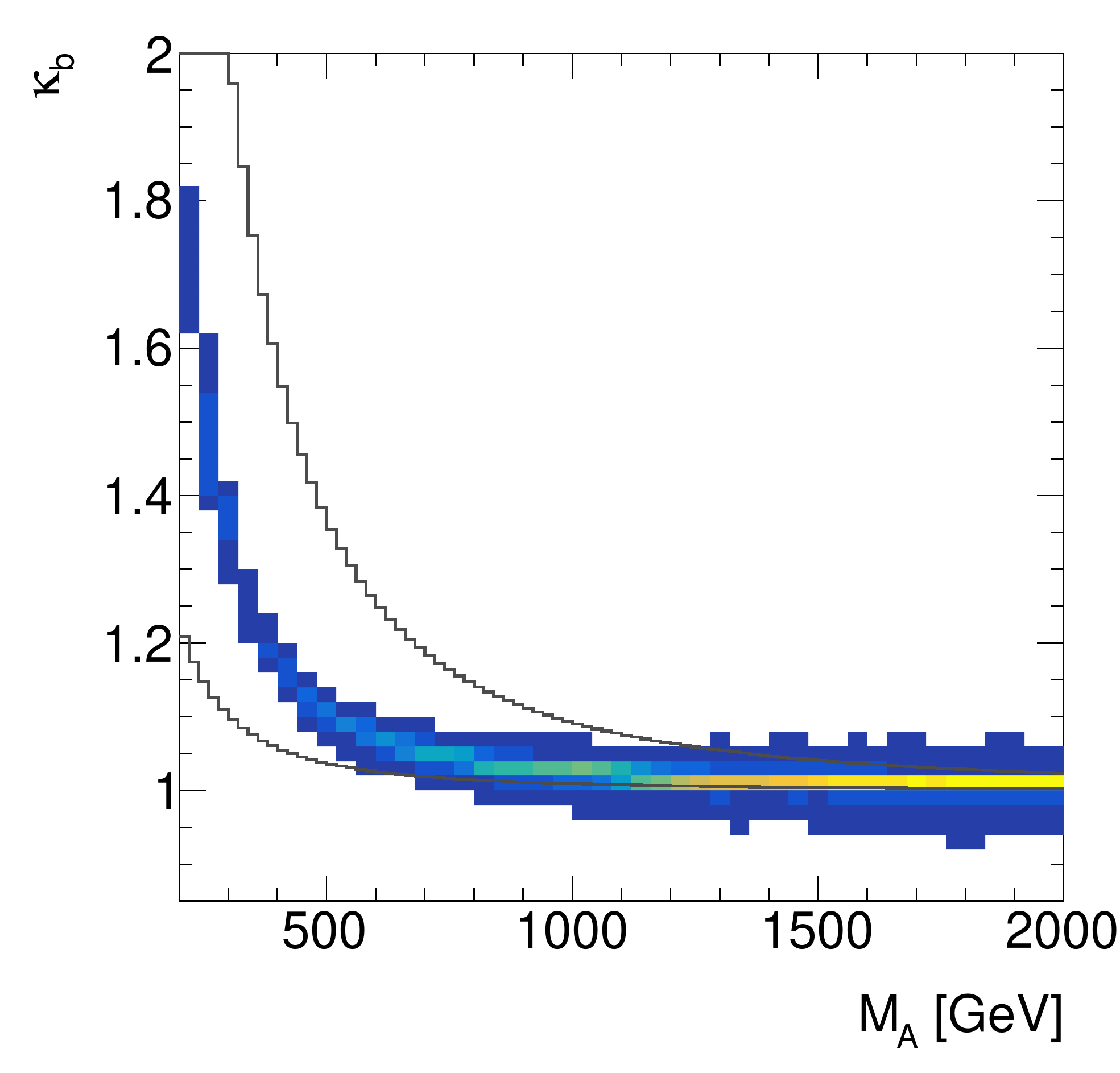} \\
  \includegraphics[width=0.40\textwidth]{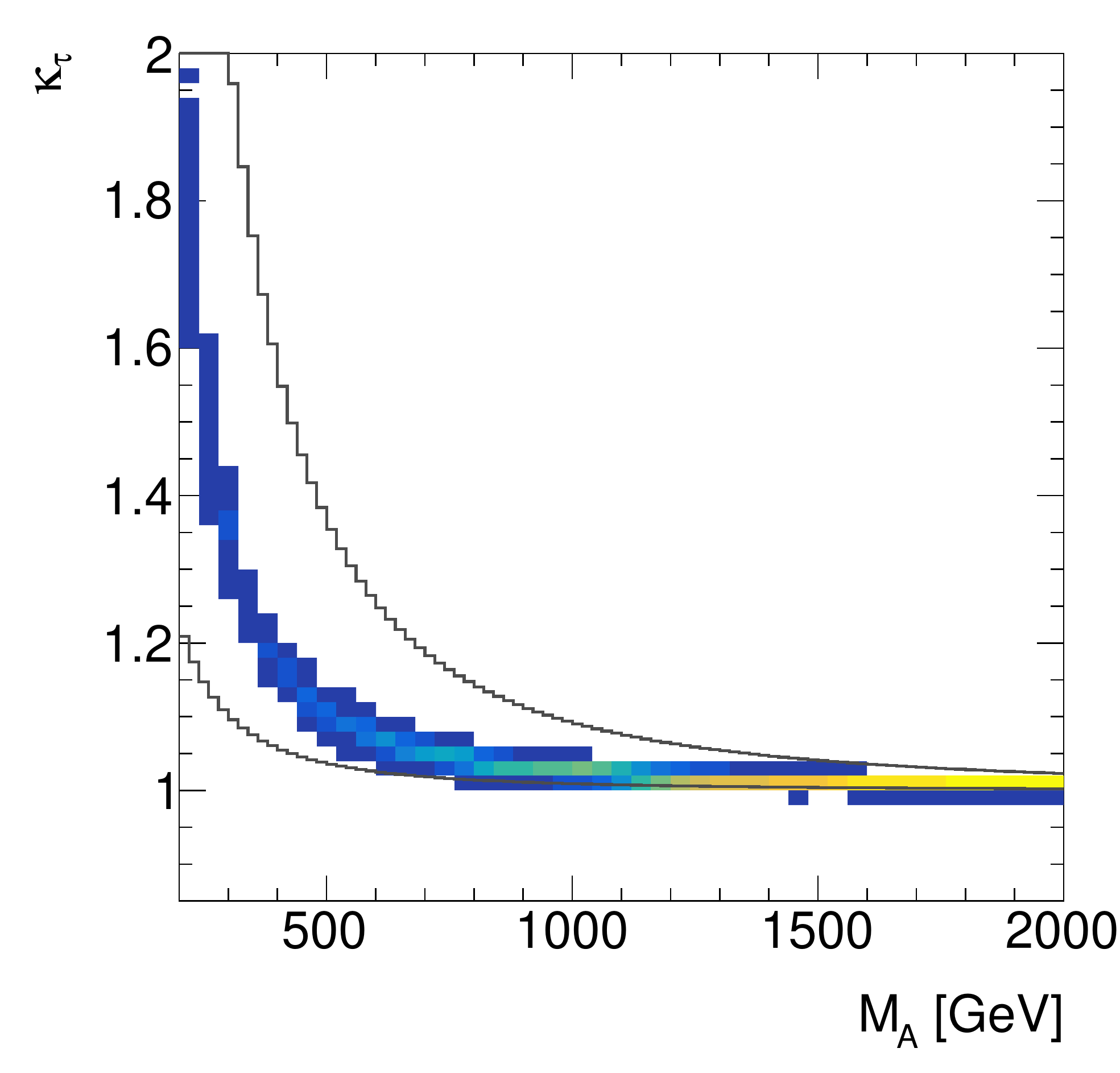} \\    
\end{center}
\caption{The variation of the $h^0$  $b \bar b$ (upper) and $\tau \tau$
(lower panel) coupling modifiers as a function of $M_A$. The histograms
are obtained with {\tt HDECAY} for the valid pMSSM scan points. The lines
show the scaling of Eq.~(\ref{gHff:decoup}) with $M_A$ for
$M_{\mathrm{S}}$ = 1000~GeV, $X_t$ = 1000~GeV, and $\tan \beta$ = 2
(upper line) and 20 (lower line). The colour scale ranges from dark to light according to the increasing fraction of scan points in each bins.}
\label{fig:CouplVsMA}
\end{figure}

Similarly to the coupling to gauge bosons, one can write the couplings of the $h$ state to isospin $\frac{1}{2}$ and $-\frac{1} {2}$ fermions in the approach to the decoupling limit as: 
\begin{equation}  
\begin{array}{llll}
c_t^0 = & \sin(\beta-\alpha)+ \cot\beta \cos(\beta-\alpha) \nonumber \\  
& \stackrel{\small M_A \gg M_Z} \longrightarrow \   1 + \chi \; \cot \beta  
& \to 1  \nonumber \\ 
c_b^0 = & \sin(\beta-\alpha)- \tan\beta \cos(\beta-\alpha)   \nonumber \\  
& \stackrel{\small M_A \gg M_Z} \longrightarrow \  1 - \chi \; \tb  
\  & \to 1
\end{array}
\label{gHff:decoup}
\end{equation}  
where the expansion parameter $\chi \propto 1/M_A^2$ is the same as
the one given in Eq.~(\ref{gHVVdecoup}).  In the approach to the
decoupling limit $M_A \gg M_Z$, the $h$ couplings to bottom (top)
quarks have an additional $\tb$ (cot$\beta)$ factor. Hence, at high
$\tb$ values, the lightest Higgs couplings to bottom quarks can be
different from the SM--one even at high $M_A$ values, contrary to the
$h$ couplings to gauge bosons and top quarks. At low $\tb$ values,
significant deviations from the decoupling can be observed for all $h$
couplings even for a relatively heavy $A$ state.
The couplings to $bb$ and $\tau \tau$ as a function of $M_A$ from our pMSSM scans are compared to the scaling of Eq.~(\ref{gHff:decoup}) with $M_A$ in Figure~\ref{fig:CouplVsMA}. 

The simple picture of the $h$MSSM is altered by additional direct
radiative corrections. These contributions can be effectively mapped into few parameters and these can, in principle, be isolated
experimentally. The most important direct correction occurs in the case
of $b$-quarks where additional one-loop vertex terms modify the
tree-level $h b \bar b$ coupling outside the decoupling regime. They
grow as $\mu \tan\beta$ and are thus very large at high $\tb$. The
dominant component comes from the SUSY-QCD corrections in which
sbottoms and gluinos are exchanged in the loops, but an important
electroweak contribution comes from stop and chargino loops in
addition. In the case of the $hb\bar b$ coupling, this correction appears only outside the decoupling regime and results in an effective correction term:
\beq
\kappa_b = \tilde g_b^h = c_b^0 \times  \frac{1- \Delta_b \cot\alpha \cot\beta } {1+\Delta_b} \;,
\label{cb-def}
\eeq
where the resummation of Eq.~(\ref{eq:gqresum}) has been
included.  In the decoupling regime, $ \tan\alpha$ approaches $-1/\tb$
when $\alpha \to \beta -\frac{\pi}{2}$, so that the SM-like $hb\bar
b$ coupling is recovered: $\kappa_b \to c_b^0 \to 1$ for $M_A \gg
M_Z$. The $\Delta_b$ correction is extremely important as it would
significantly alter the
partial width of the decay $h \to b\bar b$ that is by far the dominant
one and, hence, affect the branching fractions of all other $h$ decay
modes in an anti-correlated way.

Similarly to the bottom quark case, a direct  correction $\Delta_\tau$
from stau-Higgsino loops can affect the $h \tau \tau$ vertex,
shifting the coupling $c_\tau^0$ to $\kappa_\tau$ as in
Eq.~(\ref{cb-def}), and is given in Eqs.~(\ref{eq:deltatau0}) and
(\ref{eq:deltatau}).
Being proportional to the electroweak coupling instead of the strong coupling, this correction is much smaller than $\Delta_b$.

\begin{figure}[t!]
\begin{center}
\includegraphics[width=0.45\textwidth]{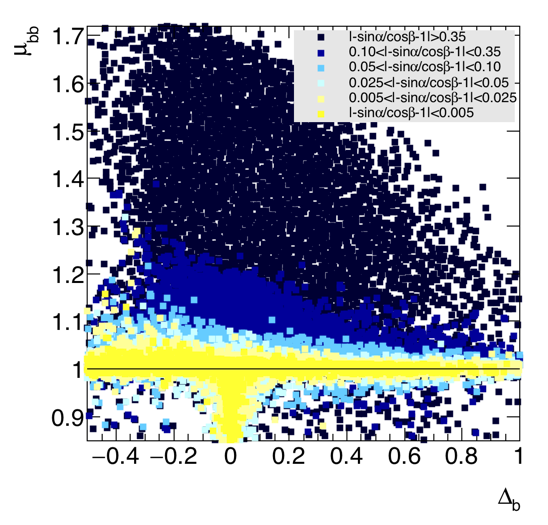} \\
\end{center}
\caption{$h \rightarrow b \bar b$ branching fraction normalised to the SM value, $\mu_{bb}$, shown as a function of $\Delta_b$ for pMSSM points at different distances from the decoupling limit expressed by   $- \sin \alpha / \cos \beta \rightarrow 1$. The convergence of $\mu_{bb}$ towards one independent of the value of $\Delta_b$ for pMSSM points approaching the decoupling limit (represented by the lighter colour squares on the plot) is evident.}
\label{fig:MubbVsDeltab}
\end{figure}

For the light scalar Higgs couplings to bottom quarks in the decoupling
regime, the typical scaling of the Yukawa couplings is described by
\begin{eqnarray}
\kappa_b & = & c_b^0 \left\{ 1 +
\frac{\Delta_b}{1+\Delta_b}~\frac{2\chi}{s_{2\beta}} \right\} \nonumber \\
& = & c_b^0 \left\{ 1 +
\frac{\Delta_b}{1+\Delta_b}~\chi~\frac{1+\tan^2\beta}{\tan\beta} \right\}\;,
\label{eq:kappa_b_decoup}
\end{eqnarray}
with $\chi$ given in Eq.~(\ref{gHVVdecoup}). The corresponding scaling
of the coupling factor $c_b^0$ has been provided in
Eq.~(\ref{gHff:decoup}).  For positive values of $\Delta_b$ the
additional direct corrections to the bottom Yukawa coupling compensate
the deviation of the indirect contributions of Eq.~(\ref{gHff:decoup})
in the decoupling limit to a large extent such that the overall
deviations to the SM limit are reduced. This is confirmed by
the scan results presented in Fig.~\ref{fig:MubbVsDeltab} that shows the
signal strength $\mu_{bb}$ of the partial width $\Gamma(h\to b\bar b)$ as a function of $\Delta_b$ for different
ranges of the deviations of the tree-level-like coupling $c_b^0$ of
Eq.~(\ref{Eq:MSSMlaws}) from the SM limit. This conclusion emerges from
the additional fact that the SUSY remainder in the partial width
$\Gamma(h\to b\bar b)$ is tiny as exemplified in Ref.~\cite{deltabr1}.
For small and negative values of $\Delta_b$, the deviation of the direct corrections from the SM limit is suppressed as well, since it is
proportional to $\chi \Delta_b$. This explains our findings displayed in
Fig.~\ref{fig:MubbVsDeltab}. The analogous considerations are valid for
the $\tau$ Yukawa coupling, too. This implies that with the present
constraints on the light scalar MSSM Higgs couplings the sensitivity to
$\Delta_b$ effects is suppressed so that it will not allow for a
fit of $\mu\tan\beta$.

Direct corrections also affect the $ht\bar t$ couplings with the major difference that these are suppressed by $\tb$ and hence, can be important only at very low $\tb$, $\tb \approx 1$. The $ht\bar t$ coupling is modified according to: 
\beq
\kappa_t = c_t^0 \times  \frac{1- \Delta_t \tan\alpha \tan\beta } {1+\Delta_t} \;,
\label{ct-def}
\eeq
where $\Delta_t$ is the correction of Eq.~(\ref{eq:deltat}).
Again, in the decoupling limit, one would have $ \tan\alpha \to
-1/\tb$ and thus, recover the SM-like coupling.
However, in this case, we have ignored all these direct corrections as they turned out to be small.

In the case of the charm quark, the same direct correction holds but
the term $\propto \lambda_s$ can be neglected, since the
strange Yukawa coupling is negligible in all
cases. The direct corrections in the case of the charm quark and the
tau lepton can be neglected in general for two simple reasons. First,
their impact should be smaller than in the case of the $\Delta_b$
and $\Delta_t$ corrections, and if a small deviation is observed
in the two vertices, a very large distortion of the $hbb$ or $ht\bar t$
vertices should be first noticed. In addition, these  couplings appear
only in the branching ratios for the decays $h \to c\bar c$ and $h
\to \tau^+ \tau^-$ which, as will be seen later, are rather small being
below the 5\% level.  As the direct corrections cannot be very large
(these are radiative corrections after all and should be at most at the level of 10\%), they will affect the total Higgs decay width, and hence, the other Higgs branching ratios, only marginally, i.e., with deviations much below the 1\% level. Hence, as their impact is far less important than the $\Delta_b$ correction, for instance, these direct corrections should be taken into account only if the branching fractions  BR($h \to c\bar c)$ and BR($h \to\tau^+ \tau^-)$ are measured at the percent level. 

Finally, we should note that because the direct corrections in the
case of the $h \to WW^*$ and $h\to ZZ^*$ decays are small, one can use the approximation,
$$\kappa_V = c_V^0\;,$$ which is good in the most relevant cases.

\section{Theoretical and parametric uncertainties in the study of Higgs properties}
\label{sec:3}
 
We discuss now the theoretical and parametric uncertainties in the
Higgs decay branching ratios and in the production cross
sections.\footnote{It has been advocated that a large part of the
theoretical uncertainties that affect the Higgs production rates (in
particular in the main gluon fusion mechanism) and some ambiguities that
affect the branching ratios (such as the one entering in the total decay
width like invisible decay channels) would cancel when measuring ratios
of signal strengths for the same production process \cite{ratios}.  This aspect should be taken into account in the experimental analyses by the full correlation matrix of the various measurements which have been performed. We will therefore not discuss it further in this work.}

 \subsection{Theoretical and parametric uncertainties in decays}

In this analysis, the light scalar Higgs boson primarily decays into
bottom and to a lesser extent into charm quarks, tau leptons and gluons. Decays into four-fermion final
states mediated by off-shell $W$ and $Z$ boson pairs and into photon pairs
are also considered. The corresponding uncertainties of their predictions
emerge from the theoretical uncertainties due to uncalculated
higher-order corrections and from the parametric uncertainties induced by the uncertainties of the input parameters, i.e.,~primarily the quark mass values and the strong coupling constant for fixed supersymmetric
para\-meters.

The pure QCD corrections to the Higgs decays into quarks are known up to
N$^4$LO \cite{hffqcd} thus leaving a residual uncertainty of about 0.3\%
for the QCD part. On the other hand SUSY-QCD corrections are only
included up to NLO exactly \cite{hffselw,hffsqcd} and the NNLO
corrections to the dominant part $\Delta_b$ of the SUSY-QCD corrections
\cite{deltab,deltabr,deltabr1,deltab2}. This yields an estimate of up to 3--4\%
residual uncertainty for the full SUSY-QCD part of these decay modes
depending on the MSSM scenario. The (SUSY-)electroweak corrections are known up
to NLO \cite{hffelw,hffselw} but are not included in our analysis
completely. Only the dominant SUSY-electroweak~corrections contained in the $\Delta_{b/\tau}$ approximation are included for the Higgs decays into bottom quarks and tau leptons. This approximation generates an
uncertainty of about 3--5\% for the missing electroweak~effects in general.
Combining all these individual uncertainties, the theoretical
uncertainties for the partial widths of the decays into quarks and
leptons can be estimated to be about 5\%. A reduction of these
uncertainties to the percent level requires, on the one hand, the
complete inclusion of all known corrections and on the other hand the
calculation of the full NNLO SUSY-QCD corrections at least. The latter step will be a task that may require a long timescale from now, since the necessary techniques for such involved NNLO calculations
are beyond the present state-of-the-art due to the many different masses
involved in the two-loop corrections. 

The (SUSY-)electroweak~corrections to the Higgs decays into four fermions are known up to NLO \cite{hvvelw,hvvselw} but are not included in our
analysis. Their omission induces a theoretical uncertainty of about 5\%
for the decays into four fermions. The inclusion of the
(SUSY-)electroweak corrections would reduce this uncertainty to the percent level. For the Higgs decay mode into two photons the electroweak~and QCD
corrections are small \cite{hgagaqcd,hgagaelw}. The full SUSY-electroweak and 
SUSY-QCD corrections are still unknown. Only the pure NLO QCD
corrections are included in our analysis \cite{hgagaqcd}. The total
theoretical uncertainty can thus be estimated to be less than 3\%--5\%.
This uncertainty can be reduced to the percent level, once the full NLO
corrections become available. 

On the other hand, the parametric uncertainties induced by the input
parameters, i.e.,~the strong coupling constant $\alpha_s(M_Z)$, the
$\overline{\rm MS}$ bottom quark mass $\overline{m}_b(\overline{m}_b)$,
and the top quark mass $m_t$ dominantly and the $\overline{\rm MS}$ charm quark mass $\overline{m}_c(3~{\rm GeV})$ to a lesser extent are relevant. The values used in our analysis including their uncertainties are collected in Table~\ref{tb:params} \cite{pdg,mc3gev}. It should be
noted that the $\overline{\rm MS}$ mass values for the bottom and charm
quarks develop a correlation with the value of the strong coupling
constant $\alpha_s(M_Z)$ that can be neglected \cite{mc3gev}.  Future
developments on the lattice may reduce the hadronic parameter
uncertainties, i.e.,~the ones of the bottom and charm masses and of the
strong coupling constant $\alpha_s$, significantly within the next 10
years, i.e., by factors of 2--5 typically.

The corresponding analysis \cite{brpus,yr4} for the SM Higgs boson leads to
total uncertainties of the branching ratios of about 2\% for $H\to b\bar
b, \tau^+\tau^-$, 7\% for $H \to c\bar c$, 3\% for $H\to \gamma\gamma$, and 2\% for $H\to WW,ZZ$ by using the updated values of Table \ref{tb:params} for the input parameters. These uncertainties correspond to the same uncertainties for the branching ratios of the light 
MSSM Higgs boson in the decoupling limit of large pseudoscalar masses
$M_A$, where it becomes SM-like, since then also the dominant SUSY
effects emerging from the $\Delta_{b/\tau}$ corrections to the
$b$/$\tau$ Yukawa coupling vanish.  Away from this decoupling limit
the total uncertainties become larger due to the larger uncertainties
induced by the genuine SUSY-QCD and -electroweak~corrections. For the heavy Higgs bosons the additional uncertainties related to these
SUSY-specific corrections have to be included in all accessible scenarios
thus adding another 3\%--5\% to the total uncertainty depending on
the MSSM scenario, while for the light Higgs these corrections are suppressed in most cases.

\begin{table}[hbt]
\renewcommand{\arraystretch}{1.5}
\begin{center}
\caption{\label{tb:params} Values of the input parameters used in this study and their uncertainties \cite{pdg,mc3gev}.}
\begin{tabular}{|c|c|c|} \hline
Parameter & Value \\ \hline \hline
$\alpha_s(M_Z)$ & $0.1181 \pm 0.0011$ \\
$m_t^{pole}$ & $(173.1 \pm 0.6)$ GeV \\ 
$\overline{m}_b(\overline{m}_b)$ & $(4.18 \pm 0.03)$ GeV \\ 
$\overline{m}_c(3~{\rm GeV})$ & $(0.986 \pm 0.026)$ GeV \\
\hline
\end{tabular}
\end{center}
\renewcommand{\arraystretch}{1}
\end{table}

\subsection{Theoretical and parametric uncertainties in Higgs production}
\label{sec:3-2}
\subsubsection{QCD uncertainties.}
The main SM-Higgs production chan\-nel is the top- and bottom-loop
mediated gluon-fusion mechanism and, at $\sqrt s$ $= 7$--14 TeV, the
three other mechanisms contribute at a level below 15\% within the SM
when their rates are added and  before kinematical cuts are applied. The majority of the signal events presently observed at the LHC, in
particular in the main search channels  $h \! \to \!  \gamma \gamma, h \! \to \!   ZZ^*  \to 4\ell,  h  \! \to \!  WW^*  \to 2 \ell 2\nu$ and, to a lesser extent, $h \!   \to \!   \tau\tau$, thus originate from the
gluon-fusion mechanism, which is significantly affected by theoretical
uncertainties.  Even the VBF channel, for which the theoretical
uncertainty is smaller when the inclusive cross section is considered,
is contaminated by the gluon-fusion channel and the uncertainties become
relevant once kinematical cuts are applied. The impact of these
theoretical uncertainties is summarised below. 

Until quite recently, the inclusive cross section for the process $gg\to h$ has been known up to next--to--next--to--leading order (NNLO) in the
heavy-top limit (HTL) in perturbative QCD \cite{gghnlo,gghnnlo},
increasing the cross section at $\sqrt s=14$ TeV by a factor of $\approx
2$ compared to the leading-order (LO) approximation and resulting in a
residual scale uncertainty of $\Delta_\mu \approx \pm 8\%$ when the
central values of the renormalisation and factorisation scales are
taken to be at half of the Higgs mass, $\mu_R=\mu_F= \frac12 M_h$, and
varied within a factor of 2 from this central value, $[\frac14 M_h,
M_h]$. This is the scale uncertainty that has been assumed by the ATLAS
and CMS Collaborations in the signal strengths from the dominant
gluon-fusion channel.  However, the ``tour de force" of deriving $\sigma
(gg\to h)$ at next-to-next--to--next--to--leading order (N$^3$LO) has
been achieved recently in the HTL \cite{gghn3lo}. The new corrections
increase the NNLO result of the total cross section slightly by 2\%
for $M_h=125$ GeV at $\sqrt s=14$ TeV for $\mu_R=\mu_F= \frac12 M_h$
and the residual scale dependence is drastically reduced from 8\% to
$\Delta_\mu^{\rm ggh} =+0.2\%, -2.4\%$ at N$^3$LO when the two scales
are varied in the adopted range $[\frac14 M_h, M_h]$. This accuracy is
underlined by the small effect of soft-gluon resummation beyond N$^3$LO
\cite{gghn3ll}.

Nevertheless, the calculation has some drawbacks. As the partonic cross
section at N$^3$LO has been inconsistently folded with the parton
distribution functions (PDFs) at NNLO, while  the calculation has not
yet been performed at the required order for the latter, the impact of
this inconsistency has been estimated at the percent level \cite{yr4}.
The dominant effect is induced by the uncertainties originating from the
PDF fits inherently. Indeed, as the $gg \to h$ process is of ${\cal
O}(\alpha_s^2)$ already at leading order and is initiated by gluons,
there are sizeable uncertainties due to the gluon parton distribution
function and the value of the  $\alpha_s$ coupling which when combined give at 68\% C.L., $\Delta^{\rm ggh}_{{\rm PDF}+\alpha_s}\! =\!  \pm
3.2\%$ for $M_h\! =\! 125$ GeV at $\sqrt s\!=\!14$ TeV \cite{yr4}.

In addition, there is another source of theoretical uncertainties
originating from the use of an effective field theory (EFT) approach to
calculate the radiative corrections beyond the NLO approximation
\cite{gghnlo}, NNLO, and N$^3$LO in QCD \cite{gghnnlo,gghnnlom,gghn3lo}
and NNLO for the  mixed electroweak--QCD corrections \cite{gghmix}. In
both cases, only the top-quark loop in the large top-quark mass limit
has been considered beyond NLO and the corrections to the $b$--quark
loop (which gives a contribution of approximately 5\% to the process at
NLO \cite{gghnlo}) which cannot be obtained in the EFT approach have
been included at NLO only. However, the calculation at NLO that has been
done keeping the exact quark mass dependence, has shown that the QCD
corrections are significantly smaller if expressed in terms of the pole
quark masses. The uncertainty of using the EFT approach beyond NLO has
been estimated to be at the level of 2\% \cite{yr4}. This
uncertainty will be slightly reduced due to the recent calculation of
the full top-mass effects at NNLO \cite{gghnnlom} that turn out to be
small. The total uncertainties on the inclusive cross section have been
estimated as $\Delta_{\rm ggh} \approx \pm 4.5\%$ \cite{yr4} that we
will adopt here.\footnote{Note that other, less optimistic, estimates of
these uncertainties have been made \cite{Baglio-AD} but we will
follow the official recommendations by the LHC Higgs Working Group.}

In addition, significant uncertainties arise when the $gg\! \to\! h$ cross section is broken into the jet categories $h\! +\! 0j, h\! +\! 1j$ and $ h\! +\! 2j$, as is experimentally done in order to enhance the search sensitivities in some important channels, such as $h\to WW^*$ and $h\to \tau\tau$. Indeed for these non--inclusive observables the QCD corrections are known at next--to--leading order and beyond in most
cases and generate large logarithms that lead to a large residual scale
variation of the cross sections as well as a strong dependence on the
jet cuts \cite{yr2}. This problem has partly been overcome by including
resummation of soft-gluon effects \cite{yr4}.

These kinds of uncertainties will also affect the other production
processes. The VBF channel, providing the dominant contribution to the $qq \to hqq$ final state, for which the inclusive cross section is known up to N$^3$LO \cite{vbfqcd,vbfnnlo} in QCD and up to
NLO-electroweak \cite{vbfelw}, has only a few percent combined scale
and PDF uncertainty according to Ref.~\cite{yr4} and develops a total
theoretical uncertainty of $\Delta_{\mu +{\rm PDF}}^{\rm VBF} \approx
\pm 2.5\%$ for $M_h=125$ GeV at $\sqrt s=14$ TeV. The contamination by
the gluon-fusion process $gg\!  \to\! h\!+\!2j$, which leads to the same
final state (and which is known at NLO in the HTL \cite{gghjjnlo}) makes
the total uncertainty in the $h\!+\!2j$ final ``VBF" sample a bit
larger. It is included in the uncertainty estimate above.

The situation is similar for the Higgs--strahlung process where for the
inclusive total cross sections, also known at NNLO QCD \cite{vhvqcd,vhnnlo}
and NLO electroweak~\cite{vhelw}, the total uncertainty is at the level of  a few percent, $\Delta_{\mu +{\rm PDF}}^{Wh} = \pm 2.5\%$ and $\Delta_{\mu
+{\rm PDF}}^{Zh} = \pm 5\%$ for $M_h=125$ GeV at $\sqrt s=14$ TeV
\cite{yr4} (in the latter case, there is an additional contribution from
the $gg \to Zh$ process which is known at NLO in the HTL \cite{ggzhnlo}
and introduces an extra scale and PDF uncertainty so that the total
uncertainty is larger than for the $Wh$ case). Finally, the associated
$pp \to t\bar th$ process is known at NLO QCD \cite{tthnlo} and NLO electroweak \cite{tthelw} supplemented by soft and collinear gluon resummation up to
NNLL \cite{tthnll}. The total uncertainty is estimated to be
$\Delta_{\mu +{\rm PDF}}^{\rm tth} = +9.5\%, -12.7\%$ \cite{yr4} for
$M_h=125$ GeV at $\sqrt s=14$ TeV.  In both cases, Higgs-strahlung and
$t\bar th$ production, if boosted topologies are selected, the
uncertainties of these exclusive rates might be considerably larger. 

Within the MSSM the associated Higgs boson production with a $b\bar b$
pair is dominant for large values of $\tan\beta$. The QCD corrections to
the inclusive cross section evaluated in the four-flavour scheme (4FS)
are large \cite{bbhqcd4}, while within the five-flavour scheme (5FS)
they are of more moderate size \cite{bbhqcd5}. The reasons for this
difference are the massless and on-shell treatment of the bottom quarks
within the 5FS and the non-resummation of PDF-related logarithms in the 4FS. Both calculations differ by about 20\% and a combination is
desirable. This has been solved quite recently by proper matchings
between both schemes \cite{bbhmatch}. The residual uncertainties of the
inclusive cross sections range at the level of about 20\% \cite{yr4}.

In summary, the cross sections for the various channels are presently
known at the level of
\begin{eqnarray}
&& \Delta_{\rm ggh} \approx 4.5\% ,~\Delta_{\rm VBF}
\approx \Delta_{\rm WH}\approx 2.5\% , \nonumber \\
&& \Delta_{\rm ZH}\approx
5\%,~\Delta_{\rm t\bar th} \approx  10\%,~\Delta_{\rm
b\bar bh} \approx  20\%
\end{eqnarray}
in the SM limit which are the numbers that we will adopt
in our analysis for illustration for the QCD-initiated uncertainties. It
is clear that this know\-ledge will improve and at the time of the
high-luminosity LHC option, the uncertainties could be reduced. It
should be noted that these uncertainties have to be extended by the genuine SUSY uncertainties to which we turn now.

\subsubsection{Genuine SUSY uncertainties.}
Genuine SUSY--QCD corrections are also known for the gluon-fusion
processes $gg\to h/H/A$ \cite{gghsqcd,gghsqcd2}. They are sizeable for
small values of $\tan\beta$ where the top loops provide the dominant
contribution.  However, at large $\tan\beta$ values the SUSY--QCD
corrections are large since the $\Delta_b$ contributions are sizeable
and dominant \cite{gghsqcd2}. If the leading $\Delta_b$ contributions
are absorbed in the effective bottom Yukawa couplings as outlined in
Section \ref{sc:hcoup} the remainder of these contributions is of
moderate size \cite{gghsqcd2}.  Since this contribution is not included
in our analysis it has to be treated as an additional uncertainty. Using
the conventions of Section \ref{sc:hcoup} this can be estimated as
$\Delta_{\rm SUSY}^{\rm ggh} \approx \delta \sigma\{g_t,\tilde
g_b[\Delta_b(1\pm 5\%)]\} / \sigma\{g_t,\tilde g_b\}$, if the
$\Delta_b$-resummed bottom Yukawa couplings of Eq.~(\ref{eq:gqresum})
are used for the bottom contributions. This notation is meant as a
variation of the pure $\Delta_b$ terms inside the cross-section
prediction by 5\% and taking the impact on the cross section as the
uncertainty related to SUSY effects.

The genuine SUSY corrections to the VBF and Higgs-strahlung processes
are small \cite{vbfsqcd,vbfsqcd2}. Only in very exceptional
regions of the MSSM parameter space they can reach the 10\% level. The
QCD corrections are easy to include in the analysis \cite{vbfsqcd}. The
residual uncertainties can be estimated at the subpercent level on top of the pure QCD corrections.

The genuine SUSY-QCD corrections to the $t\bar t\phi$ cross sections,
with $\phi=h,H,A$, are known to be of moderate size \cite{tthsqcd} on
top of the pure QCD corrections. The residual uncertainties induced by
these corrections can be estimated to about $5\%$ conservatively. On the
other hand, the genuine SUSY-QCD corrections to $b\bar b\phi^0$
production can be absorbed in the resummed bottom Yukawa couplings of
Eq.~(\ref{eq:gqresum}). The remainder beyond this approximation is
small, i.e.~of subpercent level \cite{tthsqcd,bbhsqcd}. The resummed
bottom Yukawa coupling of Eq.~(\ref{eq:gqresum}) includes, in our study, SUSY-electroweak corrections.

\section{Analysis}
\label{sec:4}

In order to relate the measurements of the Higgs properties at the LHC
and an $e^+e^-$ collider to the fundamental SUSY parameters, it is
necessary to choose an explicit model. In this study, we adopt the
pMSSM, a generic implementation of the MSSM with the neutralino being
the LSP, having all the soft SUSY-breaking mass terms and trilinear
couplings as free parameters.

\subsection{Scans in the pMSSM}

We vary the 19 pMSSM free parameters in an uncorrelated way through flat scans within the ranges given in Table~\ref{tab:paramSUSY}.
The scan range is explicitly chosen to include the so-called ``maximal mixing'' region~\cite{Mh-max}, at $X_t \sim \sqrt{6}M_{\mathrm{SUSY}}$, which corresponds to the larger values of $M_{h}$ achievable in the MSSM, and to reach SUSY masses beyond the reach of direct SUSY searches at the LHC.   
The details of the pMSSM scans and the tools used for the computations
of the spectra and relevant observables have been presented
elsewhere~\cite{Arbey:2011un}. Here, we mention only those most relevant to this study. SUSY mass spectra are generated with {\tt SOFTSUSY 3.2.3}~\cite{Allanach:2001kg}, which gives results that are comparable to other spectrum generators such as {\tt Suspect} \cite{suspect}.  As already mentioned, the decay branching fractions of Higgs bosons are obtained using {\tt HDECAY 6.53} \cite{hdecay}, including gaugino and sfermion loop corrections, and cross-checked with {\tt FeynHiggs 2.8.5}~\cite{feynhiggs}. The widths and decay branching fractions of the other SUSY particles are computed using {\tt SDECAY 1.3}~\cite{sdecay}. The flavour observables and dark matter relic density are calculated with the programs {\tt SuperIso Relic v3.2}~\cite{flavor} and {\tt micrOMEGAs} \cite{Belanger:2008sj}. 

As for the Higgs production rates,  the $gg$ and $bb$ production cross sections are computed using {\tt HIGLU 1.2}~\cite{higlu} and {\tt FeynHiggs 2.8.5} \cite{feynhiggs}, respectively. The Higgs production cross sections and the branching fractions for decays into $\gamma \gamma$ and $WW$, $ZZ$ from {\tt HIGLU} and {\tt HDECAY} are compared to those predicted by {\tt FeynHiggs}. 

In the SM, both the $gg \to H_{SM}$ cross section and the branching fractions agree within  $\sim$3\%. Significant differences are observed in the SUSY case, with {\tt HDECAY} giving values of the branching fractions to $\gamma \gamma$ and $WW$ that are on average 9\% lower and 19\% larger, respectively, than those predicted by {\tt FeynHiggs} and have a root mean square (r.m.s.) spread of the distribution of the relative difference between the two programs of 18\% and 24\%, respectively. In this study, we adopt the {\tt HDECAY} results throughout the analysis. The parametric uncertainties have been discussed in section~\ref{sec:3}.

\begin{table}[t!]
\begin{center}
\caption{pMSSM parameter ranges adopted in the scans (in GeV when applicable).\label{tab:paramSUSY}}
\begin{tabular}{|c|c|}
      \hline
~~~~Parameter~~~~ & ~~~~~~~~Range~~~~~~~~ \\
\hline\hline
$\tan\beta$ & [1, 60]\\
$M_A$ & [50, 6000] \\
$M_1$ & [-5000, 6000] \\
$M_2$ & [-5000, 6000] \\
$M_3$ & [50, 6000] \\
$A_d=A_s=A_b$ & [-10000, 10000] \\
$A_u=A_c=A_t$ & [-10000, 10000] \\
$A_e=A_\mu=A_\tau$ & [-10000, 10000] \\
$\mu$ & [-6000, 6000] \\
$M_{\tilde{e}_L}=M_{\tilde{\mu}_L}$ & [50, 6000] \\
$M_{\tilde{e}_R}=M_{\tilde{\mu}_R}$ & [50, 6000] \\
$M_{\tilde{\tau}_L}$ & [50, 6000] \\
$M_{\tilde{\tau}_R}$ & [50, 6000] \\
$M_{\tilde{q}_{1L}}=M_{\tilde{q}_{2L}}$ & [50, 6000] \\
$M_{\tilde{q}_{3L}}$ & [50, 6000] \\
$M_{\tilde{u}_R}=M_{\tilde{c}_R}$ & [50, 6000] \\
$M_{\tilde{t}_R}$ & [50, 6000] \\
$M_{\tilde{d}_R}=M_{\tilde{s}_R}$ & [50, 6000] \\
$M_{\tilde{b}_R}$ & [50, 6000] \\
\hline
\end{tabular}
\end{center}
\end{table}

The SM $\gamma \gamma$, $WW$ and $ZZ$ branching fractions receive electroweak corrections, of the order of 5\%--10\%, which are included in {\tt HDECAY}. However, their SUSY counterparts are not known. In order to make the SM and SUSY branching fractions comparable in this study, we remove the electroweak corrections to their SM values.

The ``valid'' pMSSM points are selected by requiring the neutralino to be the LSP and the lightest Higgs boson mass to be in the range 122 $< M_h <$ 128~GeV. These requirements reduce the sample to $\simeq$1.7~M points. In addition, we impose a set of constraints from flavour physics and relic dark matter data, as discussed in~\cite{Arbey:2011un}, to obtain a set of ``accepted'' pMSSM points used in the subsequent analysis. The main flavour physics measurements, summarised in Table~\ref{tab:constrScans}, are the decay $B_s \to \mu^+ \mu^-$~\cite{ATLAS:2018cur,CMS:2019bbr,LHCb:2021awg}, that can receive important SUSY contributions at large values of $\tan\beta$~\cite{Arbey:2012ax}, the inclusive rare decay $B \to X_s \gamma$~\cite{HFLAV:2019otj}, and the leptonic decay $B \to \tau \nu_\tau$~\cite{pdg}.

The relic DM density constraint is applied in a loose form, requiring the contribution by the $\tilde{\chi}$ LSP, $\Omega_{\chi}$, not to exceed the upper limit of the $\Omega_{\mathrm{CDM}}$ density determined by the PLANCK satellite~\cite{Planck:2018vyg}, i.e.\ $10^{-5} < \Omega_{\chi} h^2 < 0.163$, thus allowing for other particles to contribute to the observed cosmic DM and/or modifications to the early universe properties.

\begin{table}[h!]
\begin{center}
\caption{Summary of the flavour physics constraints applied to the accepted points in the pMSSM scans. The range for $B_s \rightarrow \mu \mu$ is based on the average reported in Ref.~\cite{Hurth:2021nsi} obtained with the technique developed by the ATLAS, CMS, and LHCb Collaborations~\cite{LHCb-CONF-2020-002}. \label{tab:constrScans}}
\begin{tabular}{|c|c|c|}
  \hline
  Constraint & Value & Ref. \\
  \hline
  \hline
  $b   \rightarrow s \gamma$ &  3.05$<$BR$<$3.59 ($\times 10^{-4}$) & \cite{HFLAV:2019otj} \\
  $B   \rightarrow \tau \nu$ &  0.71$<$BR$<$1.47 ($\times 10^{-4}$) & \cite{pdg} \\
  $B_s \rightarrow \mu \mu$  & 2.10$<$BR$<$3.60 ($\times 10^{-9}$) & \cite{ATLAS:2018cur,CMS:2019bbr,LHCb:2021awg}  \\
  \hline
\end{tabular}
\end{center}
\end{table}

The last decade has been rich in results from experiments searching for the scattering of WIMPs on solid state and gaseous detectors located in underground laboratories. Currently, the results from the XENON-1T~\cite{Aprile:2018dbl} stand as the most constraining bounds on the WIMP scattering cross section as a function of its mass. The use of these bounds in the analysis of the pMSSM scan points brings a considerable dependence on the assumptions on the dark matter profile in our galaxy. While we do not use these bounds in the pre-selection of the accepted pMSSM points, we do discuss their impact, as well as that of the next generation of direct dark matter detection experiments, on the pMSSM parameter space in relation to the Higgs invisible decay rate. 

After applying these constraints, the pMSSM scans yield $\simeq$ 0.8~M accepted points. The results discussed in Section~\ref{sec:5} depend on the distributions of MSSM variables for the valid and accepted points. Although the pMSSM scans are flat in the input variables according to the ranges given in Table~\ref{tab:paramSUSY}, the distributions of these variables for the valid and accepted points are characterised by a non flat rate of points across the range of each variable. This is an effect of the selection based on $M_h$, flavour observables and dark matter relic density.
\begin{figure*}[ht!]
\begin{center}
  \begin{tabular}{ccc}
    \includegraphics[width=0.30\textwidth]{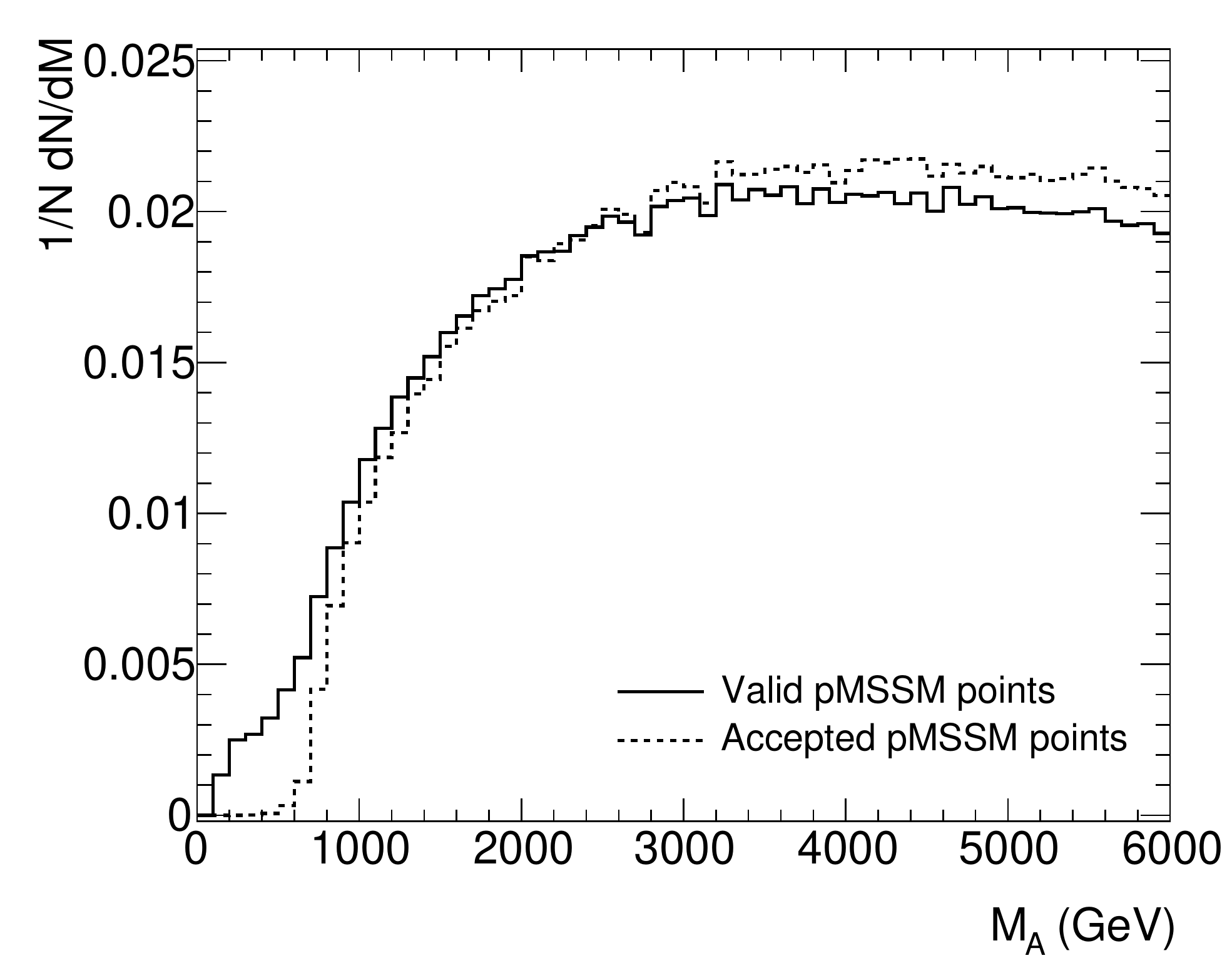} &
    \includegraphics[width=0.30\textwidth]{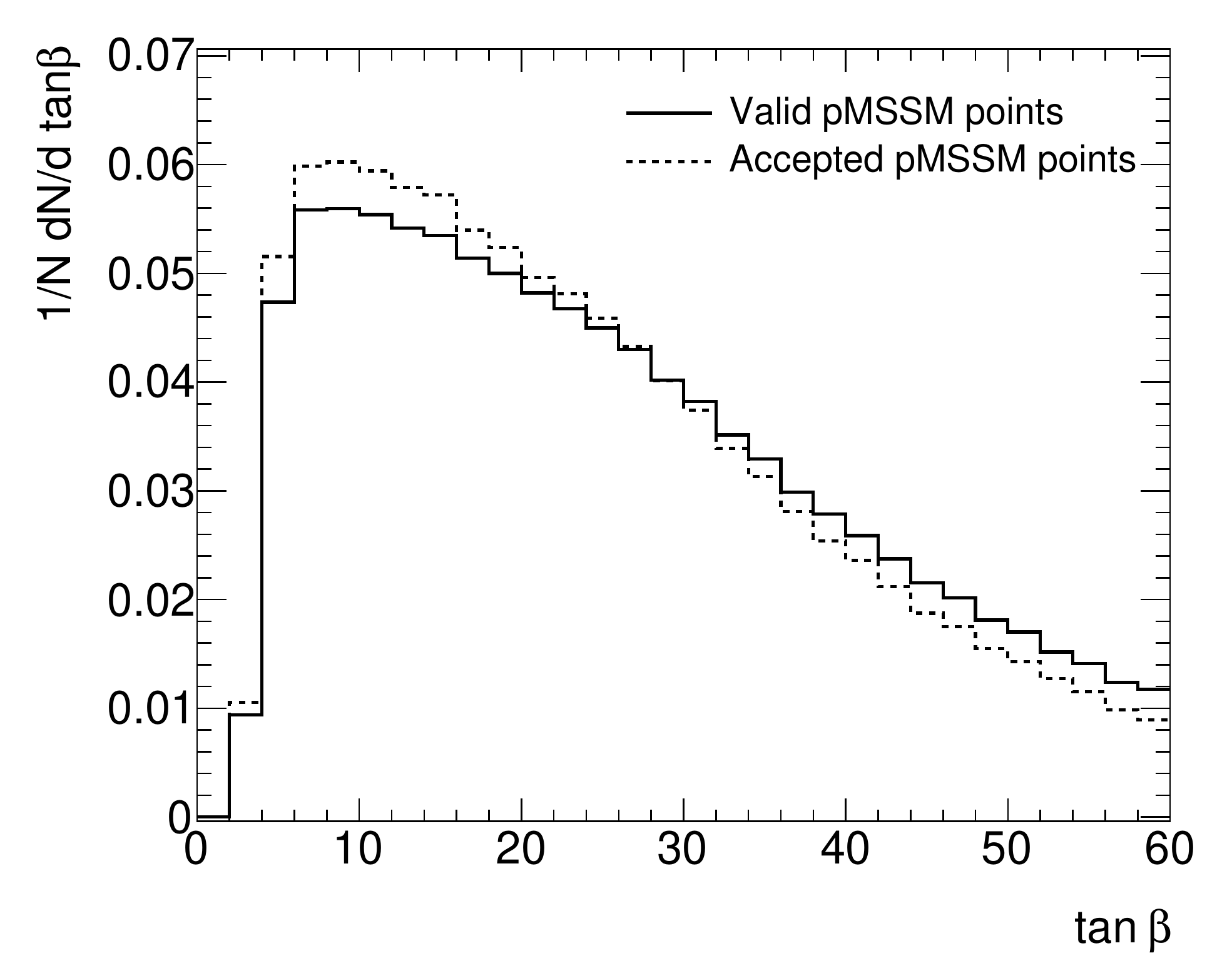} &
    \includegraphics[width=0.30\textwidth]{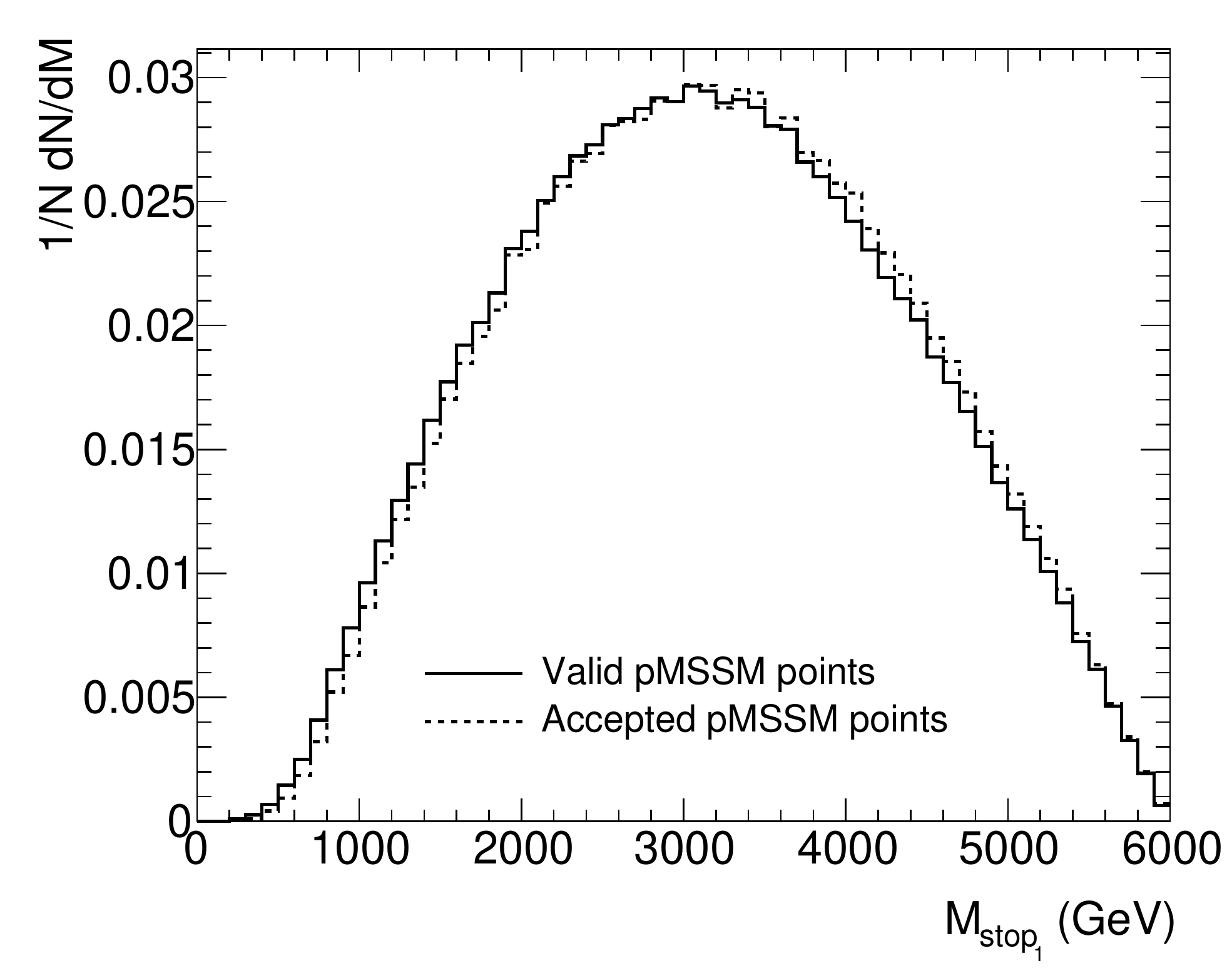} \\
    \includegraphics[width=0.30\textwidth]{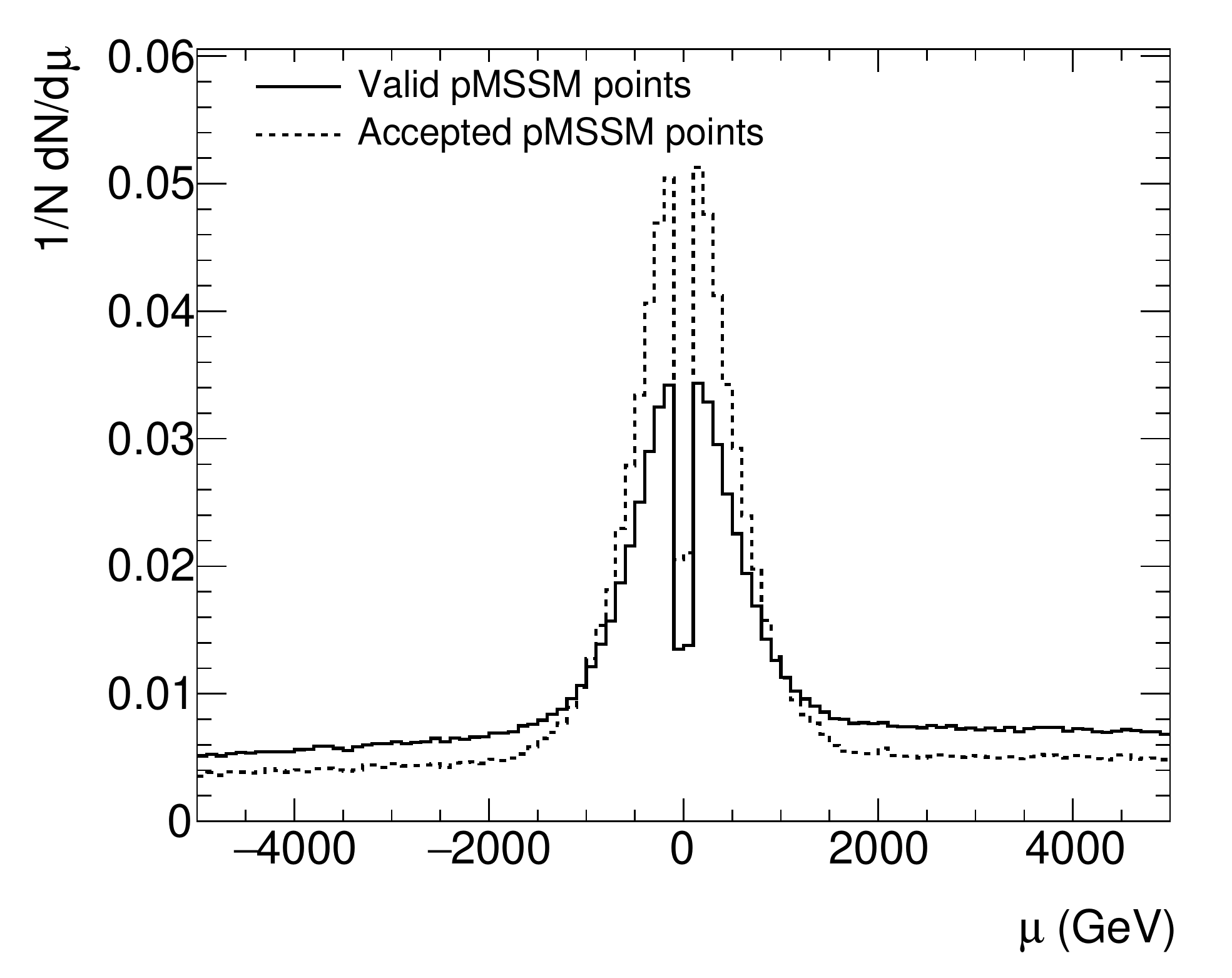} &
    \includegraphics[width=0.30\textwidth]{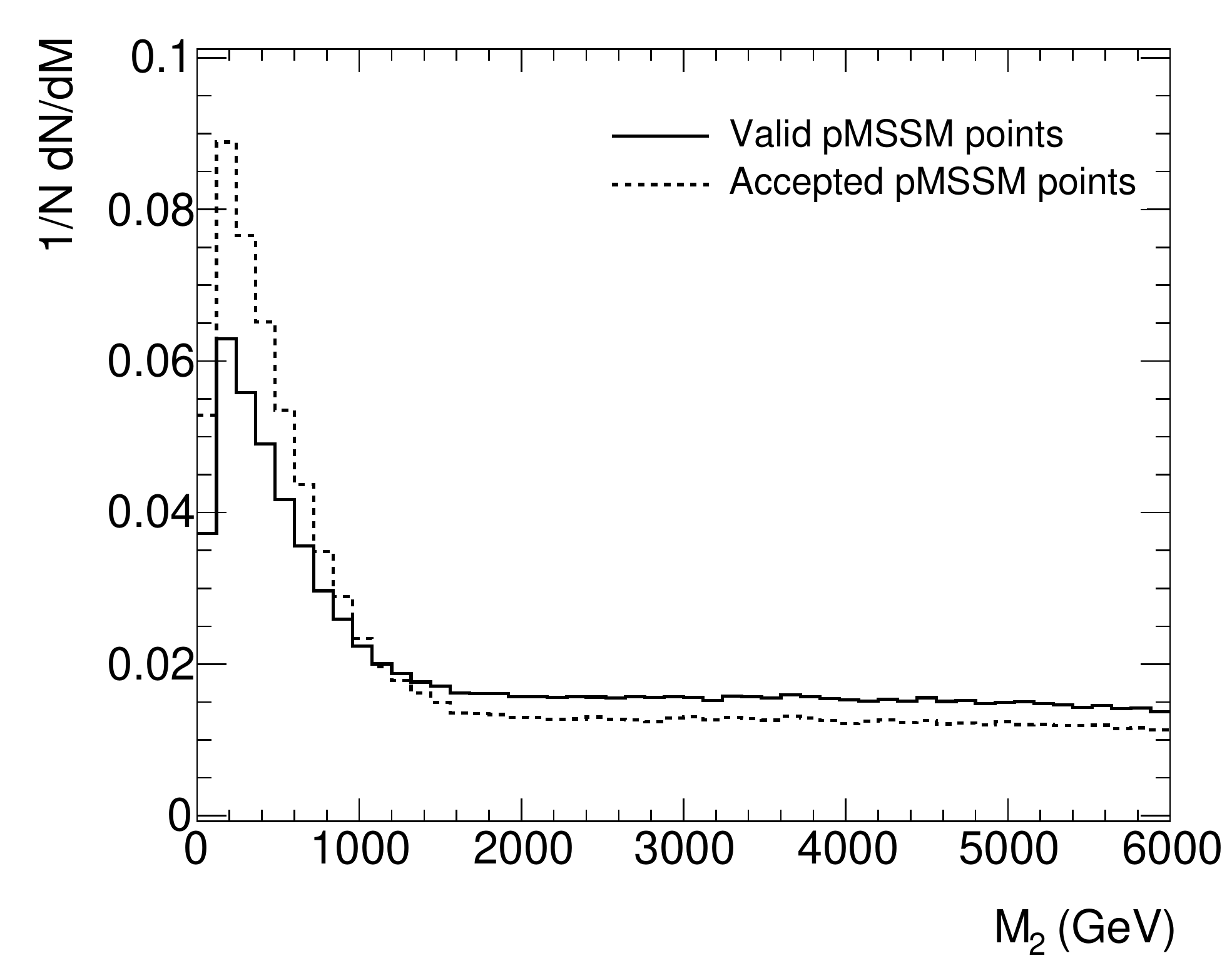} &
    \includegraphics[width=0.30\textwidth]{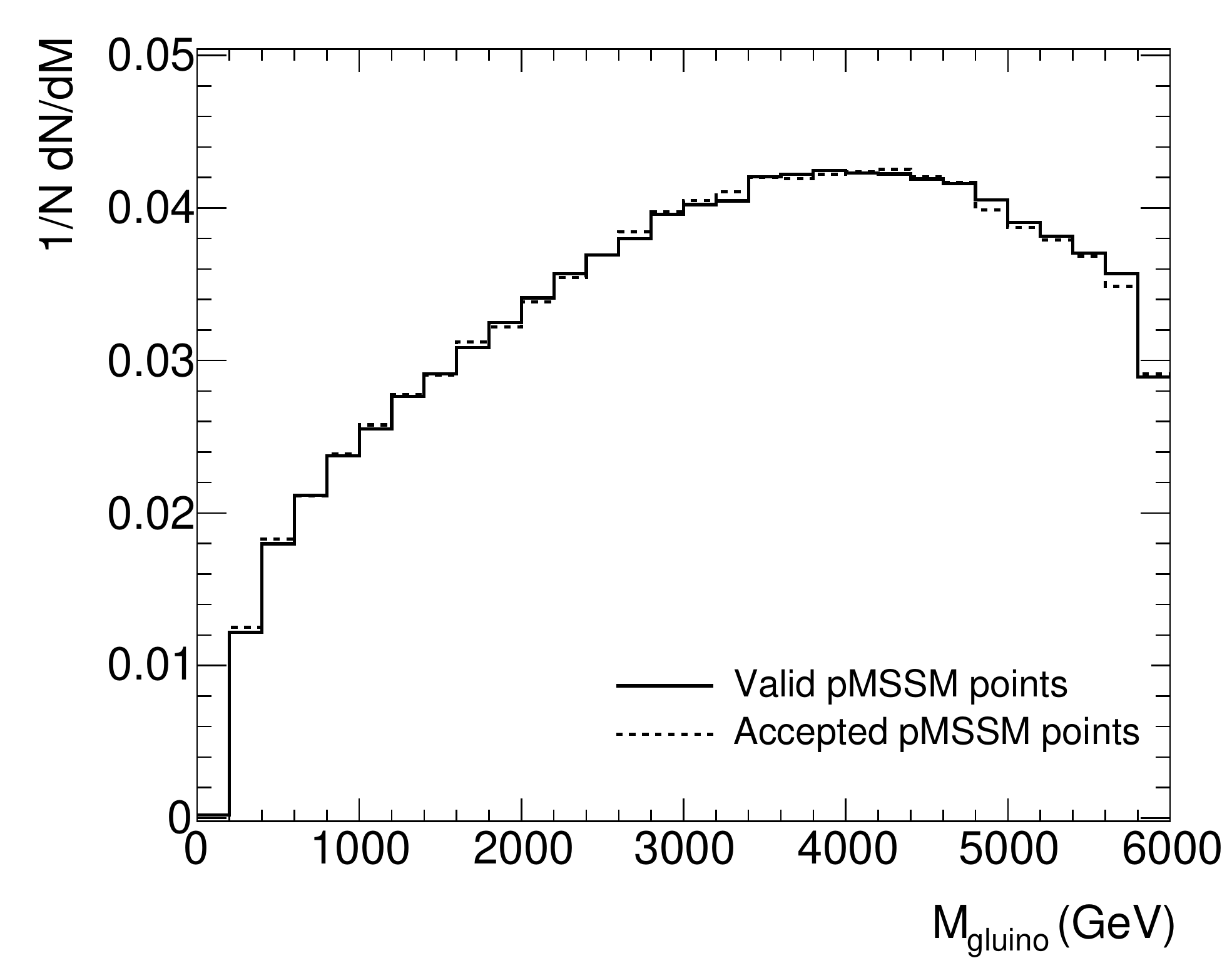} \\    
  \end{tabular}
\end{center}
\caption{Distributions of the $M_A$ (upper left), $\tan \beta$ (upper center), $M_{\tilde{t}_1}$ (upper right), $\mu$ (lower left), $M_2$ (lower center), $M_{\tilde{g}}$ (lower right panel) pMSSM parameters of valid (continuous line), and accepted (dashed line) points. The shapes of these distributions depend both on the range of the pMSSM parameters adopted in the scans (see Table~\ref{tab:paramSUSY}) and on the requirements for valid and accepted scan points. \label{fig:pMSSMPdf}}
\end{figure*}
The distribution of the pMSSM parameters most relevant to this study for the valid and the accepted scan points are given in Figure~\ref{fig:pMSSMPdf}.

\subsection{SUSY Particle Direct Searches at the LHC}
\label{sec:4-0}

The ATLAS and CMS experiments at the LHC have pursued a vast program of searches for SUSY particles. These cover both the production of particles cascading to quarks and/or leptons and the neutralino LSP, through channels with 
jets and/or leptons + MET, and that of the heavy SUSY Higgs bosons, $H$, $A$, and $H^{\pm}$.

In this study, we test the compatibility of the accepted pMSSM points with the bounds implied by the LHC sear\-ches. We consider the bounds obtained from the analyses given in Table~\ref{tab:lhc-susy}.
\begin{table}
\caption{Summary of the analyses used to assess the observability of the pMSSM points by the LHC SUSY searches.}
\begin{tabular}{|l|c|c|c|}
\hline
Channel & Int. lum. & Sensitivity & Ref. \\
        & fb$^{-1}$   &              &     \\
\hline \hline
$H$/$A \rightarrow \tau \tau$ & ~36 & $H$, $A$ & \cite{Aaboud:2017sjh} \\
$H$/$A \rightarrow ZZ$ & ~36 & $H$, $A$ & \cite{Aaboud:2017rel} \\
$H$/$A \rightarrow t \bar{t}$ & ~20 & $H$, $A$ & \cite{Aaboud:2017hnm} \\
\hline
jets + MET           & 139 & $\tilde g$, $\tilde q$ & \cite{ATLAS-CONF-2019-040} \\
jets + MET           & ~36 & $\tilde g$, $\tilde q$ & \cite{Aaboud:2017hdf} \\
1 $\ell$ + jets + MET & ~36 & $\tilde g$, $\tilde q$ & \cite{Aaboud:2017bac} \\
$\ell^{+}\ell^{+}$, $\ell^{-}\ell^{-}$ + MET & 139 & $\tilde g$, $\tilde q$ & \cite{Aad:2019ftg} \\
\hline
$b$-jets + MET       & ~36 & $\tilde t$ & \cite{Aaboud:2017ayj} \\
multiple $b$-jets + MET & ~80 & $\tilde t$, $b$ & \cite{ATLAS-CONF-2018-041} \\
\hline
2 $\ell$ + MET       & 139 & $\tilde \chi^0$, $\tilde \chi^{\pm}$, $\tilde{\ell}$ & \cite{Aad:2019vnb} \\ 
3 $\ell$ + MET       & ~36 & $\tilde \chi^0$ $\tilde \chi^{\pm}$, $\tilde{\ell}$ & \cite{Aaboud:2018jiw} \\
\hline
mono-jet + MET        & ~36 & $\tilde \chi \tilde \chi$, $\tilde q \tilde q$ & \cite{Aaboud:2017phn} \\
mono-$W$/$Z$ + MET        & 3.2 & $\tilde \chi \tilde \chi$, $\tilde q \tilde q$ & \cite{Aaboud:2016qgg} \\        
\hline
\end{tabular}
\label{tab:lhc-susy}
\end{table}
These can be divided in four classes:
i) $H$/$A \rightarrow \tau \tau$, $ZZ$ and $t \bar t$ decays;
ii) channels with jets (+$\ell$) + MET,  sensitive to gluino and scalar quark production and decays,
iii) channels with $\ell$s (+$h$) + MET sensitive to char\-gino and neutralino production and decays, and
iv) monojets + MET sensitive to production of $\tilde{\chi} \tilde{\chi}$ or $\tilde{q} \tilde{q}$, where the scalar quark is highly degenerate with the neutralino LSP. While there is an impressively large number of final states considered by the experiments, the chosen analyses provide an efficient coverage of the MSSM parameter space within a practical number of processes to be simulated and reconstructed for each pMSSM point.  Events are generated with {\tt MadGraph 5}~\cite{Alwall:2011uj} and {\tt Pythia 8.2}~\cite{Sjostrand:2014zea}. Physics observables are obtained through a parametric simulation for the detector response and event reconstruction using {\tt Delphes 3.4}~\cite{Ovyn:2009tx} fast simulation. Signal selection cuts for each of the analyses are applied to the simulated signal events. The number of SM background events in the signal regions are taken from those estimated by ATLAS for the analyses listed in Table~\ref{tab:lhc-susy}. The 95\% confidence level (C.L.) exclusion of each SUSY point in presence of background only is determined using the CLs method~\cite{Read:2002hq}. 

\begin{table}[t!]
\caption{Fractions of accepted pMSSM points excluded by the LHC
$H$/$A \rightarrow \tau \tau$, $ZZ$, $t \bar t$, and the jet/$\ell$ + MET
sear\-ches at three stages of the LHC program.}%
  \begin{tabular}{|c|c|c|c|}
  \hline
                                   & LHC               & LHC              & HL-LHC \\
                                   & 140 fb$^{-1}$      & 400 fb$^{-1}$    & 4~ab$^{-1}$ \\
    \hline \hline
     $H$/$A \rightarrow \tau \tau$, $ZZ$, $t \bar t$  & 0.10              &  0.12           & 0.19   \\                 
     + j/$\ell$s+MET               & 0.47              &  0.52           & 0.65   \\
    \hline
  \end{tabular}
  \label{tab:methiggs}
\end{table}
Results from the present Run~2 analyses are projected to the integrated luminosities reachable after Run~3 and HL-LHC operation by rescaling the signal and background event yields. Accepted pMSSM scan points yielding a rate of SUSY events incompatible with the background-only hypothesis using the CLs method for a given integrated luminosity are considered as ``excluded'' at the corresponding stage of the LHC program. The other pMSSM points are considered as ``not excluded''.

\begin{figure}[h!]
\begin{center}
\includegraphics[width=0.35\textwidth]{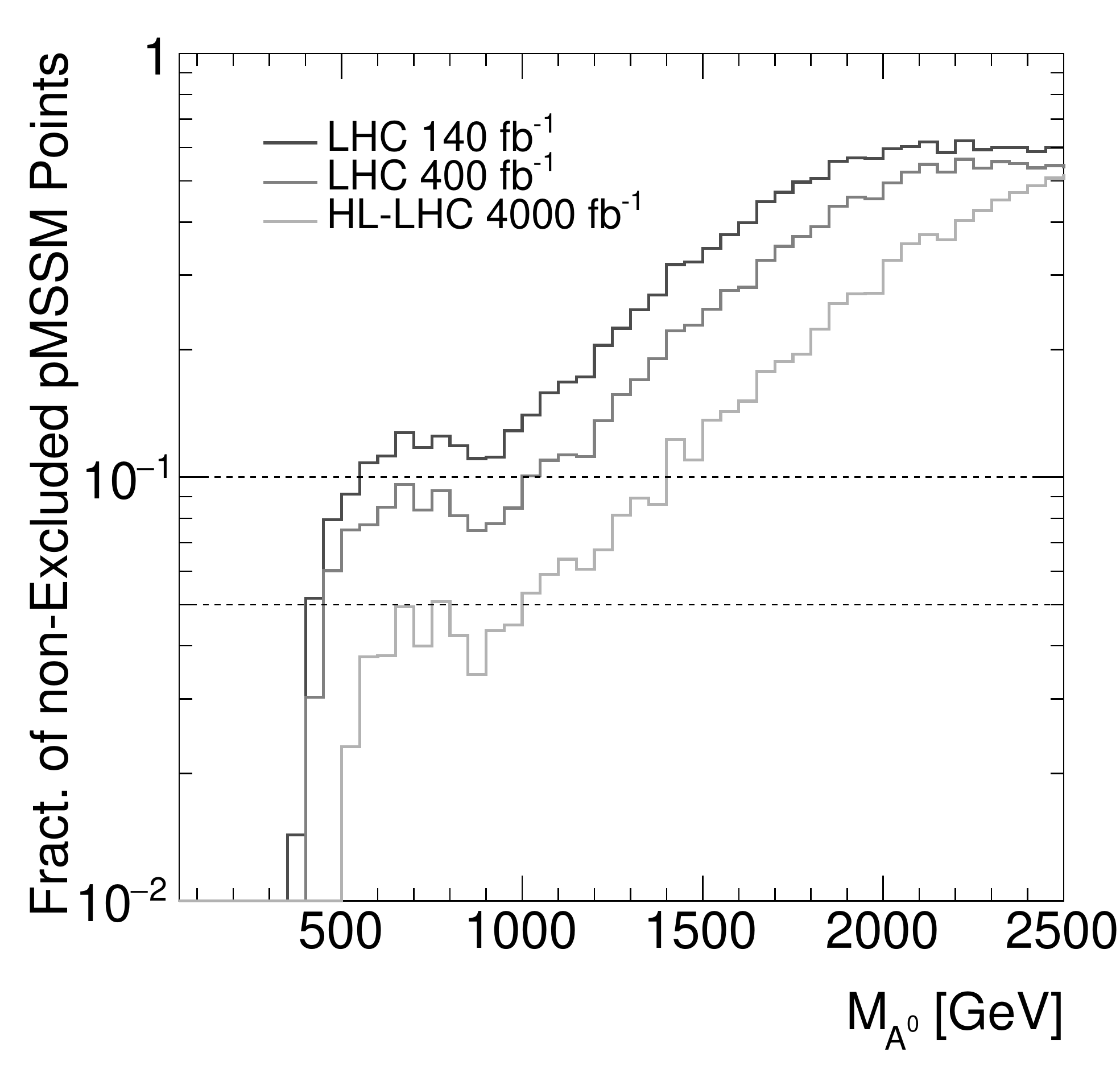}
\end{center}
\caption{Fraction of accepted pMSSM points not excluded at 95\% C.L. by the
present $H$/$A$ searches and beyond the expected sensitivity of Run~3 and HL-LHC as a function of $M_A$.}
\label{fig:malhc}
\end{figure}
\begin{figure}[h!]
\begin{center}
\begin{tabular}{c}
\includegraphics[width=0.35\textwidth]{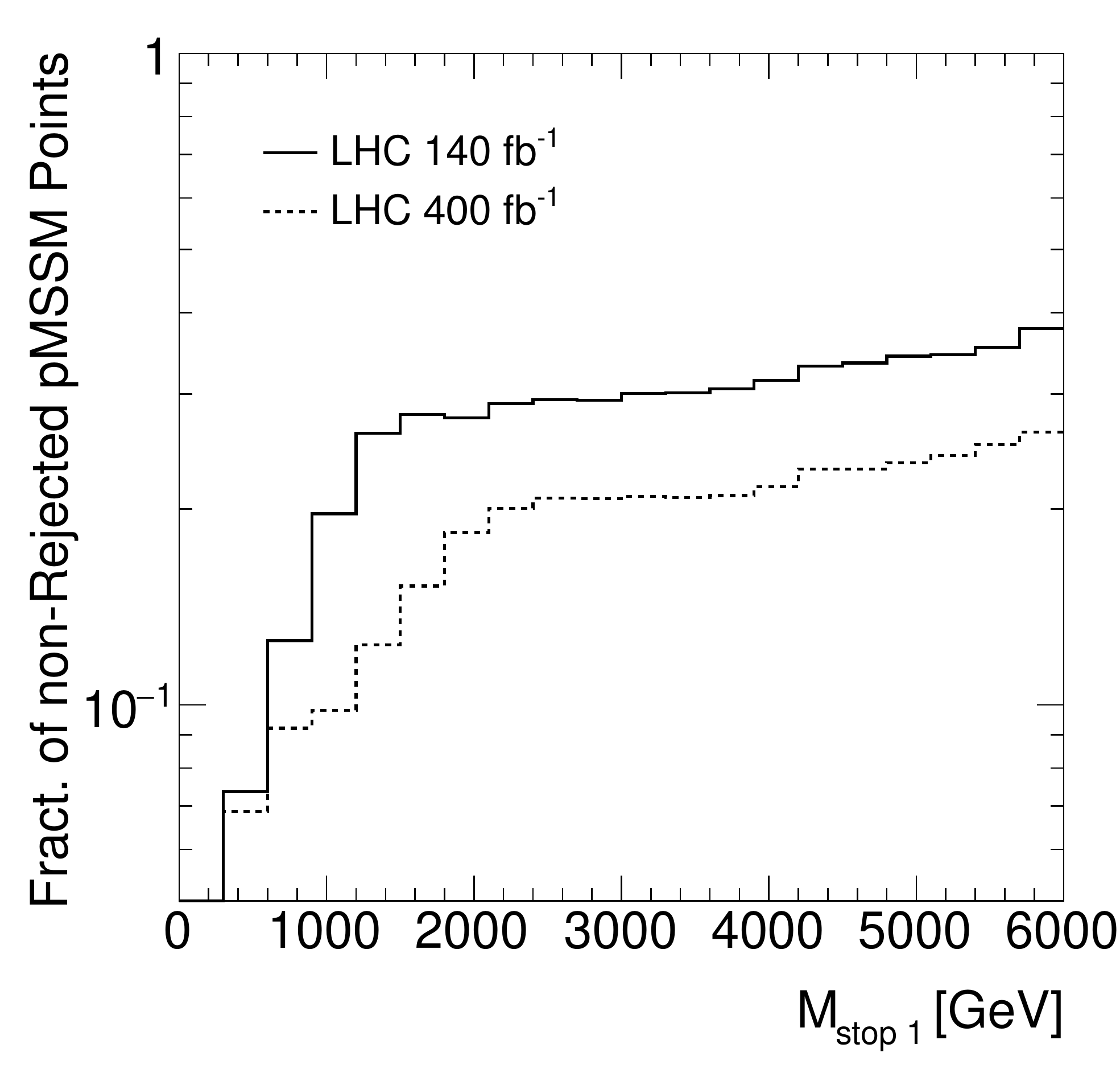} \\
\includegraphics[width=0.35\textwidth]{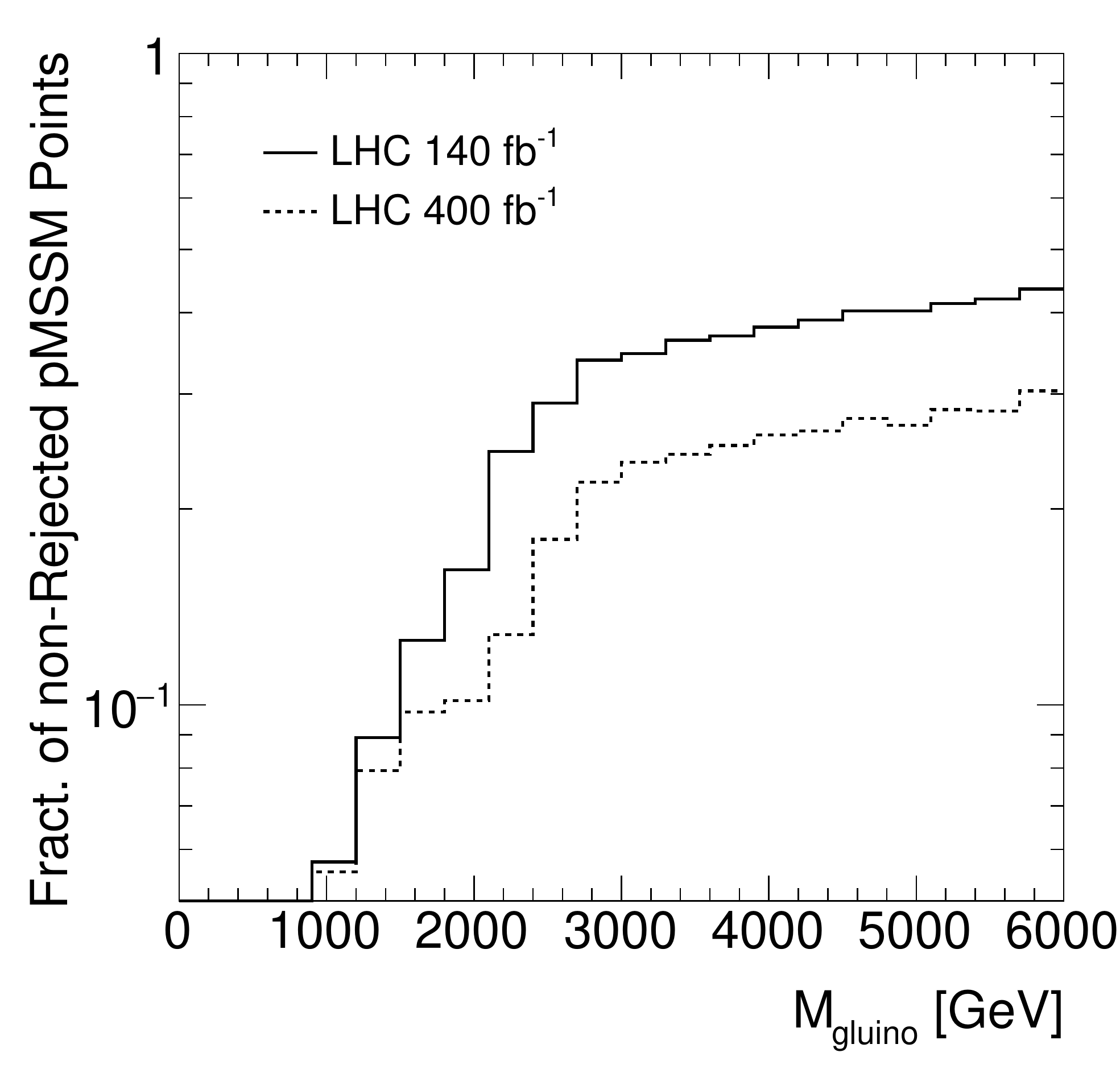} \\
\includegraphics[width=0.35\textwidth]{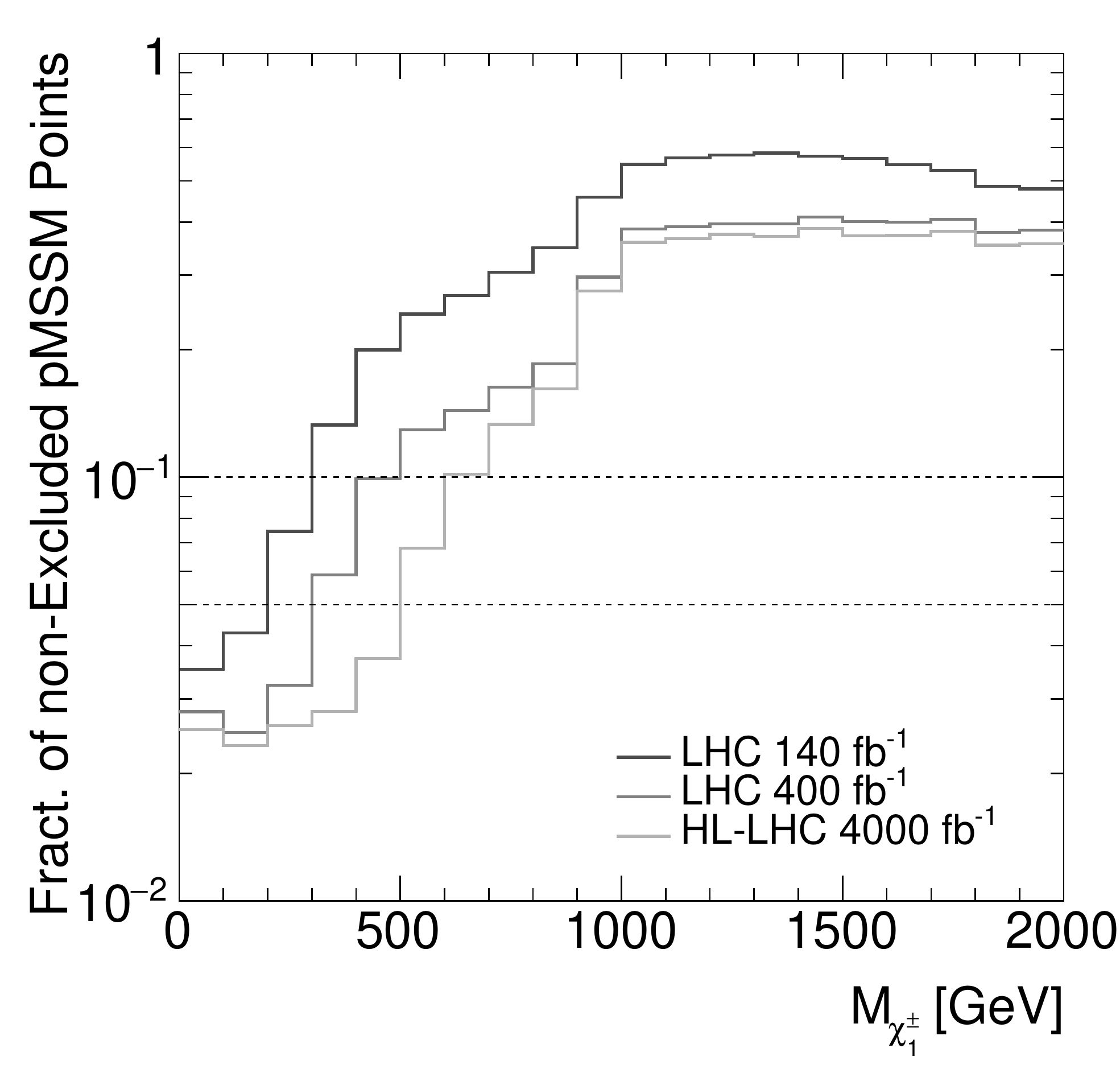} \\
\end{tabular}
\end{center}
\caption{Fraction of accepted pMSSM points not excluded at 95\% C.L. by the present jet/$\ell$+MET searches and beyond the expected sensitivity of Run~3 and HL-LHC as a function of the scalar top (top panel), gluino (centre panel), and lightest chargino (bottom panel) masses.}
\label{fig:msqlhc}
\end{figure}
The fraction of accepted pMSSM points not excluded by the present searches and beyond the sensitivity of Run~3 and the HL-LHC are shown in Figure~\ref{fig:malhc} for the $H$/$A$ search channels as a function of $M_A$ and in Figure~~\ref{fig:msqlhc} for the jet/$\ell$ + MET searches as a function of the $\tilde{t}_1$, $\tilde{g}$, and $\tilde{\chi}^0_1$ masses.

In particular, the $M_A$ values at which more than 90\% of the accepted
pMSSM points are excluded by the $H$/$A$ searches at 95\% C.L. are
$\sim$500~GeV for the present Run~2 ATLAS data, 1000~GeV for the Run~3,
and 1400~GeV for the HL-LHC statistics. The fraction of pMSSM points
excluded by the $H$/$A \rightarrow \tau \tau$ and the jets and leptons +
MET searches are summarised in Table~\ref{tab:methiggs}. The
jet/$\ell$+MET sear\-ches have already excluded more than 90\% of the accepted pMSSM points up to a gluino mass of 1400~GeV, a lighter scalar top mass of 400~GeV, and a lightest chargino mass of 200~GeV, with an expected sensitivity extending to 2000, 1200, and 500~GeV, respectively, by the end of the HL-LHC program. 

\subsection{Higgs Measurement Accuracy at the LHC and $e^+e^-$ Colliders}
\label{sec:4-1}

The measurements of the Higgs properties provide at least two sets of constraints on the MSSM. First, the prediction of the $M_h$ value in SUSY provides a constraint on the parameters when the measured value is imposed. Given the measured value of 125~GeV, the MSSM parameters need to be properly chosen to reach this mass, for example, by selecting the so-called  ``maximal mixing'' solutions. While imposing the constraint 122~GeV $< M_h <$ 128~GeV keeps only $\sim$10\% of the accepted pMSSM points, the $M_h$ value {\it per se} does not provide any discrimination between SM and SUSY Higgs.  Then, the Higgs decay yields in the accessible final states provide us with an opportunity to further constrain the MSSM parameter space and possibly tell a SUSY Higgs from the SM boson.

In this study, we consider the values of the coupling modifiers, $\kappa_X$, obtained by ATLAS combining the present analyses of Run~2 data with luminosities up to 139~fb$^{-1}$ \cite{ATL-CONF-2021-053} given in Table~\ref{tab:lhcMeasHk} and the HL-LHC program~\cite{ATL-PHYS-PUB-2018-054} given in Table~\ref{tab:EuStrKappa}. The accuracies for the determination of the Higgs couplings at future $e^+e^-$ Higgs factories are taken from the compilation in~\cite{deBlas:2019rxi} and given in Table~\ref{tab:EuStrKappa}.

\begin{table}[h!]
\renewcommand{\arraystretch}{1.5}
\begin{center}
\caption{Best fit values for the Higgs boson coupling modifiers $\kappa_X$ from the combination of the ATLAS measurements, with effective photon and gluon couplings under the assumption that the SM decay channel saturates the Higgs decay width~\cite{ATL-CONF-2021-053}.}
\begin{tabular}{|l|c|}
\hline
Channel               & 13~TeV  \\ 
                      & 25-79.8~fb$^{-1}$ \\
\hline \hline
$\kappa_W$  & 1.06 $\pm$ 0.06 \\
$\kappa_Z$  & 0.99 $\pm$ 0.06 \\
$\kappa_t$  & 0.92 $\pm$ 0.10 \\
$\kappa_b$  & 0.87 $\pm$ 0.11 \\
$\kappa_\tau$   & 0.92 $\pm$0.07 \\
$\kappa_\gamma$   & 1.04 $\pm$ 0.06 \\
$\kappa_g$  & 0.92 $^{+0.07}_{-0.06}$ \\
\hline
\end{tabular}
\label{tab:lhcMeasHk}
\end{center}
\renewcommand{\arraystretch}{1}
\end{table}

For each accepted pMSSM point, the values of the coupling modifiers, $\kappa_X$, for the accessible channels are computed using {\tt HDECAY}. The compatibility of the pMSSM with the LHC measurements is determined by computing the $\chi^2$ value as $\chi^2 = x^{T} M^{-1} x$, where $x$ is the vector, of size $N$, of the differences between the values of the coupling modifiers predicted for the pMSSM point $x_{\mathrm{pMSSM}}$ and those measured by the experiment $x_{\mathrm{exp}}$ (or the SM predictions for the future results) and $M$ is the $N \times N$ covariance matrix of the measurement, including correlations, reported by the experiment~\cite{ATL-CONF-2021-053}. This expression extends the $\chi^2$ definition to a set of $N$ correlated variables. The compatibility of each pMSSM point is determined at a given C.L.\ from the $\chi^2$ probability $\mathrm{Prob}(\chi^{2},N)$, where the number of degrees of freedom corresponds to the number of observables, $N$.
For the present results of Ref.~\cite{ATL-CONF-2021-053}, the fitted values are used as central values $x_{\mathrm{exp}}$, while for the projected performance at Run~3 and HL-LHC the $x_{\mathrm{exp}}$ are taken to be the SM values and the correlation matrix is assumed to be the same as that of the current combination~\cite{ATL-CONF-2021-053}.

At an $e^+e^-$ collider, the $e^+e^- \rightarrow Zh \rightarrow \ell \ell X$ process is available for a model-independent determination of the production cross section from the recoil mass of the $\ell \ell$ system and the branching fraction can be directly measured \cite{e+e-earlier}. The accuracy in the determination of the Higgs decay branching fractions and couplings have been studied in great details first for an $e^+e^-$ linear collider at $\sqrt{s}$ = 250 -- 500~GeV~\cite{AguilarSaavedra:2001rg,Kuhl:2007zza} and, more recently, extended to higher centre-of-mass energies. 
Results obtained first with fast simulation have been confirmed by subsequent studies based on detailed {\sc Geant-4} full simulation and reconstruction with the inclusion of (at least some of) the beam-induced backgrounds. Detector R\&D has also greatly progressed towards a validation of the unprecedented response performances assumed in the ILC studies. An important outcome of the studies of $e^+e^-$ collisions at energies above 500~GeV has been the indications of significant improvements in the determination of the Higgs properties through the $e^+e^- \to W W \nu \nu \to h \nu \nu$ fusion process, whose cross section increases $\propto \log \frac{s}{M_h^2}$ and exceeds the peak value of the $e^+e^- \to h Z$ Higgs-strahlung cross section for $\sqrt{s}$ energies $\ge~480$~GeV. 

\begin{table*}
\renewcommand{\arraystretch}{1.5}
\begin{center}
\caption{Assumed accuracies on the $h$ boson coupling modifiers
$\kappa_i$ for future collider projects (from Ref.~\cite{deBlas:2019rxi}).}
\begin{tabular}{|l|c|c|c|c|c|c|}
\hline 
Channel & HL-LHC  & ILC       & ILC     & ILC   & FCC-ee  & FCC-ee \\
        &         & 250~GeV   & 500~GeV & 1~TeV & 240~GeV & 365~GeV  \\
\hline \hline
$\kappa_W$  & 0.017 & 0.0180 & 0.0029 & 0.0024 & 0.013 & 0.0043\\
$\kappa_Z$  & 0.015 & 0.0029 & 0.0023 & 0.0022 & 0.0020 & 0.0017 \\
$\kappa_t$  & 0.033 & -- & 0.0690 & 0.016 & -- & -- \\
$\kappa_b$  & 0.036 & 0.0180 & 0.0058 & 0.0048 & 0.0130 & 0.0067 \\
$\kappa_c$  & -- & 0.025 & 0.0130 & 0.0090 & 0.018 & 0.013\\
$\kappa_{\tau}$  & 0.019 & 0.0190 & 0.0070 & 0.0057 & 0.0140 & 0.0073 \\
$\kappa_{\gamma}$ & 0.019 & 0.0670 & 0.034 & 0.019 & 0.047 & 0.039 \\
$\kappa_g$  & 0.023 & 0.0230 & 0.0097 & 0.0066 & 0.0170 & 0.0100 \\
\hline
\end{tabular}
\label{tab:EuStrKappa}
\end{center}
\renewcommand{\arraystretch}{1}
\end{table*}

\section{Results}
\label{sec:5}
In this section, we analyse the potential of measurements of the Higgs boson effective couplings at the LHC and future colliders, discussed in the previous sections, to exclude, or identify, MSSM solutions in the parameter space of the pMSSM not excluded by direct searches for heavy Higgs bosons and for SUSY particles in channels with missing $E_T$ signatures. The analysis uses as inputs the coupling modifier terms, $\kappa_X$, obtained in the context of the so-called  $\kappa$ framework, discussed in Section~\ref{sec:3}, by assuming the central values and uncertainties summarised in Section~\ref{sec:4-1}. 

\begin{figure*}[ht!]
\begin{center}
  \begin{tabular}{cc}
    \includegraphics[width=0.40\textwidth]{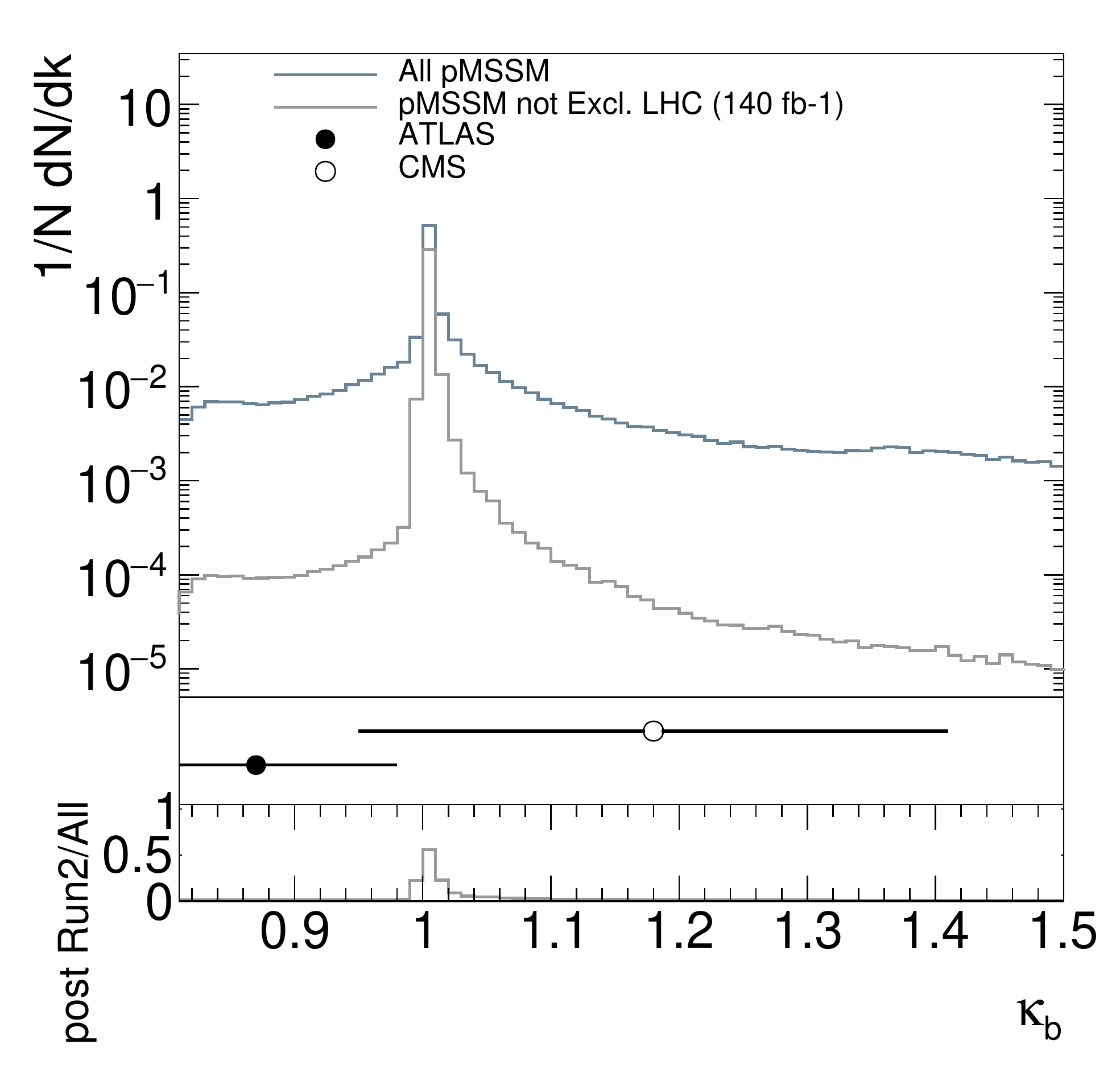} &
    \includegraphics[width=0.40\textwidth]{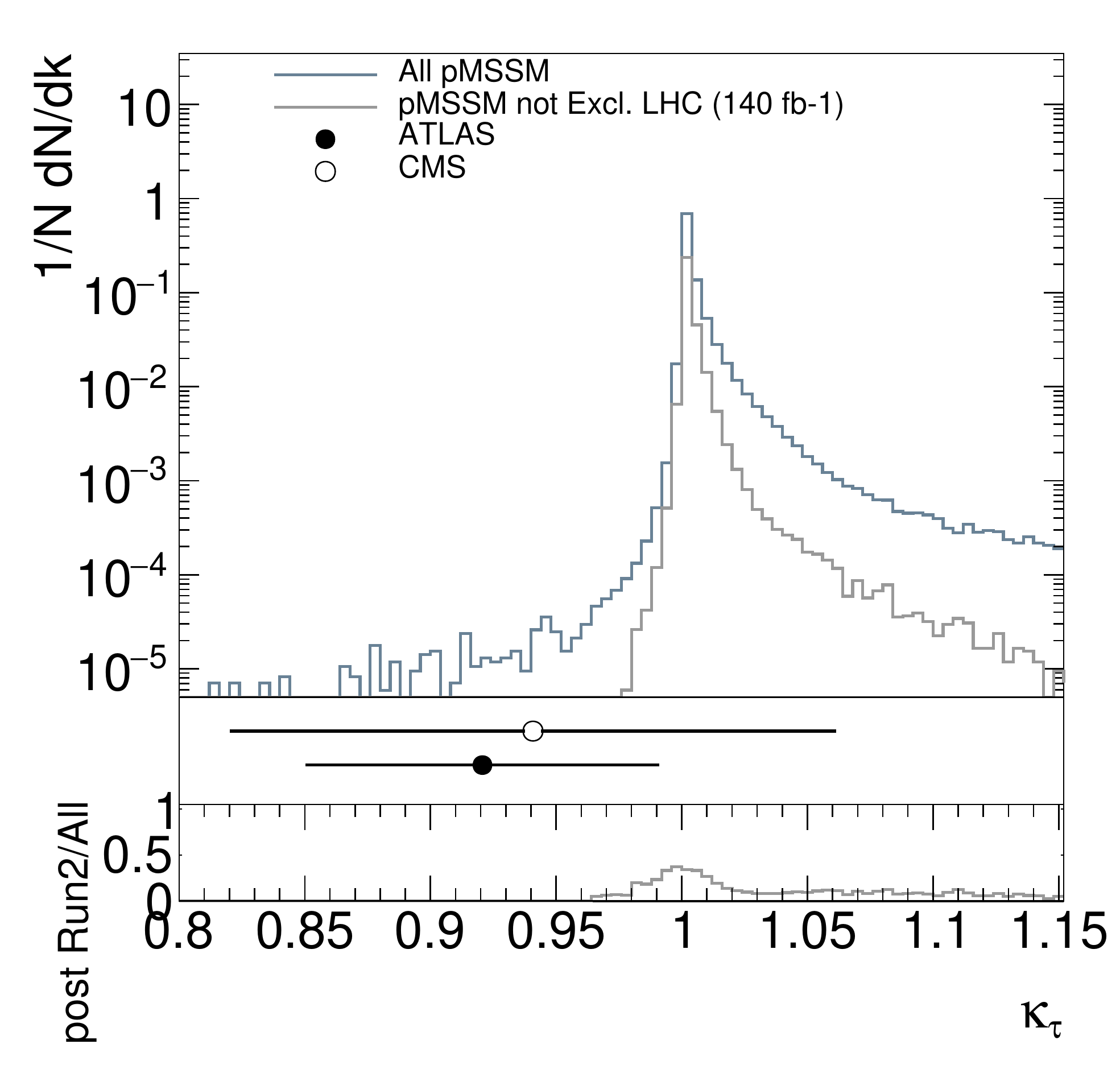} \\
    \includegraphics[width=0.40\textwidth]{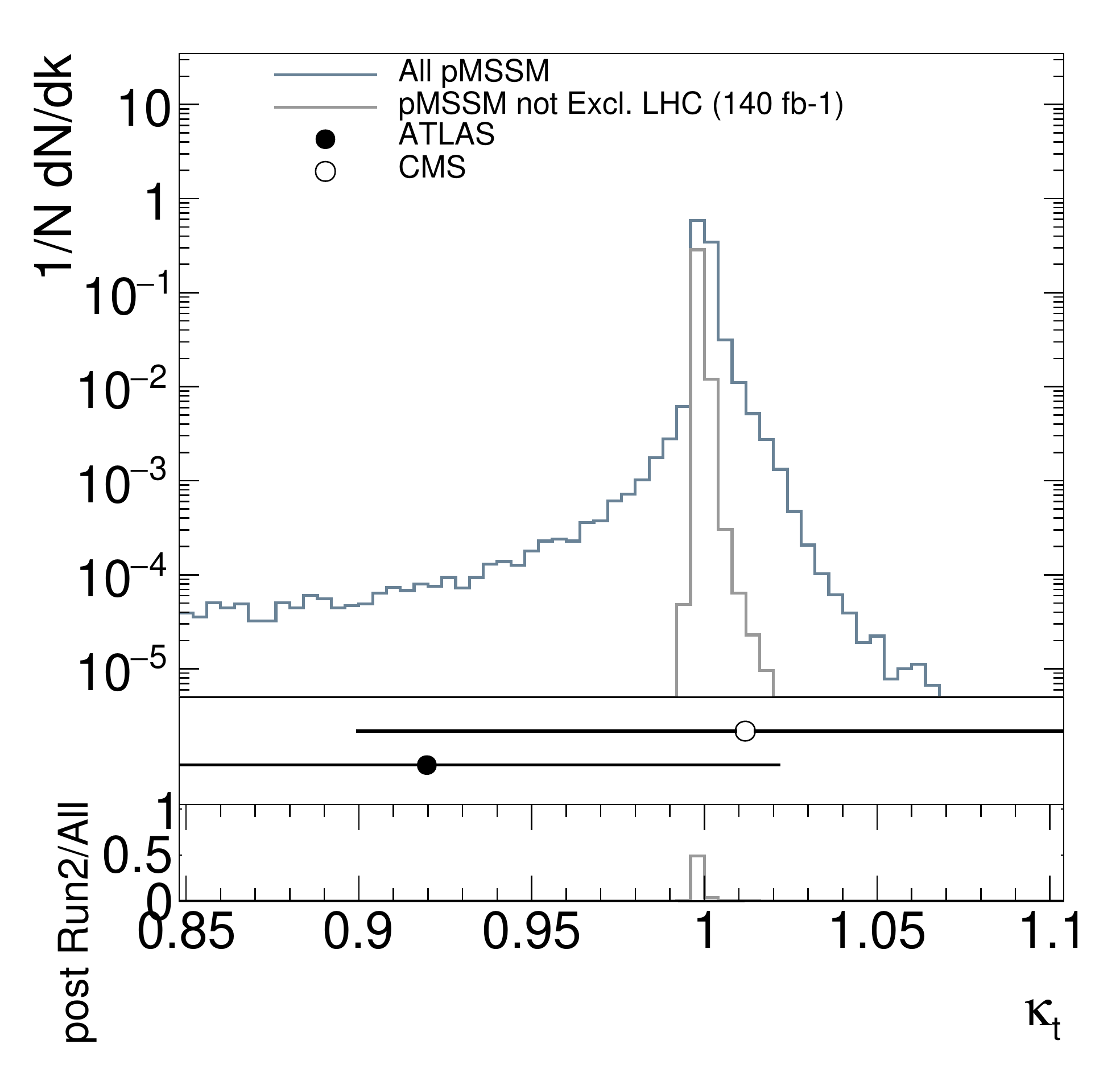} &
    \includegraphics[width=0.40\textwidth]{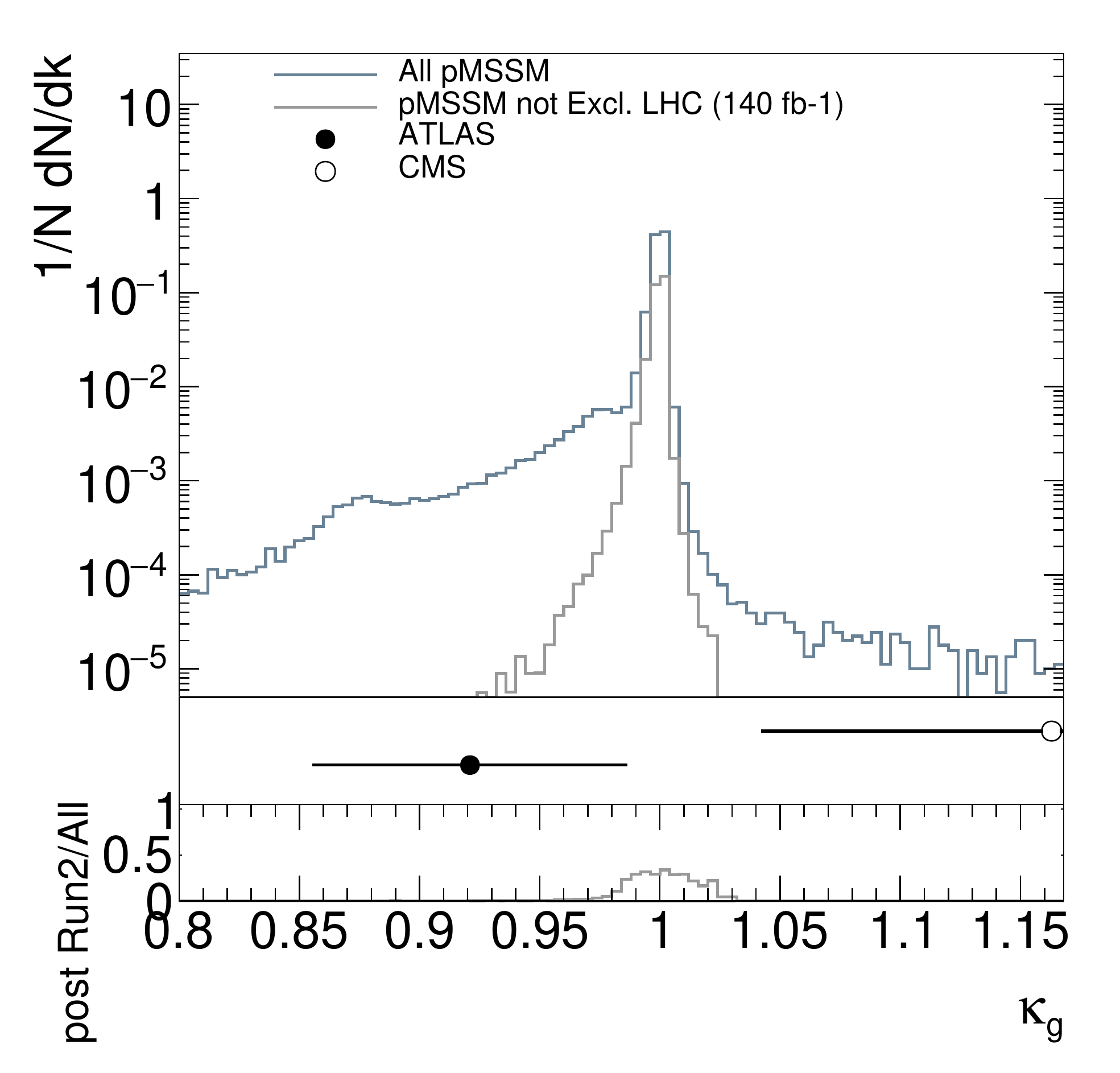} \\
  \end{tabular}
\end{center}
\caption{$h$ Higgs boson coupling modifiers, $\kappa_X$, to $b$ quarks (upper left), $\tau$ leptons (upper right), top quarks (lower left) and gluons (lower right) for all valid pMSSM points and those not excluded by the LHC Run~2 searches compared to the present measurements by the ATLAS~\cite{ATL-CONF-2021-053} and CMS~\cite{CMS-PAS-HIG-19-005} experiments. The lower panels show the fractions of non-excluded pMSSM points as a function of $\kappa_X$. \label{fig:KpMSSM}}
\end{figure*}

\begin{figure*}[ht!]
\begin{center}
  \begin{tabular}{cc}
    \includegraphics[width=0.40\textwidth]{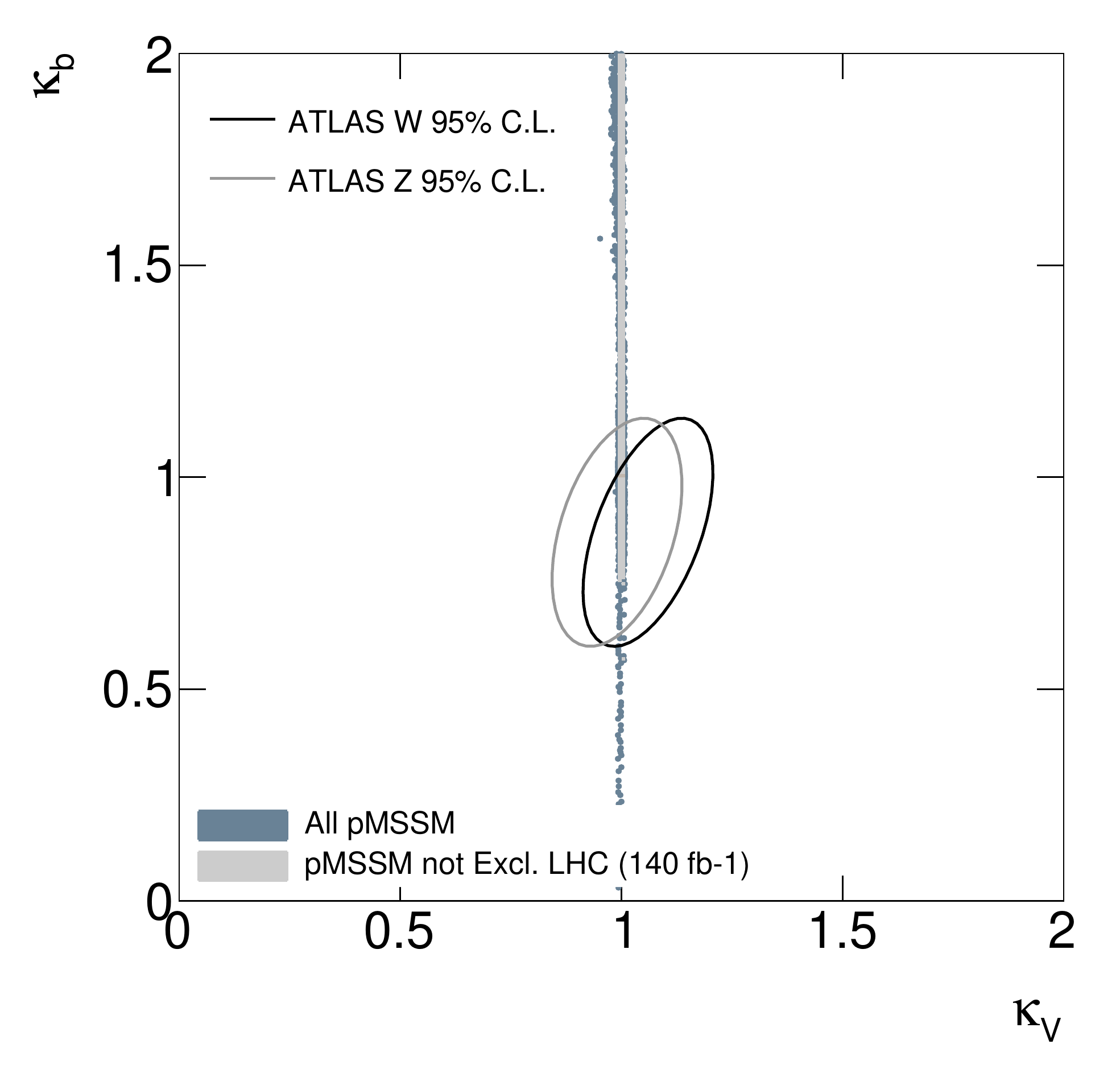} &
    \includegraphics[width=0.40\textwidth]{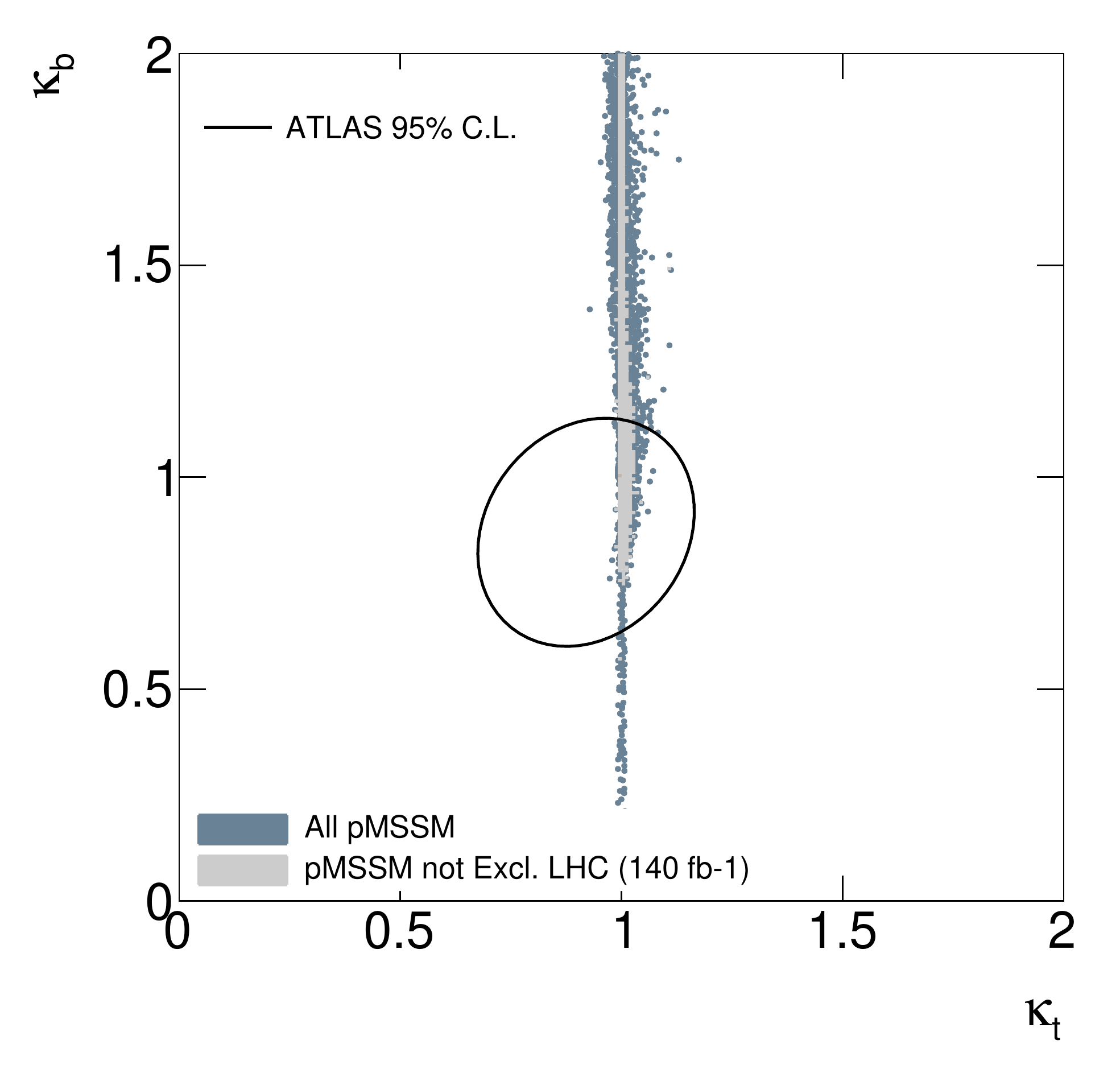} \\
    \includegraphics[width=0.40\textwidth]{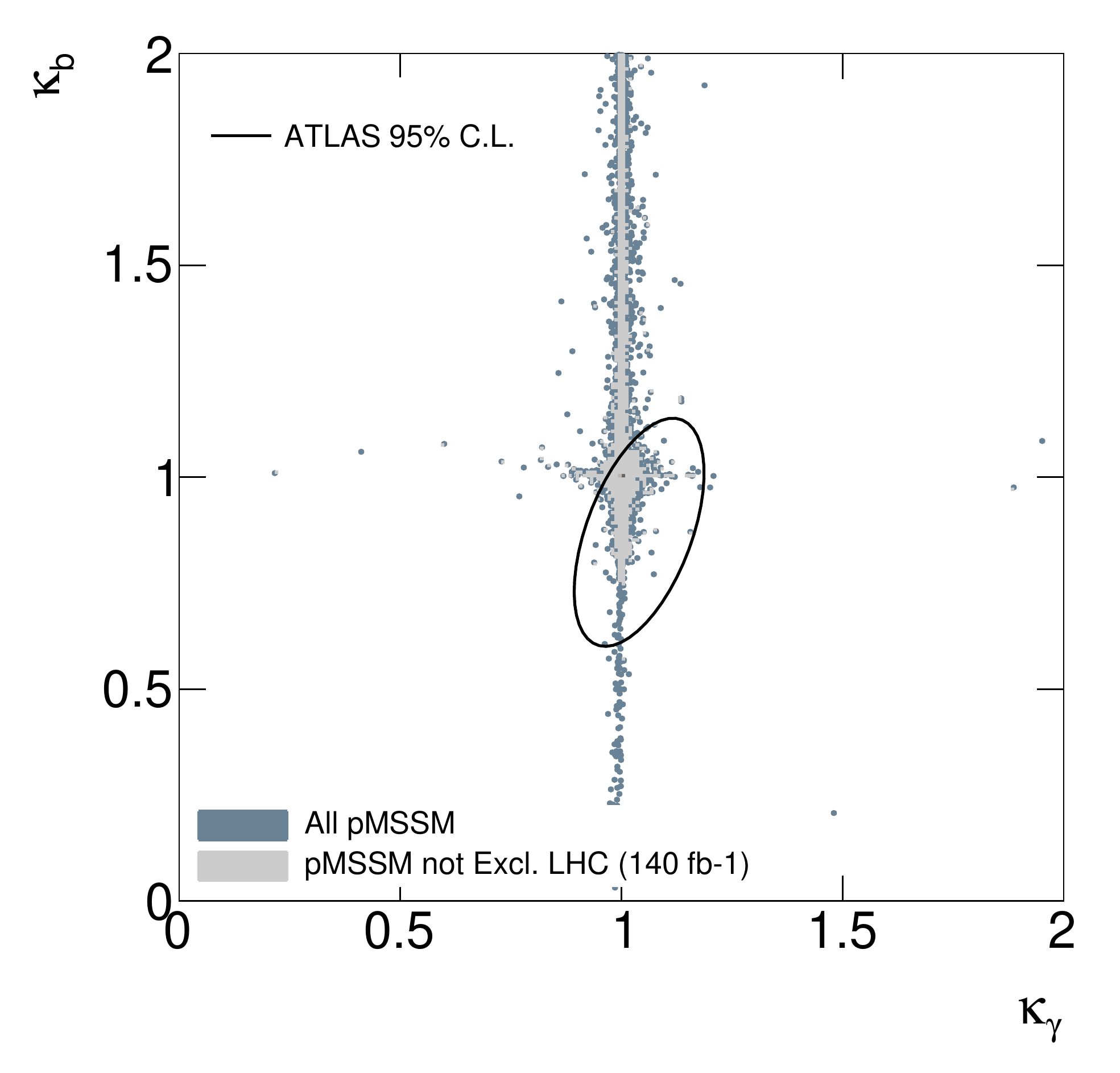} &
    \includegraphics[width=0.40\textwidth]{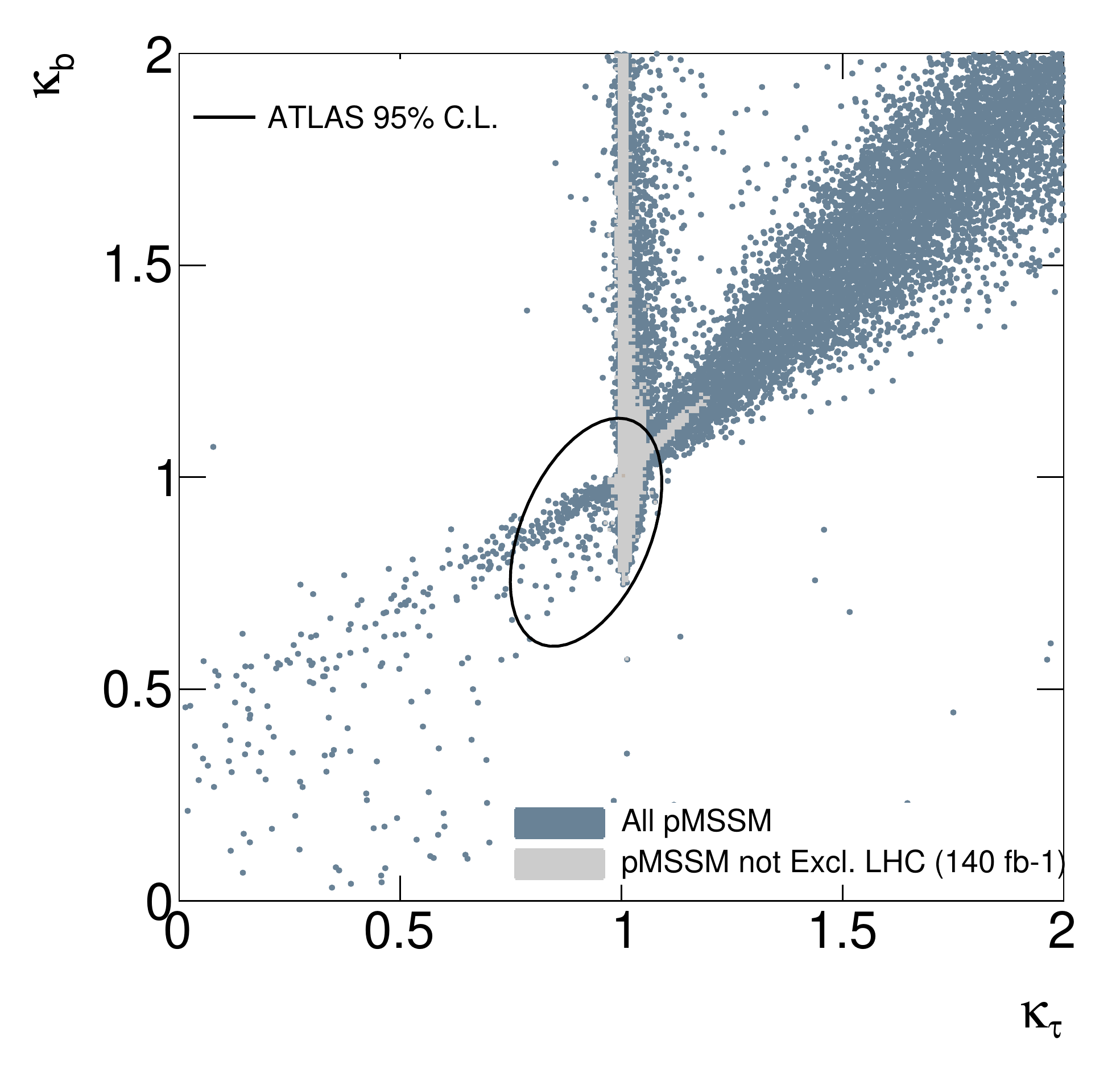} \\
  \end{tabular}
\end{center}
\caption{Correlations of the $h$ Higgs boson $\kappa$ coupling modifiers comparing the valid pMSSM points, those not excluded by the LHC Run~2 searches and the 95\% C.L. contours of the current measurements by the ATLAS experiment~\cite{ATL-CONF-2021-053}.
\label{fig:k2D}}
\end{figure*}

First, we compare the distributions of these coupling modifiers predicted for the valid pMSSM points in our scans to those obtained for the points not excluded by the direct searches in Run~2 and to the current measurements (see Figure~\ref{fig:KpMSSM}).
It is also instructive to compare the coupling modifier values, $\kappa_X$, for the pMSSM points to those obtained so far for the LHC data by correlating particles as shown in Figure~\ref{fig:k2D}.

In all these comparisons, the current experimental accuracy is close to the full spread of the pMSSM predictions. In this sense, the statement that the properties of the observed Higgs boson are SM-like could also be rephrased by saying that they are also MSSM-like. However, a fraction of the pMSSM solutions is found to be incompatible with these measurements, and more will be tested at the HL-LHC and later by a possible $e^+e^-$ Higgs factory. It is interesting to understand the Higgs coupling properties of the points that are preferentially discarded by direct searches. Comparing the $\kappa_X$ distributions for all the valid pMSSM points to those for the points not excluded by the direct LHC searches for heavy Higgs bosons and those for SUSY particles in missing $E_T$ channels, it is evident that these preferentially excluded pMSSM points are located on the tails of $k_i$ distributions at values away from the SM predictions, corresponding to $\kappa_X$=1. The large reduction of pMSSM points at $\kappa$ values significantly above the SM expectations for the $b$ quarks and $\tau$ leptons induced by the SUSY searches is mostly due to the constraint on $M_A$ set by the searches in the $H$/$A \rightarrow \tau^+ \tau^-$ channel.

These effects are discussed in quantitative terms in the next section. This is done by studying the fractions of valid and accepted pMSSM points that can be excluded by the Higgs coupling measurements with the accuracies expected for the next steps of the LHC program and for future colliders in relation to their observability at the LHC in new particle searches. In addition, the constraints that can be derived from these measurements on the relevant MSSM parameters are discussed.

\subsection{Constraints on the pMSSM}
\label{sec:5-1}

The compatibility of the Higgs couplings with the predictions for the SM or for specific MSSM scenarios defines a way to study the sensitivity for identifying SUSY and discriminate between different scenarios as a function of the collider data accuracy in the light Higgs couplings. Since increasing the integrated luminosity increases the precision of the Higgs property measurements as well as the mass bounds from direct searches, if no signal is observed, this study estimates also the complementarity between the direct (jets / $\ell$s + MET and heavy Higgs bosons) and indirect (Higgs properties) probes of new physics in the context of SUSY. In particular, we study how the indirect sensitivity evolves with increasing the accuracy of the Higgs measurements and determine the expected impact of the improved accuracy from an $e^+e^-$ Higgs factory, given the current reach of the SUSY searches by the LHC experiments.

\begin{table*}[!ht]
\renewcommand{\arraystretch}{1.5}
\begin{center}
\caption{Fraction of pMSSM points excluded at 90\% C.L. by the Higgs coupling measurements with the LHC Run 2 and the estimated HL-LHC accuracies. The first lines of each block gives the fractions for all valid and accepted pMSSM points, the following row the fractions of the pMSSM points excluded only by Higgs couplings and not by the searches for $H$/$A$ heavy Higgs bosons and those in the jets and/or leptons + MET channels with the Run~2 and the HL-LHC sensitivity.}%
\begin{tabular}{|l|c|c|}
\hline        
 pMSSM      & LHC           &  HL-LHC    \\
 Selection  & 140~fb$^{-1}$ & 3 ab$^{-1}$ \\
\hline \hline           
Valid excl. by Higgs couplings & 0.023 & 0.077 \\
Excl. only by Higgs couplings & 0.015 & 0.024 \\ \hline
Accepted excl. by Higgs couplings & 0.012 & 0.047 \\
Excl. only by Higgs couplings & 0.006 & 0.019 \\
\hline
\end{tabular}
\label{tab:HiggsLHC}
\end{center}
\renewcommand{\arraystretch}{1}
\end{table*}

\begin{table*}[h!]
\renewcommand{\arraystretch}{1.5}
\begin{center}
\caption{Fraction of pMSSM points excluded at 90\% C.L. by the Higgs coupling measurements assuming SM central values and the accuracies estimated for the HL-LHC and the $e^+e^-$ ILC and FCC-ee Higgs factory proposals. The fractions excluded by the current LHC Run-2 Higgs coupling results are also given in the first column. The first lines of each block gives the fractions of all valid (neutralino LSP+122 $< M_h <$ 128~GeV) and accepted (valid+flavour+$\Omega_{\chi} h^2$) pMSSM points, respectively, followed by the fractions for the points not excluded by the direct searches for $H$/$A$ heavy Higgs bosons and also by those in the jets and/or leptons + MET channels with the Run~2 sensitivity.}%
\begin{tabular}{|l|c|c|c|c|c|c|c|}
\hline        
 pMSSM      & LHC           &  HL-LHC    & ILC      & ILC     & ILC   & FCC-ee  & FCC-ee \\
 Selection  & 140~fb$^{-1}$ & 3 ab$^{-1}$ & 250~GeV  & 500~GeV & 1~TeV & 240~GeV & 365~GeV \\
\hline \hline           
Valid                    & 0.023 & 0.077 & 0.083 & 0.166 & 0.191 & 0.102 & 0.156 \\
$H/A$ (LHC Run-2)        & 0.017 & 0.061 & 0.068 & 0.115 & 0.133 & 0.078 & 0.108 \\
$H/A$ \& $j/\ell$ + MET (LHC Run-2) & 0.016 & 0.059 & 0.066 & 0.114 & 0.130 & 0.076 & 0.106 \\ \hline
Accepted                 & 0.012 & 0.048 & 0.054 & 0.137 & 0.161 & 0.072 & 0.125 \\
$H/A$ (LHC Run-2)        & 0.012 & 0.047 & 0.053 & 0.107 & 0.125 & 0.067 & 0.098 \\
$H/A$ \& $j/\ell$ + MET (LHC Run-2) & 0.011 & 0.046 & 0.051 & 0.106 & 0.123 & 0.065 & 0.097 \\
\hline
\end{tabular}
\label{tab:HiggsFutureColl}
\end{center}
\renewcommand{\arraystretch}{1}
\end{table*}

The test is performed by computing the $\chi^2$ probability of the Higgs observables for each pMSSM point with respect to the actual measurements or the SM hypothesis and measuring the fraction of accepted pMSSM points incompatible with the SM at the 90\% C.L. This is the fraction of pMSSM points that can be tested and excluded, assuming that the measured Higgs couplings coincide exactly with those predicted by the SM.

The fractions of points within the range of the scans performed here (see Table~\ref{tab:paramSUSY}) excluded by the SUSY direct searches (see Table~\ref{tab:lhc-susy}) are summarised in Table~\ref{tab:methiggs}. Considering the valid and accepted pMSSM points and following the fractions of those that can be excluded by the Higgs measurements at the LHC, we observe that these fractions range from 1 to 2\% and 5 to 8\% for the Run~2 and HL-LHC datasets, respectively. If we restrict ourselves to considering the points that can be excluded only by the Higgs coupling measurements and are not excluded by the direct SUSY searches (heavy Higgs bosons and scalar quarks, leptons, and gauginos) conducted on the same datasets, these fractions become approximately 1 and 2\% (see Table~\ref{tab:HiggsLHC}. Despite the increasing accuracy of the Higgs measurements, the fraction of pMSSM points that can be excluded by the Higgs coupling measurements during the LHC program remains small. However, their increase indicates that the improvement of the sensitivity obtained by the Higgs measurements with higher accuracy beats the estimated increase in sensitivity of the direct searches moving from Run~2 to the HL-LHC.

The fractions of points excluded by the Higgs couplings at the LHC and at future $e^+e^-$ colliders are presented in Table~\ref{tab:HiggsFutureColl}, in relation to the current sensitivity of the direct searches at LHC Run~2. The reduction of sensitivity of the Higgs coupling measurements observed for the accepted points, fulfilling the flavour physics constraints, and for the points not excluded by the direct searches comes almost entirely from the exclusion of the low to moderate $M_A$ scenarios in both cases. The exclusion of the pMSSM at moderate values of $M_A$ pushes the $h$ boson into the decoupling regime. The deviations of the Higgs couplings due to the $\Delta_b$ effect vanishing in the decoupling regime, the constraints from missing $E_T$ searches on the mass of scalar quarks and gauginos have only a minor effect. By improving the accuracy of these Higgs coupling measurements to the percent, or subpercent, level, as expected at future $e^+e^-$ Higgs factories, the ILC and the FCC-ee, can test up to about 10\%--12\% of the pMSSM points not excluded by the current LHC direct SUSY searches and flavour physics data (see Table~\ref{tab:HiggsFutureColl}), provided they can operate at a large enough energy to perform a full study of the Higgs profile.

\subsection{Extracting MSSM Parameters from the Higgs Profile}
\label{sec:5-2}
By studying the Higgs couplings as a function of the MSSM parameters for the accepted pMSSM points, we observe four main groups of parameters to which the Higgs couplings are sensitive. In general these are: $M_A$, $M_{\tilde{g},\tilde{b},\tilde{t},\tilde{\tau}}$, $\mu \tan \beta$ and $M_{\tilde{\chi}^0_1}$, but this sensitivity is strongly reduced when only the pMSSM points compatible with the LHC Run~2 searches are considered. 

The small fractions of pMSSM points viable after the Run~2 searches that can be excluded by future precision measurements of the Higgs couplings,
discussed in the previous section, highlight the decoupling properties of the lightest MSSM Higgs boson, once the value of $M_A$ is constrained to sufficiently large values. In this regime, the modifications of the Higgs coupling to $b \bar b$ as a function of
$M_{\tilde{g},\tilde{b},\tilde{t},\tilde{\tau}}$ and $\mu \tan \beta$ through the $\Delta_b$ term are suppressed. Among the pMSSM parameters inducing effects on the Higgs couplings, discussed in Sections~\ref{sec:2} and \ref{sec:3}, only $M_{\tilde{\chi}^0_1}$, through invisible decays if $M_{\chi} < 0.5 M_h$, and $M_A$ remain viable.

\subsubsection{$M_{\tilde{\chi}^0_1}$}

\begin{figure}[h!]
\begin{center}
\includegraphics[width=0.35\textwidth]{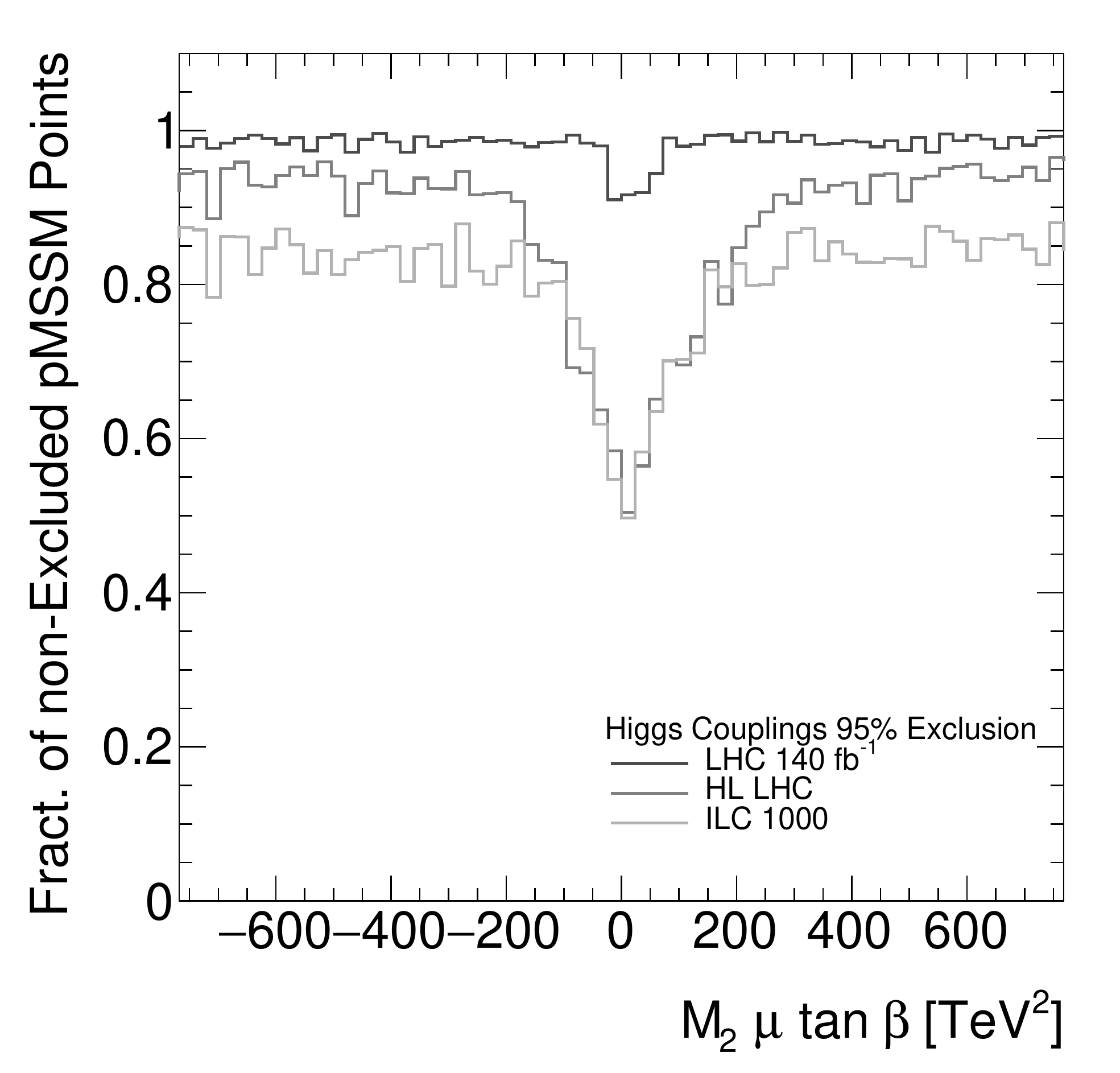} \\
\end{center}
\caption{Fraction of accepted pMSSM points not excluded at the 95\% of C.L. by the Higgs couplings as a function of $M_2 \mu \tan \beta$ for the present Run~2 ATLAS results (dark grey) and the expected HL-LHC (medium grey) and ILC-1000 (light grey) accuracies, assuming SM central values.}
\label{fig:H95M2MuTanb}
\end{figure}

If $M_{\chi} < 0.5 M_h$ the $h$ can decay to neutralino pairs, as discussed in Section~\ref{sec:2-5}. The fraction of accepted pMSSM points compatible with the Higgs couplings is shown as a function of $M_2 \mu \tan \beta$, relevant to the determination of the Higgs invisible decay rate, in Figure~\ref{fig:H95M2MuTanb}.

\begin{figure}[h!]
  \begin{center}
    \begin{tabular}{c}
      \includegraphics[width=0.40\textwidth]{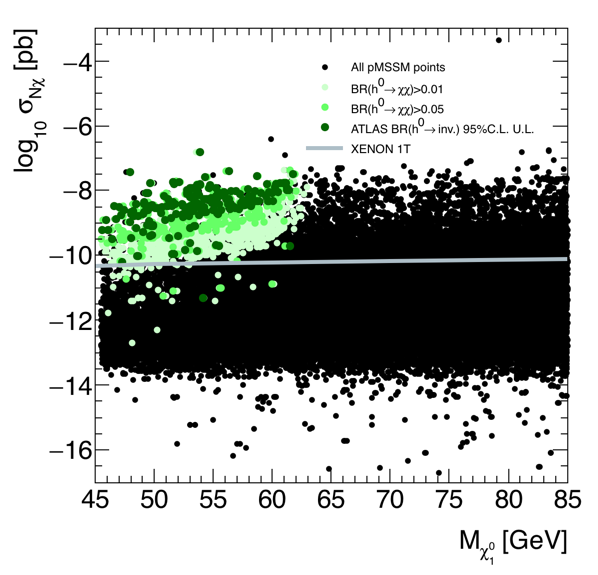} \\
      \includegraphics[width=0.40\textwidth]{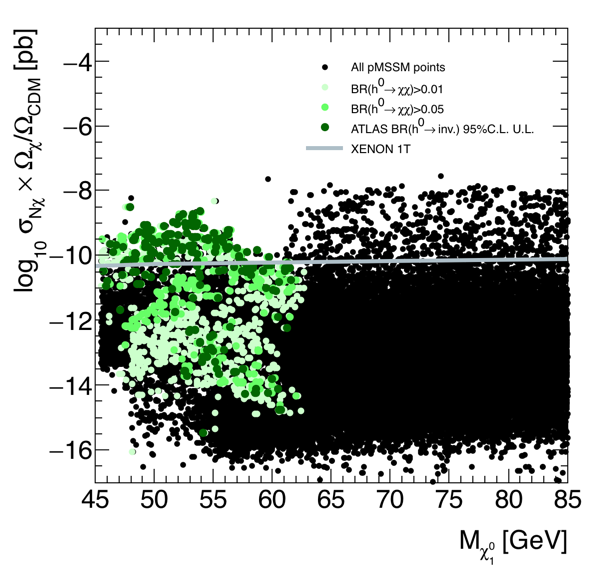} \\
    \end{tabular}
  \end{center}
\caption{Predicted spin-independent $\chi N$ scattering cross section on
nucleons, as a function of $M_{\tilde{\chi}^0_1}$ values with highlighted pMSSM points with sizeable  $h \rightarrow \chi \chi$ branching fractions. Points with sizeable $h \rightarrow \chi \chi$ branching fractions are shown in colour, the darker shade indicating those with branching fractions exceeding the ATLAS upper bounds on invisible Higgs boson decays~\cite{ATLAS-CONF-2020-052}. The line represents the upper bound from the XENON-1T data. The lower panel has the  $\chi N$ scattering cross section values rescaled by $\Omega_{\chi}/\Omega_{\mathrm{CDM}}$.} 
\label{fig:DDSIMN1inv}
\end{figure}
Limits on the invisible decay rate can be used to constrain the
value of the neutralino LSP mass. As we have already pointed out, the
$h \chi \chi$ coupling also controls the $\chi p$ scattering cross section
thus introducing a correlation between the rate of $h \rightarrow \chi \chi$,
the scattering cross section and the neutralino relic density,
$\Omega_{\tilde{\chi}^0_1}$.
A large rate for invisible Higgs decays implies a large $\chi p$
scattering cross section, because they are both due to an enhanced $h
\chi \chi$ coupling. This is shown in Figure~\ref{fig:DDSIMN1inv}
visualising the predicted spin-independent $\chi N$ scattering cross section on nucleons in the portion of our pMSSM points with low $M_{\tilde{\chi}^0_1}$ values highlighting the points with sizeable $h \rightarrow \chi \chi$ branching fractions. Bounds on the $\chi$ scattering cross section place non-trivial constraints on the Higgs invisible rate as shown in the upper panel in Figure~\ref{fig:DDSIMN1inv}. These bounds are relaxed but not invalidated even when the predicted scattering cross section is rescaled by the ratio $\Omega_{\chi}/\Omega_{\mathrm{CDM}}$ for points having neutralino relic density significantly lower that the current PLANCK result for $\Omega_{\mathrm{CDM}}$ (see the lower panel in Figure~\ref{fig:DDSIMN1inv}). 
In particular, the XENON-1T upper limit on the scattering cross section on nucleons for neutralino masses below $M_h/2$ from the 1.0~ton-yr exposure~\cite{Aprile:2018dbl} removes almost all of the MSSM solutions having BR($h \rightarrow \chi \chi$) above 0.01 and provides a competitive constraint compared to the direct current upper bound on Higgs invisible decays at 0.11~\cite{Sirunyan:2018owy,Aaboud:2019rtt,ATLAS-CONF-2020-008, ATLAS-CONF-2020-052} within the MSSM with the lightest neutralino as the only source of dark matter.

This reduces the invisible Higgs rate likely below the sensitivity at the LHC, determined either directly through $ZH$ and VBF production or indirectly through the sum of the Higgs rates, and makes it unique to the $e^+e^-$ collider program. In our pMSSM scenario with $M_{\chi}$ = 58.3~GeV, the fit to the Higgs branching fractions makes it possible to indirectly reconstruct the neutralino mass with better than 10\% relative statistical accuracy.

\subsubsection{$M_A$}

\begin{figure}[hbtp]
\begin{center}
\includegraphics[width=0.35\textwidth]{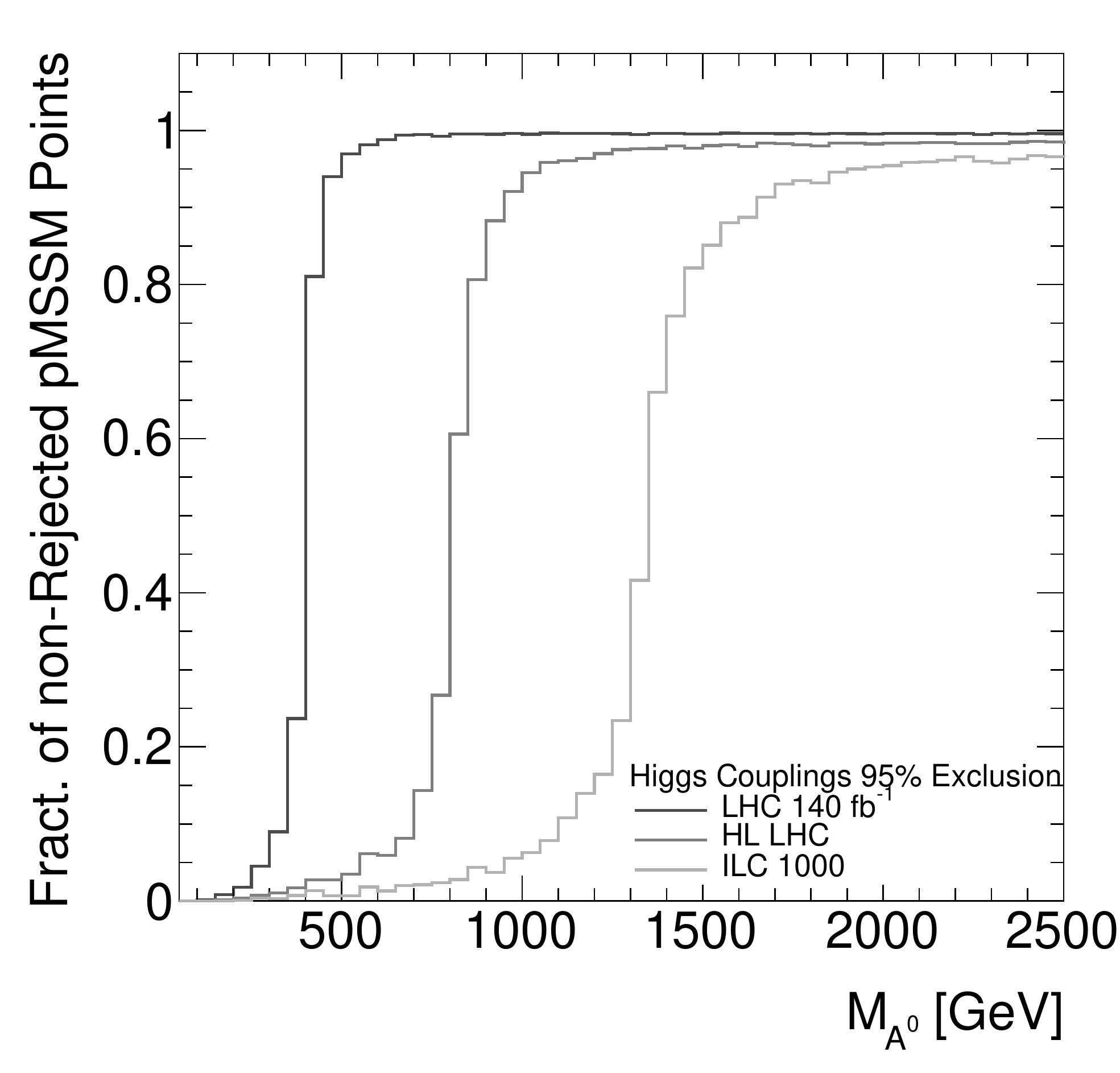}
\end{center}
\caption{Fraction of accepted pMSSM points not excluded at the 95\% of C.L. by the Higgs couplings as a function of the $M_A$ mass for the present Run~2 ATLAS results (dark grey) and the expected HL-LHC (medium grey) and ILC-1000 (light grey) accuracies, assuming SM central values.}
\label{fig:H95MA}
\end{figure}

The sensitivity to the pseudoscalar Higgs boson mass, $M_A$, is the main MSSM benchmark for the light Higgs coupling measurements. The scaling of the coupling deviations with $M_Z^2/M_A^2$ (see section \ref{sec:3-1}) offers us with an opportunity to discriminate a MSSM $h$ from the SM $H$ boson and to infer the mass of the pseudoscalar state from the precision measurements of the properties of the lighter state. The fraction of accepted pMSSM points excluded by the $h$ coupling determination for different accuracies is shown as a function of $M_A$ in Figure~\ref{fig:H95MA}. We have verified that our scans contain points corresponding to the so-called ``alignment scenario'', where small $M_A$ values correspond to SM-like Higgs couplings~\cite{alignment}. However these points in our scans are removed by flavour data and direct searches at the LHC.

\begin{figure*}[hbtp]
  \begin{center}
    \begin{tabular}{ccc}
      \includegraphics[width=0.30\textwidth]{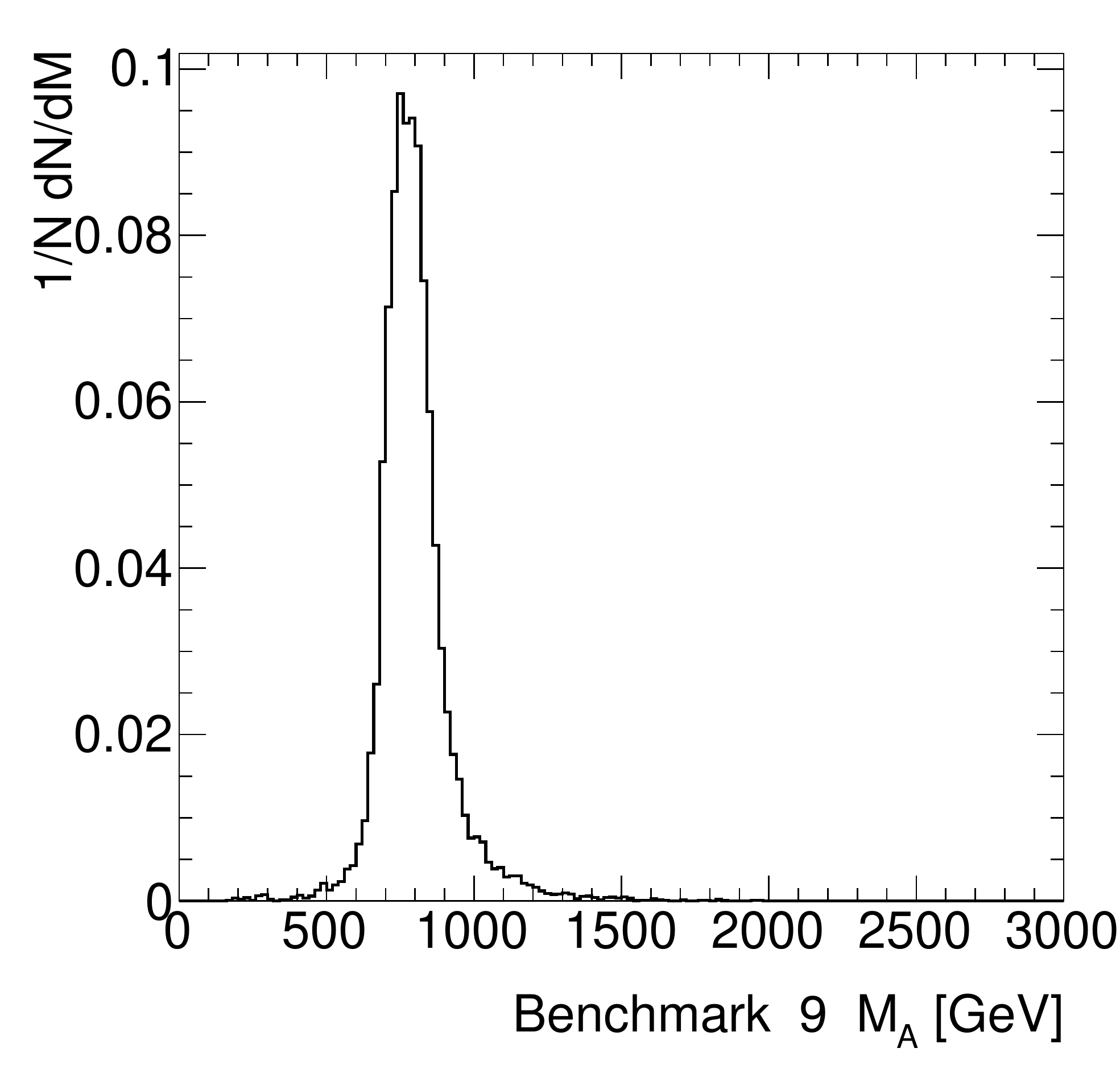} &
      \includegraphics[width=0.30\textwidth]{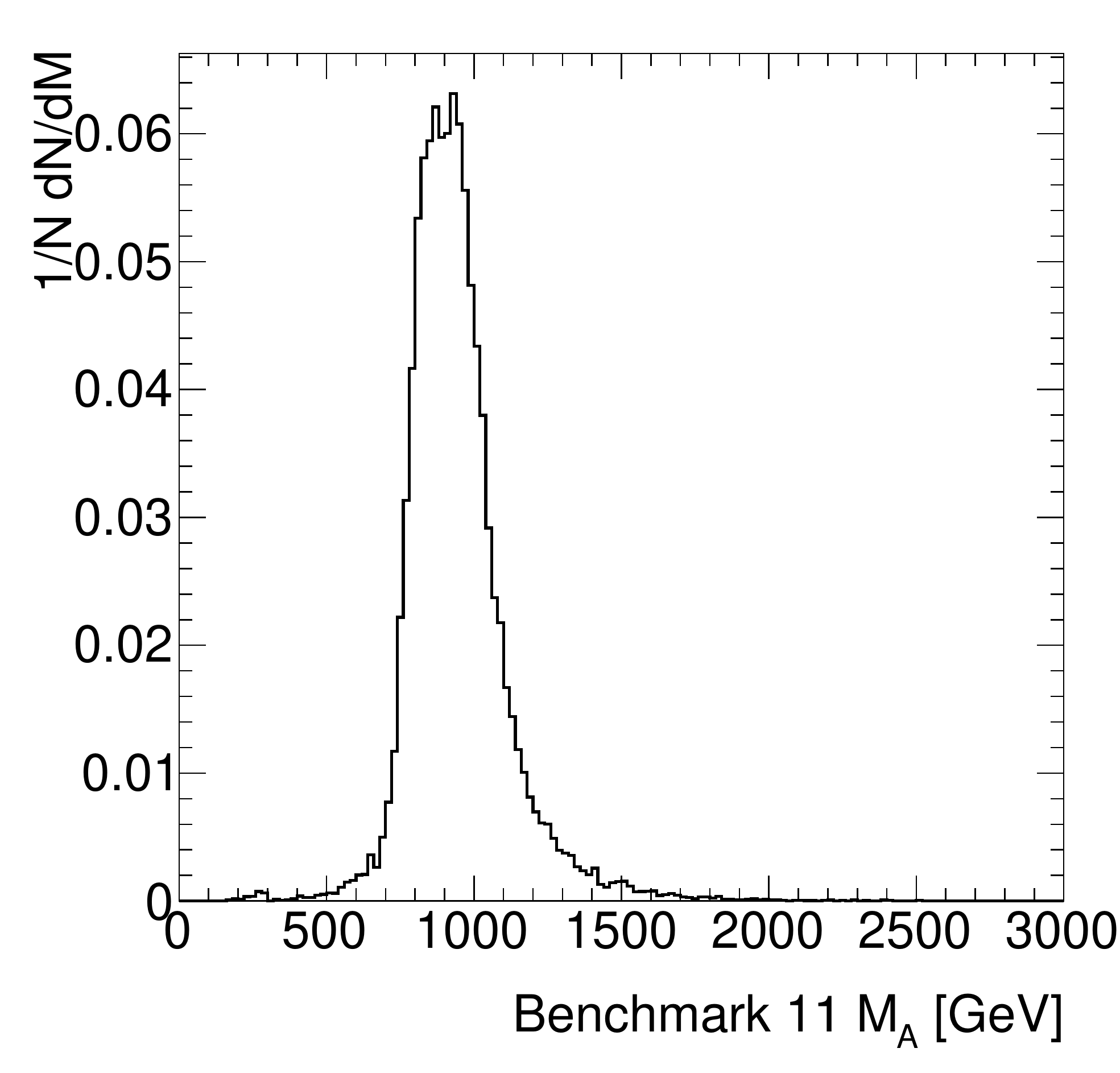} &
      \includegraphics[width=0.30\textwidth]{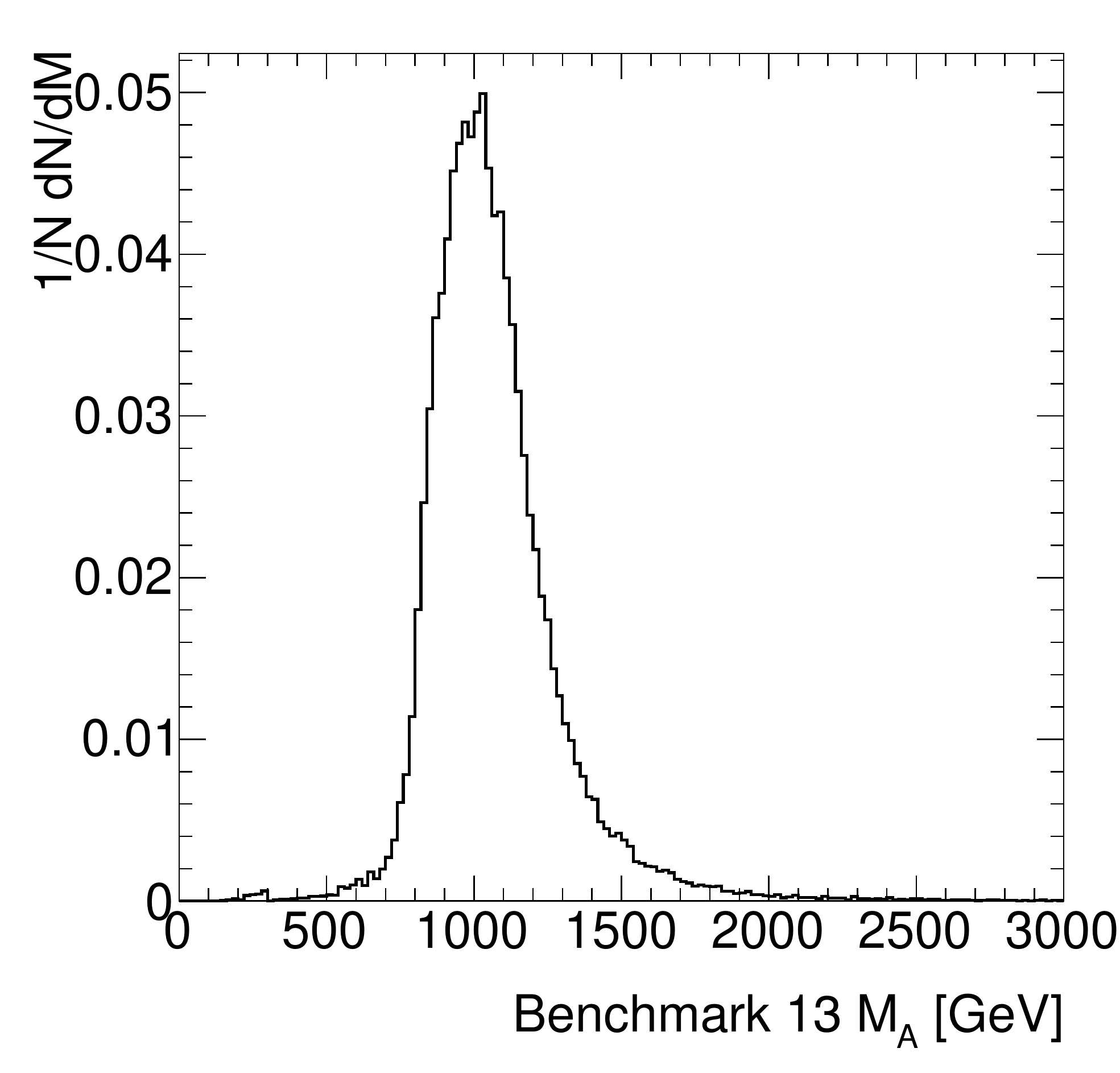} \\
    \end{tabular}
  \end{center}
\caption{Probability density function for $M_A$ corresponding to the ILC 1000 accuracy obtained for the benchmark points 9, 11, and 13.}
\label{fig:mapdf}
\end{figure*}

The $M_A$ values at which more than 90\% (95\%) of the accepted pMSSM
points are excluded by the Higgs couplings at 95\% C.L. are 300
(275)~GeV for the present Run~2 ATLAS data, 675 (525)~GeV and 1075
(925)~GeV for the expected HL-LHC and ILC-1000 accuracies assuming SM
central values. This picture changes when only the pMSSM points not
excluded by the jet/$\ell$+MET searches on the LHC Run~2 data are
considered. These searches remove a significant fraction of points with
relatively light pseudoscalars, as discussed in Section~\ref{sec:4-0}.
These points see different contributions to the Higgs couplings, from
$M_A$ as well as from the lighter SUSY particles, resulting in a spread
of the values of the couplings. Once these points are largely removed,
the dependence of the Higgs couplings on $M_A$ becomes dominant, and the
bounds on $M_A$ are improved. The $M_A$ values at which more than 90\%
(95\%) of the pMSSM points not excluded by the SUSY searches as listed
in Table~\ref{tab:lhc-susy} are excluded by the Higgs couplings at 95\%
C.L. are 350 (325)~GeV for the present Run~2 ATLAS data, 750 (725)~GeV
and 1225 (1165)~GeV for the expected HL-LHC and ILC-1000 accuracies
assuming again SM central values. By the end of the LHC program, if no
deviation from the SM prediction is observed, the Higgs couplings will
probe $M_A$ up to 750~GeV. Owing to its improved accuracy, an $e^+e^-$
Higgs factory is expected to extend this indirect sensitivity up to
heavy Higgs boson masses of $\simeq$1200~GeV (see
Figure~\ref{fig:mapdf}). This reach is comparable to that of the direct
$H$/$A$ searches at the HL-LHC, discussed in Section~\ref{sec:4-0} (see Figure~\ref{fig:malhc}).

\begin{table*}
\caption{pMSSM scenarios adopted in the study of the extraction of $M_A$.} 
\begin{center}
\begin{tabular}{|r|r|r|r|r|r|r|r|r|}
\hline
Scenario & $M_A$ & $\tan \beta$ & $\mu$ & $M_{\tilde{t}_1}$ & $M_{\tilde{\tau}_1}$ & $M_{\tilde{\chi}^0_1}$ & $M_{\tilde{\chi}^{\pm}_1}$ & $M_{\tilde{g}}$ \\
         & (GeV) &              & (GeV) & (GeV)           & (GeV) & (GeV)             & (GeV)     & (GeV) \\   
\hline \hline
1  &  285 &  6.4  &  -12.2 & 4023 & 1546 & 3.3 &  14.9 & 5204  \\
2  &  434 &  5.6  &  -549  & 1493 & 744 & 562 &  564 & 5059  \\
3  &  472 &  6.6  &  1994  & 1652 & 1407 & 315 &  1459 & 390  \\
4  &  510 &  5.5  &  -181  & 3749 & 1999 & 185 &  186 & 4585  \\
5  &  554 &  6.5  &  2351  & 4613  & 2294 & 737 & 2012 & 4379  \\
6  &  606 &  6.3  &  369  & 3688 & 1510 & 380 &  381 & 3571  \\
7  &  649 &  5.7  &  -1411 & 1886 & 1178 & 444 &  445 & 2961  \\
8  &  704 &  5.2  &  480  & 4924 & 865 & 170 &  493 & 3436  \\
9  & 747 &  4.5  &  -3596  & 3483 & 3080 & 1072 &  1073 & 2324  \\
10  & 798 &  5.8  &  -3302  & 2712 & 2258 & 1329 &  3284 & 2085  \\
11  & 844 &  9.1  &  -1679  & 1902 & 2045 & 1695 &  1696 & 4502  \\
12  & 893 &  7.1  &  -368  & 5038 & 2095 & 379 &  381 & 2889  \\
13  & 947 &  7.7  &  -4268  & 2941 & 2276 & 364 &  3677 & 441 \\
14  & 1002 &  5.4  &  716  & 3037 & 827 & 732 &  733 & 3234 \\
15  & 1095 &  12.0  & 1330 & 3068 & 4060 & 1351 & 1352 & 5722 \\
\hline
\end{tabular}
\label{tab:points}
\end{center}
\end{table*}

\begin{figure}[h!]
\begin{center}
\includegraphics[width=0.35\textwidth]{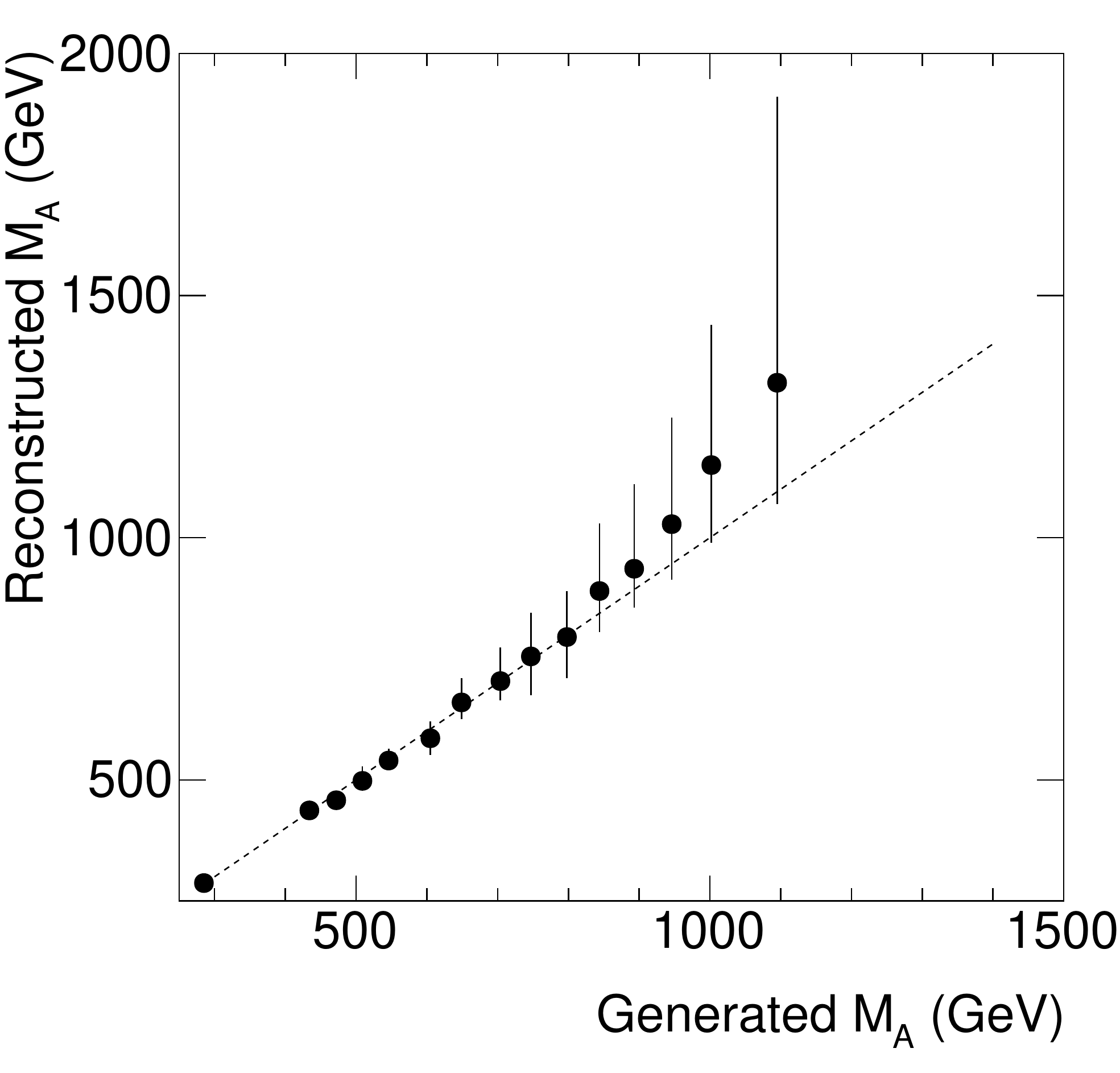}
\end{center}
\caption{Reconstructed most probable value vs.\ generated value of $M_A$ for the study points. The deviation of the reconstructed most probable value from the diagonal at large values of $M_A$ is due to the loss of sensitivity on the lower side of the $M_A$ probability density distribution.}
\label{fig:magenfit}
\end{figure}
As expected, the errors on the reconstructed mass increase with the $M_A$ value.

Further, the accurate determination of the $h$ couplings can be used to estimate $M_A$, in the case the measured values significantly deviate from the SM predictions. We evaluate the accuracy of this estimate as a function of the value of $M_A$ by considering a set of benchmark
scenarios, not excluded by Run~2 data, summarised in Table~\ref{tab:points}. The $h$ couplings and branching fractions are computed for these benchmarks and compared to the value for our accepted pMSSM points in the scan. Each point is assigned a weight defined as the $\chi^2$ probability of the point with the scenario tested, where the $\chi^2$ is computed using the Higgs coupling modifiers with the accuracies for the LHC and the $e^+e^-$ Higgs factories given in Table~\ref{tab:EuStrKappa}. The values of $M_A$ of these weighted points give distributions (some of which are shown in Figure~\ref{fig:mapdf}) from which the central value and uncertainty of the $M_A$ parameters are extracted. The central values are computed as the average in an interval integrating 68\% of the entries around the most probable value. The uncertainties are obtained by determining the interval of parameter values around the central value, which integrates 68\% of the weighted entries allowing for asymmetric ranges. The relation between the generated and reconstructed values of $M_A$ for the chosen benchmark points is shown in Figure~\ref{fig:magenfit}.

\section{Conclusions}
\label{sec:6}

The study of the Higgs boson properties offers compelling perspectives for testing the effects of physics beyond the Standard Model at the LHC and at future colliders.  This is particularly the case in the context of supersymmetric theories and its minimal version, the MSSM. The Higgs couplings to SM particles, both at tree level and through loops, are sensitive to new physics effects and can be used to discriminate the MSSM $h$ from the SM $H$.

In this study, we have reviewed the SUSY corrections to the couplings and decay rates of the SM-like Higgs boson and their dependence on the MSSM parameters. The constraints and predictivity of the Higgs measurements are applied directly
on the relevant supersymmetric parameters using scans of the pMSSM parameter space and contrasted with those derived from direct searches for new particles at the LHC. Theoretical and parametric uncertainties in Higgs production and decay also need to be considered alongside the experimental accuracies. These are revisited and discussed in detail.

The sources of these corrections can be classified in three main categories: the pseudoscalar Higgs mass $M_A$, the invisible decays $h \rightarrow \tilde{\chi}^0 \tilde{\chi}^0$, $\tilde{\nu} \tilde{\nu}$, and the SUSY--QCD corrections generating the $\Delta_b$, $\Delta_t$, and $\Delta_{\tau}$ terms through scalar quarks and gluino or scalar tau and gaugino-Higgsino contributions.

The values of the lightest neutralino mass as well as the Higgsino and gaugino mass parameters, $\mu$ and $M_1, M_2$, that can generate significant rates of invisible decays are already constrained by dark matter direct detection data, even more severely than by the results of the current LHC invisible Higgs decay searches. For the range of $M_A$ values not yet probed by the ATLAS and CMS data or excluded by flavour data, the $\Delta_b$ contribution to the light $h$ boson coupling to $b \bar b$ is largely reduced, by compensation of direct and indirect contributions, and the remaining one-loop SUSY corrections are tiny. The effects of the $\Delta_t$ and $\Delta_{\tau}$ corrections are in general small. This reduces the sensitivity of Higgs couplings and decay rates to MSSM parameters other than $M_A$ as the main MSSM parameter that can be probed in the continuation of the LHC program and at future $e^+e^-$ colliders.

This study has shown that Higgs coupling measurements with the accuracies obtained on the LHC Run~2 data and those expected for the HL-LHC and future $e^+e^-$ colliders can exclude $\sim$ 2\%, 8\%, and 20\%, respectively, of the accepted pMSSM points in our scans, but only $\sim$ 1\%, 5\%, and 12\% of the points that are not yet excluded by flavour data and by the LHC heavy Higgs direct searches, while direct SUSY searches have only a mild impact. The indirect sensitivity to $M_A$ in the pMSSM through the Higgs coupling measurements will evolve from $\sim$~450~GeV for the Run~2 data to $\sim$~800~GeV at HL-LHC and $\sim$1400~GeV at future $e^+e^-$ colliders. Within this range, future $e^+e^-$ colliders of sufficient energy can indirectly determine $M_A$ to a relative accuracy ranging from $\simeq$ 8\% to 40\% for $M_A$ values from 700~GeV to 1.1~TeV, from the deviations of the measured lightest $h$ couplings with respect to their SM expectations.

Thus, large parts of the MSSM parameters are still to be probed and the statement that the properties of the observed Higgs boson are SM-like, often used when discussing the present LHC results, could also be rephrased by saying that they are also MSSM-like.

\section*{Acknowledgements}
We thank G. Robbins for his participation and contributions in the early stage of this work. A.D. is supported by the Estonian Research Council Grants No. MOBTT86 and by the Junta de Andalucia through the Talentia Senior program as well as by Grants No. A-FQM-211-UGR18, No. P18-FR-4314 with ERDF.



\end{document}